\documentclass[]{aa}
\pdfoutput=1
\usepackage{txfonts}%\usepackage[varg]{txfonts}
\usepackage{natbib}
\bibpunct{(}{)}{;}{a}{}{,} % to follow the A&A style
\usepackage{color}
\usepackage{float}
\usepackage{subfigure}
\usepackage{hyperref}
\usepackage{longtable}
\usepackage{rotating}
\usepackage{lscape}
\usepackage{soul}

\newcommand{\udft}{\textsf{udf-10}}
\newcommand{\mosaic}{\textsf{mosaic}}

\newcommand{\lya}{\mbox{Ly$\alpha$}}

\def\HI{\mbox{H\,{\sc i}}}
\newcommand{\OII}{[O\,{\sc ii}]}
\newcommand{\OIII}{[O\,{\sc iii}]}
\begin{document}

   \title{The MUSE Hubble Ultra Deep Field Survey}
    \subtitle{XI\hspace{-.1em}V. The evolution of the Ly$\alpha$ emitter fraction from $z=3$ to $z=6$}

   \author{Haruka Kusakabe\inst{\ref{inst1}\thanks{e-mail: haruka.kusakabe@unige.ch}}
   \and J{\'e}r{\'e}my Blaizot\inst{\ref{inst2}}%OK
   \and Thibault Garel\inst{\ref{inst1},\ref{inst2}}
   \and Anne Verhamme\inst{\ref{inst1},\ref{inst2}}
   \and Roland Bacon\inst{\ref{inst2}}
   \and Johan Richard\inst{\ref{inst2}}
   \and Takuya Hashimoto\inst{\ref{inst3}}  
   \and Hanae Inami\inst{\ref{inst4}}
   \and Simon Conseil\inst{\ref{inst2},\ref{inst_simon}}
   \and Bruno Guiderdoni\inst{\ref{inst2}}
   \and Alyssa B. Drake\inst{\ref{inst5}} 
   \and Edmund Christian Herenz \inst{\ref{inst6}} 
   \and Joop Schaye \inst{\ref{inst7}} 
   \and Pascal Oesch \inst{\ref{inst1}} 
   \and Jorryt Matthee\inst{\ref{inst8}}%OK
   \and Raffaella Anna Marino\inst{\ref{inst8}}
   \and Kasper Borello Schmidt \inst{\ref{inst9}}%OK
   \and Roser Pell\'{o} \inst{\ref{inst10}}
   \and Michael Maseda \inst{\ref{inst7}}
   \and Floriane Leclercq \inst{\ref{inst1},\ref{inst2}}
   \and Josephine Kerutt \inst{\ref{inst1}, \ref{inst9}}   
   \and Guillaume Mahler \inst{\ref{inst11}}%[0000-0003-3266-2001]
   }

   %\fnmsep\thanks{Just to show the usage  of the elements in the author field}

\institute{Observatoire de Gen\`{e}ve, Universit\'e de Gen\`{e}ve, 51 chemin de P\'egase, 1290 Versoix, Switzerland\label{inst1}
\and%2
Univ. Lyon, Univ. Lyon1, Ens de Lyon, CNRS, Centre de Recherche Astrophysique de Lyon UMR5574, F-69230, Saint-Genis-Laval, France
\label{inst2}
\and%3
Tomonaga Center for the History of the Universe (TCHoU), Faculty of Pure and Applied Sciences, University of Tsukuba,  Tsukuba, Ibaraki 305-8571, Japan \label{inst3} 
\and%4
Hiroshima Astrophysical Science Center, Hiroshima University 1-3-1 Kagamiyama, Higashi-Hiroshima, Hiroshima 739-8526, Japan,\label{inst4}
\and
Gemini Observatory/NSF’s OIR Lab, Casilla 603, La Serena, Chile,\label{inst_simon}
\and%5
Max-Planck-Institut f\"{u}r Astronomie, K\"{o}nigstuhl 17, D-69117 Heidelberg, Germany\label{inst5}
\and%6
ESO Vitacura, Alonso de C{\'o}rdova 3107,Vitacura, Casilla 19001, Santiago de Chile, Chile \label{inst6}
\and%7
Leiden Observatory, Leiden University, PO Box 9513, 2300 RA Leiden, The Netherlands \label{inst7}
\and%8
Department of Physics, ETH Z\"urich, Wolfgang-Pauli-Strasse 27, 8093 
Z\"urich, Switzerland\label{inst8}
\and%9
Leibniz-Institut f\"{u}r Astrophysik Potsdam, An der Sternwarte 16 
14482 Potsdam, Germany\label{inst9}
\and%10
Aix Marseille Universit\'e, CNRS, CNES, LAM (Laboratoire d’Astrophysique de Marseille), UMR 7326, 13388, Marseille, France\label{inst10}
\and%11
Department of Astronomy, University of Michigan, 1085 South University Ave, Ann Arbor, MI 48109, USA \label{inst11}
}
%\thanks{The university of heaven temporarily does not       accept e-mails}}
\date{Received December 18, 2019; accepted March 26, 2020}% \date{Received September 15, 1996; accepted March 16, 1997}

% \abstract{}{}{}{}{} 
% 5 {} token are mandatory
 
  \abstract
   {The Ly$\alpha$ emitter (LAE) fraction, $X_{\rm LAE}$, is a potentially powerful probe of the evolution of the intergalactic neutral hydrogen gas fraction. However, uncertainties in the measurement of $X_{\rm LAE}$ are still debated.
  }
   {Thanks to deep data obtained with the integral field spectrograph MUSE (Multi-Unit Spectroscopic Explorer), we can measure the evolution of the LAE fraction homogeneously over a wide redshift range of $z\approx3$--$6$ for \emph{UV-faint} galaxies (down to UV magnitudes of $M_{1500}\approx-17.75$). This is significantly fainter than in former studies ($M_{1500}\leq-18.75$), and allows us to probe the bulk of the population of high-redshift star-forming galaxies. }
   {We construct a UV-complete photometric-redshift sample following UV luminosity functions and measure the Ly$\alpha$ emission with MUSE using the latest (second) data release from the MUSE Hubble Ultra Deep Field Survey. }
   {   
   We derive the redshift evolution of $X_{\rm LAE}$ for $M_{1500}\in [-21.75;-17.75]$ for the first time with a equivalent width range $EW(\rm Ly\alpha)\geq65$ \AA\ and find low values of $X_{\rm LAE}\lesssim30$\% at $z\lesssim6$. The best fit linear relation is $X_{\rm LAE} = 0.07^{+0.06}_{-0.03}z -0.22^{+0.12}_{-0.24}$. For $M_{1500}\in [-20.25;-18.75]$ and $EW(\rm Ly\alpha)\geq25$ \AA, our $X_{\rm LAE}$ values are consistent with those in the literature within $1\sigma$ at $z\lesssim5$, but our median values are systematically lower than reported values over the whole redshift range. In addition, we do not find a significant dependence of $X_{\rm LAE}$ on $M_{1500}$ for $EW(\rm Ly\alpha)\geq50$ \AA\ at $z\approx3$--$4$, in contrast with previous work. The differences in $X_{\rm LAE}$ mainly arise from selection biases for Lyman Break Galaxies (LBGs) in the literature: UV-faint LBGs are more easily selected if they have strong Ly$\alpha$ emission, hence $X_{\rm LAE}$ is biased towards higher values when those samples are used.       }
   {Our results suggest either a lower increase of $X_{\rm LAE}$ towards $z\approx6$ than previously suggested, or even a turnover of $X_{\rm LAE}$ at $z\approx5.5$, which may be the signature of a late or patchy reionization process. We compared our results with predictions from a cosmological galaxy evolution model. We find that a model with a bursty star formation (SF) can reproduce our observed LAE fractions much better than models where SF is a smooth function of time.} 
  
   \keywords{Cosmology:reionization, observations, early Universe, - Galaxies: high-redshift, evolution, intergalactic medium }
   \maketitle
%
%________________________________________________________________

\section{Introduction}\label{sec:intro}

In the early Universe, the first objects formed and filled the Universe with light. They ionized the neutral gas in the intergalactic medium (IGM) via a phenomenon called ``cosmic reionization''. One of the candidates for the main source of reionization is star-forming galaxies, whose ionizing radiation, called ``Lyman Continuum'’ (LyC, $\lambda<912$ \AA), emitted from massive stars, is expected to leak into the IGM \citep[e.g.][]{Bouwens2015a,Bouwens2015b,Finkelstein2015a, Robertson2015, Livermore2017}. Another candidate is active galactic nuclei \citep[AGNs, e.g.][]{Madau2015}. However they have recently been reported to contribute less than $\approx10$\% of the ionizing photons needed to keep the IGM ionized \citep[over a UV magnitude range of $-18$ to $-30$ mag; ][see also \citealt{Parsa2018}]{Matsuoka2018}. Previous studies using the Gunn-Peterson absorption trough seen in quasar spectra \citep[e.g.][see however, \citealt{Bosman2018}]{Gunn1965, Fan2006, McGreer2015} and in gamma-ray burst spectra \citep[e.g.][]{Totani2006,Totani2014} suggest that cosmic reionization was completed by $z\approx6$. 
The Thomson optical depth of the cosmic microwave background measured by Planck suggests that the midpoint redshift of reionization (i.e. when half the IGM had been reionized) is at $z\approx7.7\pm0.7$ \citep[1$\sigma$ confidence interval,][]{PlanckCollaboration6_2018arXiv}.

Ly$\alpha$ emission is intrinsically the strongest UV spectral feature of young star forming galaxies, and galaxies with mostly detectable Ly$\alpha$ emission or with Ly$\alpha$ equivalent widths higher than $\approx25$ \AA\  are called ``Ly$\alpha$ emitters (LAEs)''. Ly$\alpha$ emission is  scattered by neutral hydrogen gas (\HI) in the IGM, and, therefore, the detectability of LAEs is affected by the \HI\ gas fraction in the IGM. The redshift evolution of Ly$\alpha$ luminosity functions has thus been used to investigate the history of the neutral hydrogen gas fraction of the IGM \citep[e.g.][]{Malhotra2004,Kashikawa2006,Hu2010,Ouchi2010, Santos2016,  Drake2017b, Ota2017, Zheng2017, Konno2018,Itoh2018}. Ly$\alpha$ luminosity functions can be used to compute the evolution of the Ly$\alpha$ luminosity density, and its rapid decline at $z\gtrsim5.7$ compared with that of the cosmic star formation rate density derived from UV luminosity functions is interpreted to be caused by IGM absorption \citep[e.g.][]{Ouchi2010, Konno2018}.

Similarly, the fraction of LAEs among UV selected galaxies, $X_{\rm LAE}$, can also be used to probe the evolution of the \HI\ gas fraction of the IGM  \citep[e.g.][]{Fontana2010, Pentericci2011, Stark2011, Ota2012, Treu2013, Caruana2014, Faisst2014, Schenker2014}. $X_{\rm LAE}$ has been reported to increase from $z\approx3$ to $6$ and then to drop at $z>6$. This has again been interpreted as a signature of the IGM becoming more neutral at $z>6$ \citep[e.g.][]{Dijkstra2011,Jensen2013, Mason2018}.
The LAE fraction is complementary to the test of Ly$\alpha$ luminosity functions (LFs) and has some advantages:
efficient spectroscopic observations as a follow-up of continuum-selected galaxies, which is insensitive to the declining number density of star forming galaxies, and rich information obtained from the spectra such as spectroscopic redshifts and kinematics of the interstellar medium \citep[e.g.][]{Stark2010, Hashimoto2015}. It also enables us to solve the degeneracy between the Ly$\alpha$ escape fraction among star forming galaxies with different UV magnitudes and the comparison between luminosity densities of Ly$\alpha$ emission and UV continuum, which are obtained from the integration of UV and Ly$\alpha$ LFs. In addition, recently, \citet{Kakiichi2016} suggested that the UV magnitude-dependent evolution of the LAE fraction combined with the Ly$\alpha$ luminosity function can be used to constrain the ionization topology of the IGM and the history of reionization. 

Using $X_{\rm LAE}$ to set quantitative constraints on the evolution of the neutral content of the IGM remains challenging. In particular, we need to understand whether observed variations of $X_{\rm LAE}$ are exclusively due to variations in the IGM properties, or whether they can be attributed to galaxy evolution. Following the Ly$\alpha$ spectroscopic observations of Lyman break galaxies (LBGs) at $z\approx3$ by \citet{Steidel2000} and \citet{Shapley2003}, \citet{Stark2010, Stark2011} have found that $X_{\rm LAE}$ among LBGs evolves with redshift and depends on the rest-frame Ly$\alpha$ equivalent width ($EW(\rm Ly\alpha)$) cut. They also show that $X_{\rm LAE}$ depends on the absolute rest-frame UV magnitude ($M_{1500}$), so that UV-faint galaxies are more likely to show Ly$\alpha$ than UV-bright galaxies \citep[see also][]{Schaerer2011a,Forero-Romero2012, Garel2012}.  One conclusion from these studies is that the evolution of $X_{\rm LAE}$ with redshift is more prominent for UV-faint galaxies and low $EW(\rm Ly\alpha)$ cuts.

However, several recent studies show lower values of $X_{\rm LAE}$ for UV-faint galaxies ($-20.25< M_{1500}<-18.75$ mag) than those in the pioneering work of \citet{Stark2011}. At $z\approx4$ and $z\approx5$, \citet{ArrabalHaro2018} show more than $1\sigma$ lower $X_{\rm LAE}$ for the faint $M_{1500}$ and low EW cut ($25$ \AA), though their result at $z\approx6$ is consistent with that in \citet{Stark2011}. \citet{DeBarros2017} also investigate $X_{\rm LAE}$ for UV-faint galaxies with a low EW cut, at $z\approx6$. They obtain a low median value of $X_{\rm LAE}$, which is even slightly lower than the value previously found at $z\approx5$, though their $X_{\rm LAE}$ is consistent within $1\sigma$. They conclude that the drop at $z>6$ is less dramatic than previously found \citep[see also][for their recent study at $z\approx7$]{Pentericci2018}. \citet{DeBarros2017} and \citet{Pentericci2018} also suggest the possibility that the effect of an increase of the \HI\ gas fraction in the IGM is observed between $5 < z < 6$. This would be consistent with a later and more inhomogeneous reionization process than previously thought, as has also been recently suggested by observations and simulations of fluctuations in Ly$\alpha$ forest \citep[e.g.,][]{Bosman2018, Kulkarni2019,Keating2019}.
The parent LBG sample in \citet{DeBarros2017} is selected with an additional UV magnitude cut on a normal LBG selection, while the parent sample in \citet{ArrabalHaro2018} is mostly based on photometric redshift (photo-z) even though it is regarded as an LBG sample in their paper. Therefore, the results of $X_{\rm LAE}$ for the faint $M_{1500}$ are not yet conclusive, possibly due to different parent sample selections. Moreover, for the UV-bright galaxies, the redshift evolution of $X_{\rm LAE}$ for a $25$ \AA\ EW cut has not been confirmed yet  \citep[e.g.][]{Stark2011,Curtis-Lake2012,Ono2012,Schenker2014,Stark2017,Cassata2015,Mason2019}. \citet{DeBarros2017} and \citet{Pentericci2018} suggest that some previous results are affected by an LBG selection bias. As strong Ly$\alpha$ emission affects the red band, strong LAEs can be selected more easily compared to galaxies without Ly$\alpha$ emission at faint UV magnitudes. It results in a high LAE fraction of LBGs \citep[see also][]{Stanway2008,Inami2017}. \citet{ArrabalHaro2018} assess UV completeness of their parent sample using UV luminosity functions and find that their 90\% completeness magnitude is $\approx-20$ and $-21$ mag at $z\approx4$ and $z\approx5$, respectively.

To summarize, it is important to obtain a firm conclusion about the evolution of $X_{\rm LAE}$ in the post-reionization epoch in order to quantify the drop of $X_{\rm LAE}$ at $z > 6$ and to assess the reliability of using $X_{\rm LAE}$ as a good probe of reionization. However, although there are a number of observational studies of $X_{\rm LAE}$, uncertainties in the measurement and interpretation of $X_{\rm LAE}$ are still a matter of debate \citep[e.g.][]{Stark2011, Garel2015, DeBarros2017, Caruana2018, Mason2018, Hoag2019aApJ, Hoag2019bMNRAS}. One of the biggest problems is the LBG selection bias due to the different depths of selected bands in previous studies. It is worth pointing out that none of the previous studies were based on complete parent samples of UV faint galaxies ($-20.25< M_{1500}<-18.75$ mag). Completeness in terms of UV magnitudes, as well as homogeneously selected samples over a wide redshift range are essential for the determination of $X_{\rm LAE}$. In addition, we also need deep and homogeneous spectroscopic observations of Ly$\alpha$ emission over a wide redshift range. 

In this study, we use a combination of deep Hubble Space Telescope (HST) and Very Large Telescope (VLT)/Multi-Unit Spectroscopic Explorer (MUSE) data, to overcome these limitations and improve our knowledge of the evolution with redshift of $X_{\rm LAE}$ among a homogeneous parent sample of UV faint galaxies. We use HST bands that are not contaminated by Ly$\alpha$ emission to measure UV magnitudes to avoid a selection bias. Deep and homogeneous spectroscopic Ly$\alpha$ observations at a wide redshift range have been achieved in the {\it Hubble} Ultra Deep field (HUDF) by VLT/MUSE \citep[][]{Bacon2010} in the guaranteed time observations (GTO), MUSE-HUDF survey \citep[e.g.][]{Bacon2017}. The LAE fraction has already been investigated with MUSE and HST data, using the MUSE-Wide GTO survey in the Cosmic Assembly Near-infrared Deep Extragalactic Legacy Survey (CANDELS) Deep region in \citet{Caruana2018}. Their sample are constructed with an apparent magnitude cut of $F775W \leq26.5$ mag for an HST catalog, which is roughly converted to $M_{\rm 1500}\approx-19$--$-20$ mag at $z\approx3$--$6$. However, the MUSE HUDF data enable us to measure faint Ly$\alpha$ emission even for faint UV sources ($-17.75$ mag) in existing HUDF photometric catalogs.

The paper is organized as follows. In Sect. 2, we describe the data, methods, and samples: our UV-selected samples, the MUSE data, the detection and measurement of Ly$\alpha$ emission, the calculation of the LAE fraction, and its uncertainties. Sect. 3 presents the LAE fraction as a function of redshift and UV magnitude. In Sect. 4, we discuss our results: the differences in LAE fraction from previous results, a comparison with predictions from a model of galaxy formation, and implications for reionization. Finally, the summary and conclusions are given in Sect. 5. 

Throughout this paper, we assume a flat cosmological model with a matter density of $\Omega_{\rm m} = 0.3$, a cosmological constant of $\Omega_{\Lambda} = 0.7$, and a Hubble constant of $H_0 = 70$ km s$^{-1}$ Mpc$^{-1}$ ($h_{100} = 0.7$). Magnitudes are given in the AB system \citep{Oke1983}.

\section{Data, methods, and samples}\label{sec:data_sample}
In Sect. \ref{subsec:parent_sample}, we first discuss the construction of a volume-limited UV-selected sample of galaxies from the HUDF catalog of \citet[][hereafter R15]{Rafelski2015}\footnote{\url{https://archive.stsci.edu/hlsp/uvudf}}. In Sect. \ref{subsec:CountingLAEs} we explain how we use MUSE data to detect and measure Ly$\alpha$ emission from galaxies of our UV sample. In Sect. \ref{subsec:uncertainties} we lay out our calculations of $X_{\rm LAE}$ and discuss the error budget. In Sect. \ref{subsec:slope} we present our measurement of slopes of $X_{\rm LAE}$ as a function of $z$ and $M_{1500}$. We discuss briefly in Sect. \ref{subsec:haloDiscussion} the effects of extended Ly$\alpha$ emission. 

\subsection{UV-selected samples}\label{subsec:parent_sample}

We build our parent sample of high-redshift UV-selected galaxies using the latest HUDF catalog from R15. Sources in this catalog are detected in the average-stacked image of eight HST bands: four optical bands from $ACS/WFC$ ($F435W$, $F606W$, $F775W$, and $F850LP$), and four near infrared (NIR) bands from $WFC3/IR$ ($F105W$, $F125W$, $F140W$, and $F160W$). In total, out of the $9969$ sources in R15's catalog,$1095$ and $7904$ objects are within the footprints of the \udft{} and \mosaic{} regions of the MUSE HUDF Survey\footnote{The survey consists of two layers of different depths: a shallower area with $9$ MUSE pointings (\mosaic) and a deeper area with $1$ MUSE pointing (\udft) within the \mosaic. More details are given in Sect. \ref{subsec:muse}.} \citep[][]{Bacon2017}, respectively (the duplicated region in \udft{} is removed). We note that the $F140W$ image only covers $6.8$ arcmin$^2$ of the $11.4$ arcmin$^2$ footprint of the R15's catalog. The $F140W$ photometry is only used when it is available in R15. As discussed in footnote 2 of \citet{Hashimoto2017b}, the lack of $F140W$ may affect the detection of sources in R15's catalog. Moreover, the footprint of $F140W$ is also covered by deeper NIR images ($F105W$, $F125W$, and $F160W$). Indeed, the fraction of sources identified by R15 at $z\gtrsim6$ within the footprint of the $F140W$ image is higher than where there is no $F140W$ data.

In order to avoid contamination from neighboring objects, we follow \citet{Inami2017} and discard all HST sources which have at least one neighbour within $0\farcs6$. Such associations cannot be resolved in our MUSE observations where the full width at half maximum (FWHM) of the average seeing of DR1 data is $\approx0\farcs6$ at $7750$ \AA{}. This procedure excludes $\approx 20\%$ of the sources. We assume that this does not result in a significant bias as it is effectively only decreasing the survey area. This assumption is true if interacting systems are not more often LAEs than isolated systems.  

R15 provide photometric redshifts and associated errors for all objects. These are obtained via spectral energy distribution (SED) fitting of photometric data in $11$ HST bands using either the Bayesian Photometric Redshift (BPZ) algorithm \citep{Benitez2000,Benitez2004,Coe2006} or the EAZY software \citep{Brammer2008}. In the present paper, we choose to use the results from BPZ because they are found to be more accurate \citep[][]{Rafelski2015,Inami2017,Brinchmann2017}. 
 Note that R15 do not include Spitzer/IRAC data in their SED fitting. We state in Appendix \ref{ap:irac} that this addition would not improve the photometric redshifts of the faint galaxies studied here. Below, $z_{\rm p}$ denotes the photometric redshift given in R15\footnote{The photometric redshift value is the one which maximizes the likelihood estimate in BPZ.}, and we use it to define redshift selections of all our UV samples (see Table \ref{tbl:criteria}). In R15's catalog, the $95$\% lower and upper limits of $z_{\rm p}$ are provided as uncertainties on $z_{\rm p}$. We use these to construct an {\it inclusive parent sample}, that we will use for Ly$\alpha$ searches. This sample includes all sources with photometric redshift estimates ($95$\% confidence interval) overlapping with the redshift range $(2.91$-- $6.12)$ where Ly$\alpha$ can be observed with MUSE. Note that we remove sources at $z_{\rm} > 6.12$ from our sample because the parent photometric-redshift sample may be affected by selection bias (see Sect. \ref{subsec:parent_sample}), and because Ly$\alpha$ detectability in MUSE spectra is strongly reduced by sky lines \citep[][]{Drake2017b}. In the end, this inclusive parent sample consists of $3233$ and $402$ sources in the \mosaic{} and \udft, respectively (without duplication). 

 We derive the absolute rest-frame UV magnitude using two or three HST photometric points to fit a power-law to the UV continuum. The power-law describes the spectral flux density $f_\nu$ as  $f_{\nu}= f_{0} (\lambda/\lambda_0)^{\beta+2}$, where $\lambda_0 = 1500$ \AA, $f_0$ is the spectral flux density at $1500$ \AA\ (in $\mathrm{erg\ s^{-1}\ cm^{-2}\ Hz^{-1}}$), and $\beta$ is the continuum slope. We then simply have $M_{\rm 1500}=-2.5\log{(f_0)}-48.6 -5\log{(d_{\rm L}/10)} + 2.5\log{(1 + z_{\rm p})}$, where $d_{\rm L}$ (pc) is the luminosity distance. We choose the HST filters following \citet{Hashimoto2017b} so that Ly$\alpha$ emission and IGM absorption are not included in the photometry: we use HST/$F775W$, $F850LP$, and $F105W$ for objects at $2.91\leq{z_{p}}\leq4.44$; $F105W$, $F125W$, and $F140W$ for objects at $4.44< z_{p} \leq5.58$; and $F125W$, $F140W$, and $F160W$ for objects at $5.58< z_{p} \leq 6.12$. While \citet{Hashimoto2017b} use the MUSE spectroscopic redshifts, we use  the values of the photometric redshifts from \citet{Rafelski2015}. Our derived $M_{\rm 1500}$ values are consistent with those of \citet{Hashimoto2017b} for the sources which we have in common (LAEs). The standard deviation of the relative difference in $M_{\rm 1500}$ for the sources included in both studies is $\approx3$\% without a systematic offset.

\begin{figure}
 \begin{flushright}
   \includegraphics[width=9.3cm]{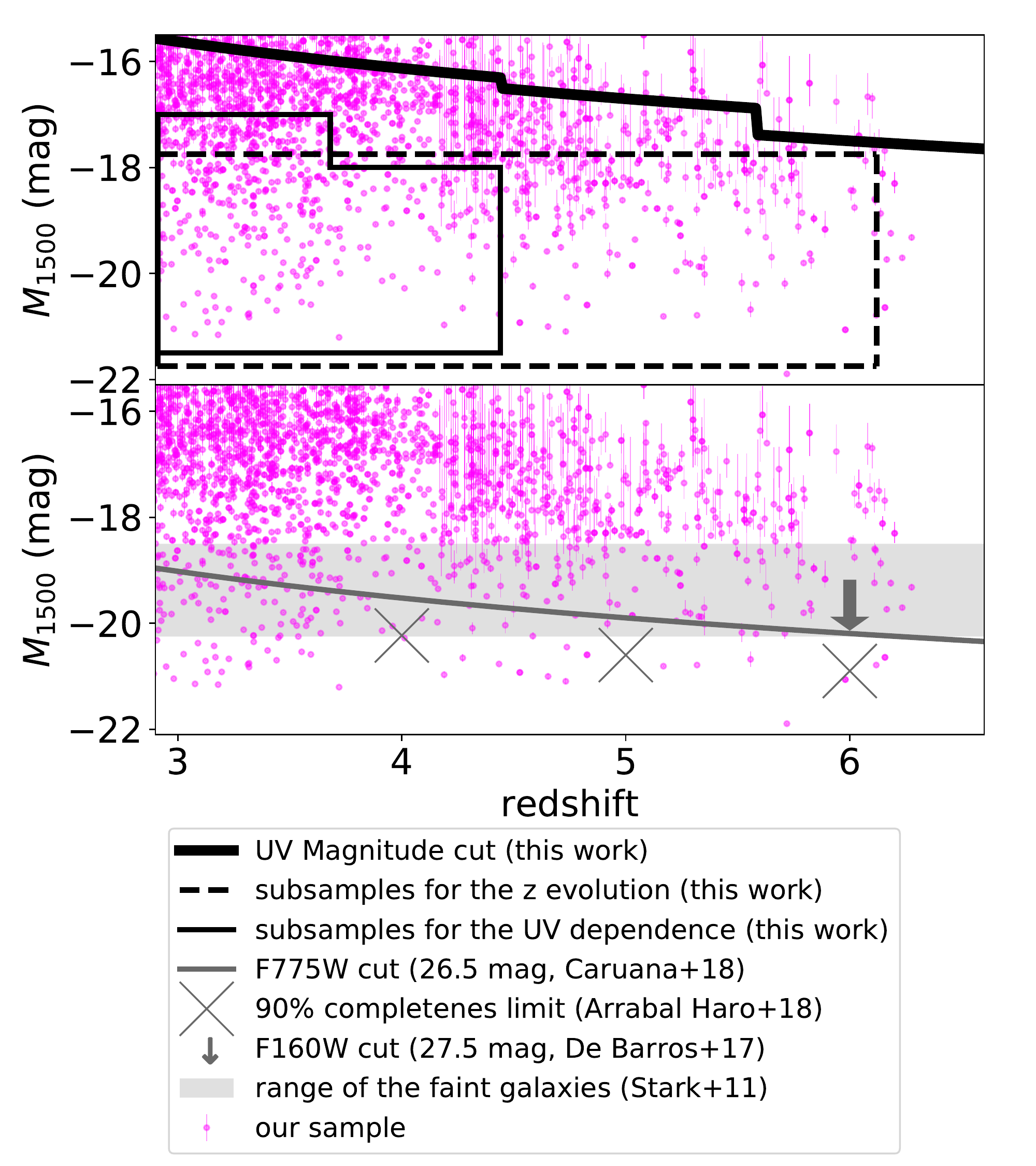}\\
      \caption[$M_{1500}$ versus $z_{p}$ for our sample.]{
            $M_{1500}$ versus $z_{p}$ for our sample and the literature. The $M_{1500}$ and $z_{p}$ of our parent sample from \citet{Rafelski2015} are shown by magenta filed circles (identical in the two panels).  In the upper panel, the $M_{1500}$ cut ($N_{1500}^{\rm lim}$) for our sample defined from the rest-frame UV HST bands is indicated by a black thick solid line. The parameter space studied here, $X_{\rm LAE}$ vs. $M_{\rm 1500}$ at $z\approx3$--$4$ and $X_{\rm LAE}$ vs. $z$ at $z\approx3$--$6$, is shown by a black solid polygon and a black dashed rectangle, respectively. In the lower panel, the UV magnitude cut of $F160W \leq27.5$ mag at $z\approx6$ in \citet[][]{DeBarros2017} and UV magnitudes corresponding to $\approx90$\% completeness at $z\approx4$, $5$, and $6$ in \citet[][]{ArrabalHaro2018} are represented by a dark-grey arrow and dark-grey crosses, respectively. The apparent magnitude cut of $F775W \leq26.5$ mag in \citet{Caruana2018} is shown by a grey solid line. The parameter space for the faint galaxies studied in \citet{Stark2011} is indicated by a light-grey shaded region. \label{fig:muv_z} 
            }
\end{flushright}
 \end{figure}

 In Figure\,\ref{fig:muv_z}, we show the distribution of $M_{\rm 1500}$ as a function of $z_{\rm p}$ for sources in our parent sample that have $2.91<z_{\rm p}\leq6.12$. In order to construct a complete parent sample in terms of $M_{\rm 1500}$, we define a limiting magnitude $M_{1500}^{\rm lim}$ so that objects brighter than $M_{1500}^{\rm lim}$ are detected with a signal-to-noise ratio ($S/N$) larger than two in at least two HST bands among the rest-frame UV HST bands. To compute $M_{1500}^{\rm lim}$, we again use a power-law continuum model, this time with a fixed UV slope $\beta=-2$, which is commonly used as a fixed value \citep[e.g.,][]{Caruana2018}. At each redshift, we derive the normalisation $M_{\rm 1500}^{\rm lim}$ such that the flux can be detected in at least two HST UV bands with $S/N>2$. The resulting $M_{\rm 1500}^{\rm lim}$ is shown with the thick black curve over-plotted to the data in magenta in the upper panel of Figure\,~\ref{fig:muv_z}. From this panel, we see that we can
build a complete UV-selected sample at redshifts $z\approx3$--$4$ down to $M_{\rm 1500}^{\rm lim}\approx-16.3$ mag, and even at $z\approx6$ we can achieve completeness down to $M_{\rm 1500}^{\rm lim}\approx-17.7$ mag. 

In the upper panel of Figure\,\ref{fig:muv_z}, we highlight two regions of parameter space that we select to do two complementary studies: $X_{\rm LAE}$ vs. $M_{\rm 1500}$ with $z$ in the range $[2.91;4.44]$ (the polygon marked with a solid black line) and $X_{\rm LAE}$ vs. $z$ with $z$ in the range $[2.91;6.12]$ (the dashed-line rectangle). To define the criteria for $X_{\rm LAE}$ vs. $M_{\rm 1500}$ plots, completeness simulations for Ly$\alpha$ emission (see Sect. \ref{subsec:compsimu}) and $z_{\rm p}$ bins for the $M_{1500}^{\rm lim}$ calculation are also taken into account. In the lower panel of Figure\,\ref{fig:muv_z}, we show for comparison the locus of previous studies in the $(M_{\rm 1500}-z)$ plane. The faint galaxies in \citet{Stark2011} are shown by the light-grey shaded area. \citet{DeBarros2017} use a UV magnitude cut of $F160W \leq27.5$ mag at $z\approx6$ in their sample, which is shown with a dark-grey arrow. The recent sample of \citet{ArrabalHaro2018} is shown with dark-grey crosses which indicate the UV magnitudes at which they reach $\approx90$\% completeness. The LAE fraction with the MUSE-Wide GTO survey data \citep{Caruana2018} adopt an apparent magnitude cut of $F775W \leq26.5$ mag for an HST catalog \citep[][]{Guo2013}, which we roughly convert to $M_{\rm 1500}$ for illustrative purposes, and show with the solid grey line in the lower panel of Figure\,\ref{fig:muv_z}.  Our MUSE-Deep data combined with the HST catalog from R15 allow us to probe deeper than all previous work, and to extend our study to UV-faint galaxies (i.e., $M_{\rm 1500}\geq-18.75$ mag).

\begin{figure}
   \sidecaption
   \includegraphics[width=9.3cm]{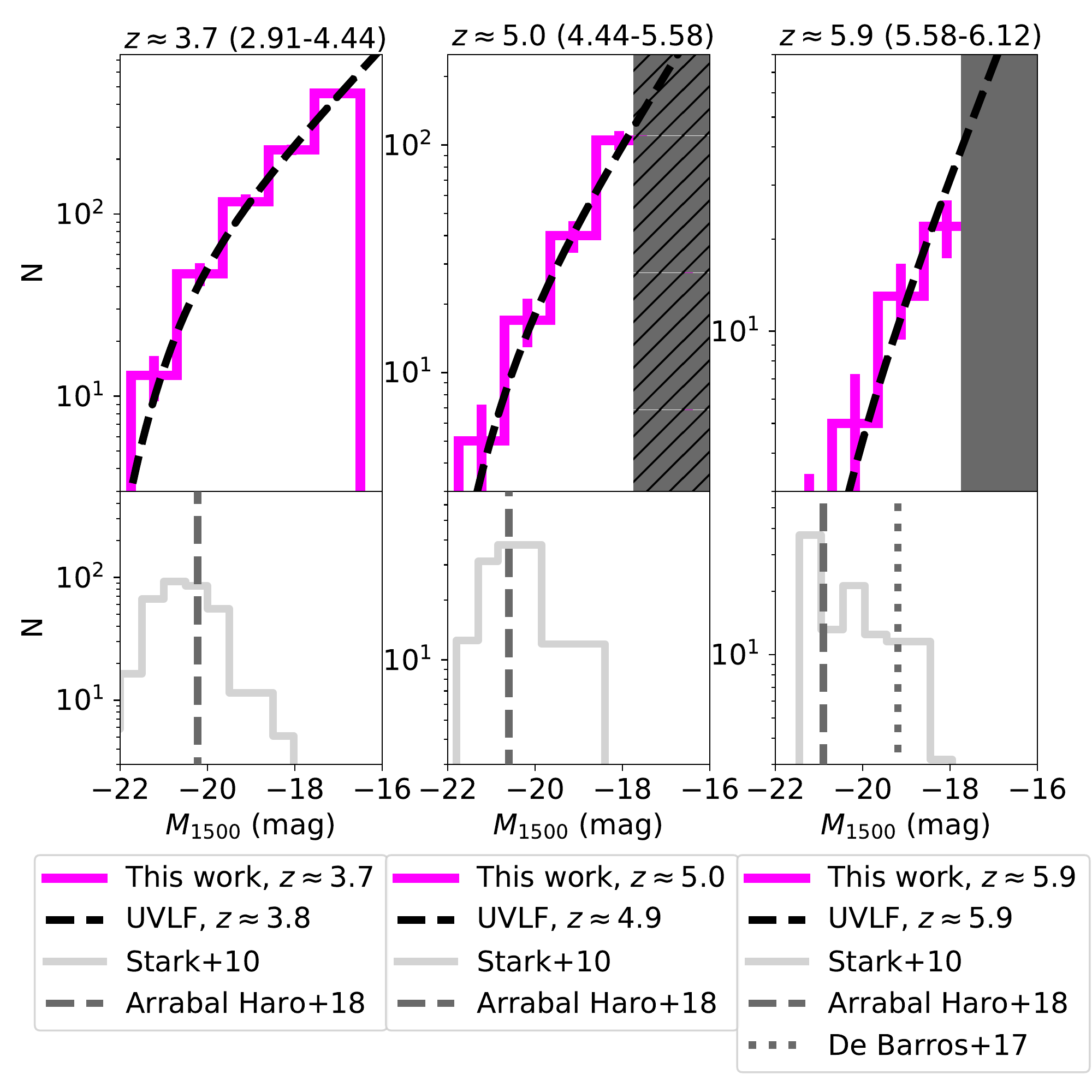}\\
      \caption{Histograms of $M_{1500}$ for our parent samples at $z\approx3.7$ (=$2.91$--$4.44$, left), $5.0$ ($=4.44$--$5.58$, middle), and $5.9$ ($=5.58$--$6.12$, right). In the upper panels, magenta histograms and black dashed lines represent the number distribution of our parent sample and that expected from the UV luminosity functions (UVLF) in \citet{Bouwens2015b} for the same effective survey area, respectively. The uncertainty of the number distribution of our parent sample is given by the Poisson error. Grey hashed areas indicate $M_{1500}$ ranges that are not used in this work. In the lower panels, light-grey histograms shows the number distribution in \citet{Stark2010} at $z\approx3.75$, $z\approx5.0$ and $z\approx6.0$. Dark-grey dashed and dotted lines indicate the $M_{1500}$ for $90$\% completeness at $z\approx4$, $5$, and $6$ in \citet{ArrabalHaro2018} and the magnitude cut at $z\approx6$ in \citet{DeBarros2017}, respectively. 
       \label{fig:n_muv}       }
         
   \end{figure}

In Figure\,\ref{fig:n_muv}, we demonstrate the completeness of our UV-selected sample at different redshifts by comparing our UV number counts to what we would expect from the UV luminosity function (UVLF) of \citet{Bouwens2015b}. We find that the distribution of $M_{\rm 1500}$ for our samples (magenta) follow well those expected from the UVLFs for the same survey area at similar redshifts (black solid lines) within 1$\sigma$ error bars. For comparison, we also show in the bottom panels of Figure\,\ref{fig:n_muv} the distribution of magnitudes of the sample of \citet{Stark2010}. Clearly, their samples are incomplete even at relatively bright magnitudes ($\approx -20.25$ mag), most likely because of the shallow depth of their data and the LBG selection. Note that the LBG samples in \citet{Stark2010,Stark2011} consist of two LBG samples. One is from their own sample \citep{Stark2009}, and the other is a sample from the literature, which is biased towards bright objects including few B and V dropouts with magnitudes fainter than 26 mag in $F850LP$ ($z$) according to \citet{Stark2010}. Generally, the $M_{\rm 1500}$ ranges used in LAE fraction studies in the literature are close to those in \citet{Stark2010, Stark2011}. Recently, \citet{ArrabalHaro2018} tested the UVLF of their LBG sample (mainly constructed from photo-$z$ samples) used in their LAE fraction study. At $z\approx4$ and $z\approx5$, their LBG samples follow the UVLF in \citet{Bouwens2015b} at $M_{\rm 1500}\lesssim-20.5$ mag (dark-grey dashed lines in Figure\,\ref{fig:n_muv}). However, it is still not complete in the UV for the faint sample ($-20.25\leq M_{\rm 1500}\leq-18.75$ mag) just as \citet{Stark2011}.These comparisons illustrate the methodological improvement of our study in terms of the $M_{\rm 1500}$ completeness of the sample of galaxies for which we estimate the LAE fraction.

For future use, $N_{\rm 1500}(z_{\rm p},\, M_{1500})$ denotes the number of galaxies with a photometric redshift $z_{\rm p}$ and absolute magnitude $M_{1500}$ within the given ranges (see Table \ref{tbl:criteria} which summarises different samples). As discussed above, our UV-selected samples are volume-limited and $N_{\rm 1500}$ is directly measured from the catalog with no need for incompleteness corrections. 

%====
%====

\begin{table*}
\caption{Subsample criteria.}\label{tbl:criteria}
\centering
\begin{tabular}{ccccc}
\hline         
\noalign{\smallskip}
	$z$  range & mean $z$ &  $M_{1500}$ range (mag) & $N_{\rm 1500}(z_{\rm p},\, M_{1500})$  &$EW(\rm Ly\alpha)$ cut (\AA) \\
\noalign{\smallskip}
\hline
\noalign{\smallskip}
\multicolumn{5}{l}{Subsamples for Figure\, \ref{fig:xevo}: $X_{\rm LAE}$ vs. $z$ }\\ 
\noalign{\smallskip}
\hline
$2.91<z\leq3.68$ & $3.3$ &  $-21.75\leq M_{1500}\leq-17.75^{\rm (a)}$  &  228 & 45$^{\rm (b)}$, 65 \\
$3.68<z\leq4.44$ & $4.1$ &  $-21.75\leq M_{1500}\leq-17.75^{\rm (a)}$ &  119 & 45$^{\rm (b)}$, 65 \\
$4.44<z\leq5.01$ & $4.7$ &  $-21.75\leq M_{1500}\leq-17.75^{\rm (a)}$   & 98 & 45$^{\rm (b)}$, 65 \\
$5.01<z\leq6.12$ & $5.6$ &  $-21.75\leq M_{1500}\leq-17.75^{\rm (a)}$ &  89 & 45$^{\rm (b)}$, 65 \\ 
\hline
\noalign{\smallskip}
\multicolumn{5}{l}{Subsamples for Figures\,\ref{fig:zevocompare}, \ref{fig:xevo_galics}, and  \ref{fig:xevo_galicscv}: $X_{\rm LAE}$ vs. $z$ for comparison with previous work} \\ 
\noalign{\smallskip}
\hline
$2.91<z\leq3.68$ & $3.3$ &  $-20.25\leq M_{1500}\leq-18.75$  &  87 & 25, 55 \\
$3.68<z\leq4.44$ & $4.1$ &  $-20.25\leq M_{1500}\leq-18.75$  &  40 & 25, 55 \\
$4.44<z\leq5.01$ & $4.7$ &  $-20.25\leq M_{1500}\leq-18.75$  & 28 & 25, 55 \\
$5.01<z\leq6.12$ & $5.6$ &  $-20.25\leq M_{1500}\leq-18.75$  &  35 & 25, 55 \\ 
\hline
\noalign{\smallskip}
\multicolumn{5}{l}{Subsamples for Figures\, \ref{fig:uvdep} and \ref{fig:uvdep_galics}: $X_{\rm LAE}$ vs. $M_{1500}$ }\\ 
\noalign{\smallskip}
\hline
$2.91<z\leq3.68$ & $3.3$ &  $-21.5\leq M_{1500}\leq-20.0$  &  31 & 25, 45, 65, 85  \\
$2.91<z\leq3.68$ & $3.3$ &  $-20.0<M_{1500}\leq-19.0$  &  58 & 25, 45, 65, 85 \\
$2.91<z\leq3.68$ & $3.3$ &  $-19.0<M_{1500}\leq-18.0$  &  106 & 45, 65, 85  \\
$2.91<z\leq3.68$ & $3.3$ &  $-18.0<M_{1500}\leq-17.0$  &  197 &  85  \\
\hline
\noalign{\smallskip}
\multicolumn{5}{l}{Subsamples for Figure\,\ref{fig:uvdepcompare}: $X_{\rm LAE}$ and $M_{1500}$ for comparison with previous work }\\ 
\noalign{\smallskip}
\hline
$2.91<z\leq3.68$ & $3.3$ &  $-21.5\leq M_{1500}\leq-20.0$  &  31 & 50 \\
$2.91<z\leq3.68$ & $3.3$ &  $-20.0<M_{1500}\leq-19.0$  &  58 & 50\\
$2.91<z\leq3.68$ & $3.3$ &  $-19.0<M_{1500}\leq-18.0$  &  106 & 50\\
$3.68<z\leq4.44$ & $4.1$ & $-21.5\leq M_{1500}\leq-20.0$  &  8 & 50 \\
$3.68<z\leq4.44$ & $4.1$ &  $-20.0<M_{1500}\leq-19.0$  &  23 & 50\\
$3.68<z\leq4.44$ & $4.1$ &  $-19.0<M_{1500}\leq-18.0$  &  56 & 50\\
\hline
\end{tabular}
\tablefoot{Subsample criteria of redshift and $M_{1500}$ for the continuum-selected parent sample. The mean redshift, sample size ($N_{\rm 1500}(z_{\rm p},\, M_{1500})$), and $EW(\rm Ly\alpha)$ cut are also shown. To increase the sample size used in Figures \ref{fig:xevo}, \ref{fig:zevocompare}, \ref{fig:xevo_galics}, and  \ref{fig:xevo_galicscv}, we combine the two highest redshift bins used to compute the UV magnitude and Ly$\alpha$ completeness. (a) We also calculate $X_{\rm LAE}$ for the $-21.75\leq M_{1500}\leq-18.75$ mag and $-18.75<M_{1500}\leq-17.75$ mag subsamples. (b) The $45$  \AA\ $EW(\rm Ly\alpha)$ cut is only applied to the $-21.75<M_{1500}\leq-18.75$ mag subsamples.} 
\end{table*}

\subsection{Counting LAEs within the UV-selected sample} \label{subsec:CountingLAEs}
\subsubsection{MUSE data}\label{subsec:muse}

The data of the MUSE HUDF Survey were obtained as part of the MUSE GTO program (PI: R. Bacon). The MUSE HUDF Survey design is presented in \citet{Bacon2017}. It consists of two layers of different depths: the \mosaic{} is composed of $9$ MUSE pointings that cover a $3'\times 3'$ area ($9.92$ arcmin$^2$) with an integration time of $\approx10$ hours; the \udft{} is a deeper integration at a $1'\times 1'$ sub field within the \mosaic{}, with an integration time of $\approx31$ hours. MUSE covers a wide optical wavelength range, from $4750$ \AA{} to $9300$ \AA, which allows the observation of the \lya{} line from $z\approx2.9$ to $z\approx6.6$. The typical spectral resolving power is $R=3000$, with a spectral sampling of $1.25$ \AA{}. The spatial resolution (pixel size) is $0\farcs2 \times 0\farcs2$ per pixel.

In the present paper, we use the latest data release (the second data release, hereafter DR2) from the MUSE HUDF (Bacon et al.\, in prep.). The improved data reduction process results in data cubes with fewer systematics and a better sky subtraction. The FWHM of the Moffat point spread function (PSF) is $0\farcs65$ at $7000$ \AA\ in the MUSE HUDF. The estimated $1$ $\sigma$ surface brightness limits are $2.8\times10^{-20}$ erg s$^{-1}$ cm$^{-2}$ \AA$^{-1}$ arcsec$^{-2}$ and $5.5\times10^{-20}$ erg s$^{-1}$ cm$^{-2}$ \AA$^{-1}$ arcsec$^{-2}$ in \udft\ and \mosaic, respectively, in the wavelength range of $7000$--$8500$ \AA\ (excluding regions of OH sky emission, see \citealt{Inami2017}, Bacon et al. in prep. for more details).  For instance, the estimated $3$ $\sigma$ flux limits are $1.5\times10^{-19}$ erg s$^{-1}$ cm$^{-2}$ and $3.1\times10^{-19}$ erg s$^{-1}$ cm$^{-2}$ in \udft\ and \mosaic, respectively, for a point-like source extracted over three spectral channels (i.e., $3.75$\AA) around $7000$ \AA\ \citep[see Figure 20 in][]{Bacon2017}. The PSF and noise characteristics are similar to the DR1 data, except in the reddest part of the wavelength range.

In order to measure the fraction of galaxies which have a strong \lya{} line, we first extract a 1D spectrum from the MUSE cube for each HST source in our parent sample. We proceed as follows. First, we convolve the HST segmentation map of the R15 catalog with the MUSE PSF, which is normalized to $1$. To obtain a spatial mask applicable to MUSE observations for each object, we apply a threshold value of $0.2$ to the normalized convolved segmentation map. The median value of the radius of the normalized mask is $\approx1\farcs0$ to $0\farcs8$ arcsecond at $z\approx3$ to $6$, which is not affected by Ly$\alpha$ halo flux \citep[][see Sect. \ref{subsec:haloDiscussion} for more detailed discussion of our choice]{Leclercq2017}. Second, we integrate the cube spatially over the extent of the mask. We note that PSF weighted or white-light weighted integrations are used to extract spectra in the DR1 catalog of \citet{Inami2017}. These provide a higher $S/N$ in the extracted spectrum. However, in the present paper, we do not use a spatial weighting. This results in slightly lower $S/N$ values, but more accurate estimates of the fluxes (i.e., conserved flux), which are needed to assess the completeness of our \lya{} detections. Third, we subtract local residual background emission from the extracted spectrum for the 1D spectra as in \citet{Inami2017}. The local background is defined in $5\farcs$ $\times$ $5\farcs$ subcubes avoiding the masks of any source.

\subsubsection{Search for Ly$\alpha$ emission}\label{subsec:marz}

In order to detect Ly$\alpha$ emission lines in the 1D spectra extracted above, we use a customized version of the \textsf{MARZ} software\footnote{The original \textsf{MARZ} in \citet{Hinton2016} is based on a cross-correlation algorithm \citep[\textsf{AUTOZ},][]{Baldry2014} and is publicly available at: \url{https://github.com/Samreay/Marz.}} \citep{Hinton2016} described in \citet{Inami2017}. \textsf{MARZ} compares 1D spectra to a list of templates and returns the best-fitting spectroscopic redshift, the best-fitting 1D template, and a confidence level for the result (called the quality operator, QOP). In our customized \textsf{MARZ} version, the list of templates consists of templates made using MUSE data, and the interface is improved \citep{Inami2017}. We use our version of \textsf{MARZ} in a similar manner to \citet{Inami2017}, except for the two following changes. First, we do not activate cosmic ray replacement in \textsf{MARZ} because (1) it affects the detectability of bright and spectrally peaky Ly$\alpha$ emission, and (2) cosmic rays are efficiently removed in the data reduction. Second, we only use template spectra with Ly$\alpha$ emission: those of IDs$=10$, $18$, and $19$, which are used in \citet{Inami2017}, and those of IDs$=25$, $26$, $27$ and $30$, which are newly built from MUSE data in Bacon et al. in prep. and show single peaked Ly$\alpha$ (see Appendix \ref{ap:temp_marz} for the template spectra). 
As in \citet{Inami2017}, we use the 1D spectra and source files including the subcubes and cutouts of HST UV to NIR images for the parent sources as input for \textsf{MARZ}.

To select robust Ly$\alpha$ detections, we keep only galaxies which \textsf{MARZ} identifies as LAEs with a high confidence level \citep[``Great'' and ``Good'' shown in Figure 1 in][]{Inami2017}\footnote{The confidence level is given as QOP, which is calculated from the peak values of the cross-correlation function \citep[figure of merit, hereafter FOM, see Sect. 5.3 in][for more details]{Hinton2016}. In our version of \textsf{MARZ} \citep{Inami2017}, QOP $=3$ (QOP $=2$) is regarded as ``Great'' (``Good'') and  corresponds to $99.55$\% ($95$\%) confidence in the original \textsf{MARZ} \citep{Hinton2016}. However, the original relation between QOP and confidence percentage is calibrated with SED templates different from ours, and the confidence percentage may not be directly applicable to our data. Note that the FOM criterion for QOP $=3$ (QOP $=2$) in our version of \textsf{MARZ} is the same as that for QOP $\geq4$ (QOP $=3$) in the original \textsf{MARZ} (see Figure1 in \citealt{Inami2017} and Figure 12 in \citealt{Hinton2016}). }. Sources with lower confidence levels are not regarded as detected LAEs. According to this selection, among the $3233$ ($402$) sources in \mosaic\  (\udft), $374$ ($70$) are LAEs. However, some of these LAE candidates are in fact \OII\ emitters or non-LAEs polluted by extended Ly$\alpha$ emission from LAE neighbors. We visually inspect all the LAE candidates as in \citet{Inami2017}: we check the entire MUSE spectra, Ly$\alpha$ line profiles, MUSE white-light images, MUSE narrow-band images of Ly$\alpha$ emission, all the existing HST UV to NIR images, and HST colors by eye using the customized \textsf{MARZ}. The MUSE white-light image is created by collapsing the $5\farcs$ $\times$ $5\farcs$ MUSE subcubes in wavelength direction \citep[see][]{Inami2017}, while the narrow-band image for the Ly$\alpha$ emission is extracted from the wavelength range around the Ly$\alpha$ emission in the subcubes for the sources \citep[see][]{Drake2017b, Drake2017a}. In contrast to \citet{Inami2017}, we also use a consistent photometric redshift ($95$\% uncertainty range) as an evidence of Ly$\alpha$ emission. As a result, we have $276$ ($58$) LAEs at $z\approx2.9$--$6.1$ among the parent sample in the \mosaic\ (\udft) field. Most of the removed sources ($\approx80$\%) have a 1D spectrum contaminated by (extended) Ly$\alpha$ emission from neighboring objects, which can be distinguished using the Ly$\alpha$ narrow-band images, MUSE white-light images, and HST images. We show an example of Ly$\alpha$ contamination in Appendix \ref{ap:contami}.

\subsubsection{Measurement of Ly$\alpha$ fluxes}\label{subsec:measurement} 

For our LAEs, we measure the Ly$\alpha$ fluxes from the 1D spectra used for Ly$\alpha$ detection and described in Sect. \ref{subsec:muse}. The aperture size is defined by the R15's segmentation map for each source convolved with the MUSE PSF (see Sect. \ref{subsec:CountingLAEs} and Sect. \ref{subsec:haloDiscussion} for more details). 
It has been reported that Ly$\alpha$ emission is often spatially offset from the stellar UV continuum \citep[e.g.][]{Erb2019arXiv,Hoag2019bMNRAS}. The typical offset values for our LAEs are however measured to be less than $0\farcs1$ \citep{Leclercq2017}, significantly less than the PSF scale of our observations. We therefore assume that all the flux of the central Ly$\alpha$ component can be captured by our apertures, centered on the continuum emission peak, and with median radius ranging from $\approx1\farcs0$ to $0\farcs8$ at $z\approx3$ to $6$. We fit the Ly$\alpha$ emission either with an asymmetric Gaussian or a double Gaussian profile. The choice between these two solutions is made after visual inspection. In practice, we use the \textsf{gauss\_asymfit} and \textsf{gauss\_dfit} methods of the publicly available software \textsf{MPDAF} \citep{Piqueras2017}\footnote{MPDAF, the MUSE Python Data Analysis Framework, is publicly available from the following link: \url{https://mpdaf.readthedocs.io/en/latest/}}. We use the spectroscopic redshift from \textsf{MARZ} as information for the center wavelength of the fit, with a fitting range of rest-frame $1190$ \AA\ to $1270$ \AA. We inspect all the Ly$\alpha$ spectral profile fits to confirm their validity.

\begin{figure}
   \sidecaption
   \includegraphics[width=9.3cm]{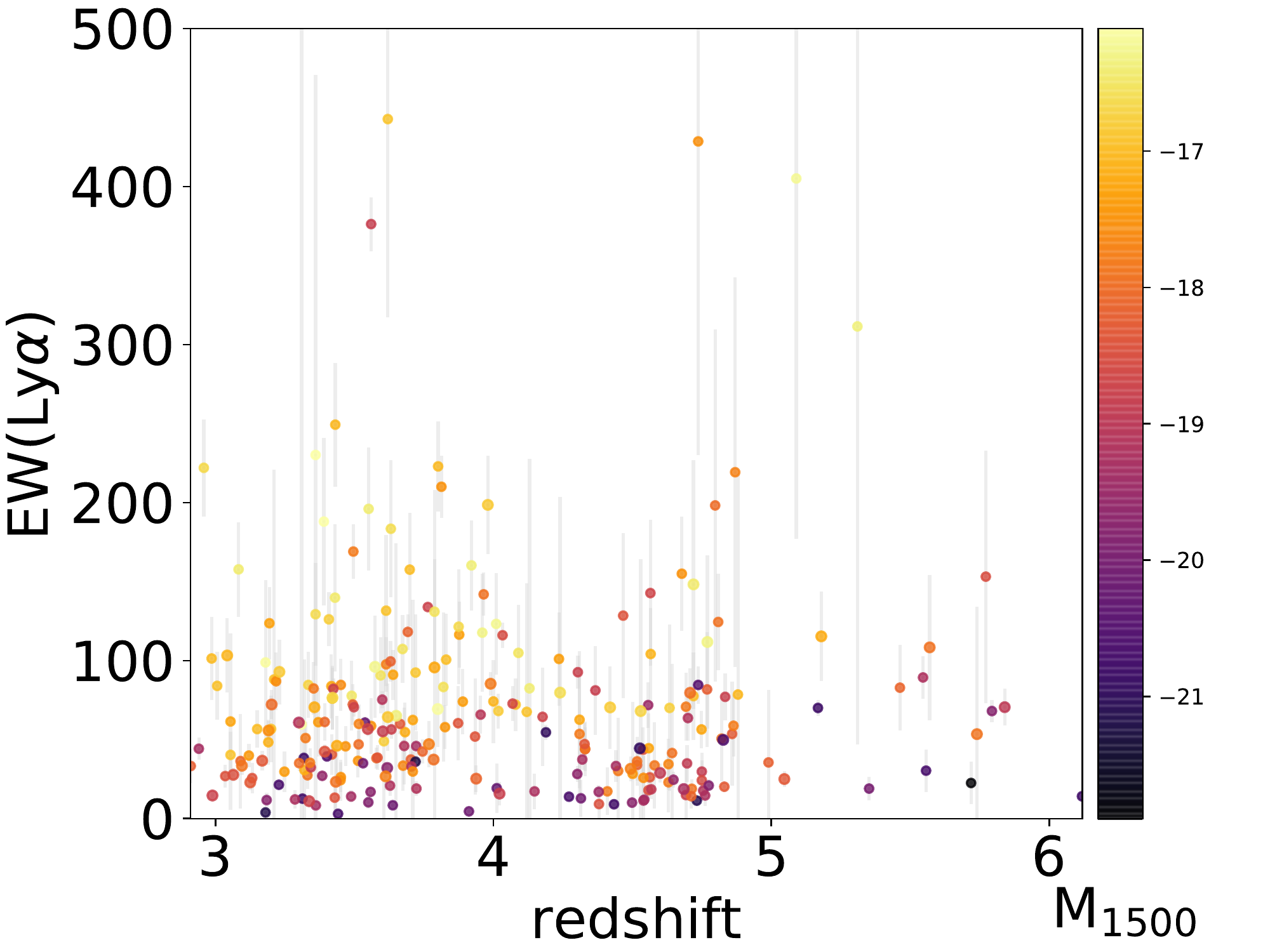}\\
      \caption{$EW(\rm Ly\alpha)$ versus $z_{\rm z}$ for our final LAE sample. Colors
show the $M_{\rm 1500}$.  \label{fig:ew_z}}
         
   \end{figure}

Once we have measured the Ly$\alpha$ fluxes, we compute the rest-frame Ly$\alpha$ equivalent width using $M_{\rm 1500}$ and $\beta$ defined in Sect. \ref{subsec:parent_sample} to estimate the continuum. The distribution of $EW(\rm Ly\alpha)$ as a function of redshift is shown in Figure \ref{fig:ew_z}.

\vspace{0.5cm}

\subsubsection{Completeness estimate and correction}\label{subsec:compsimu} 

 \begin{figure}
 \begin{flushright}
   \includegraphics[width=9.3cm]{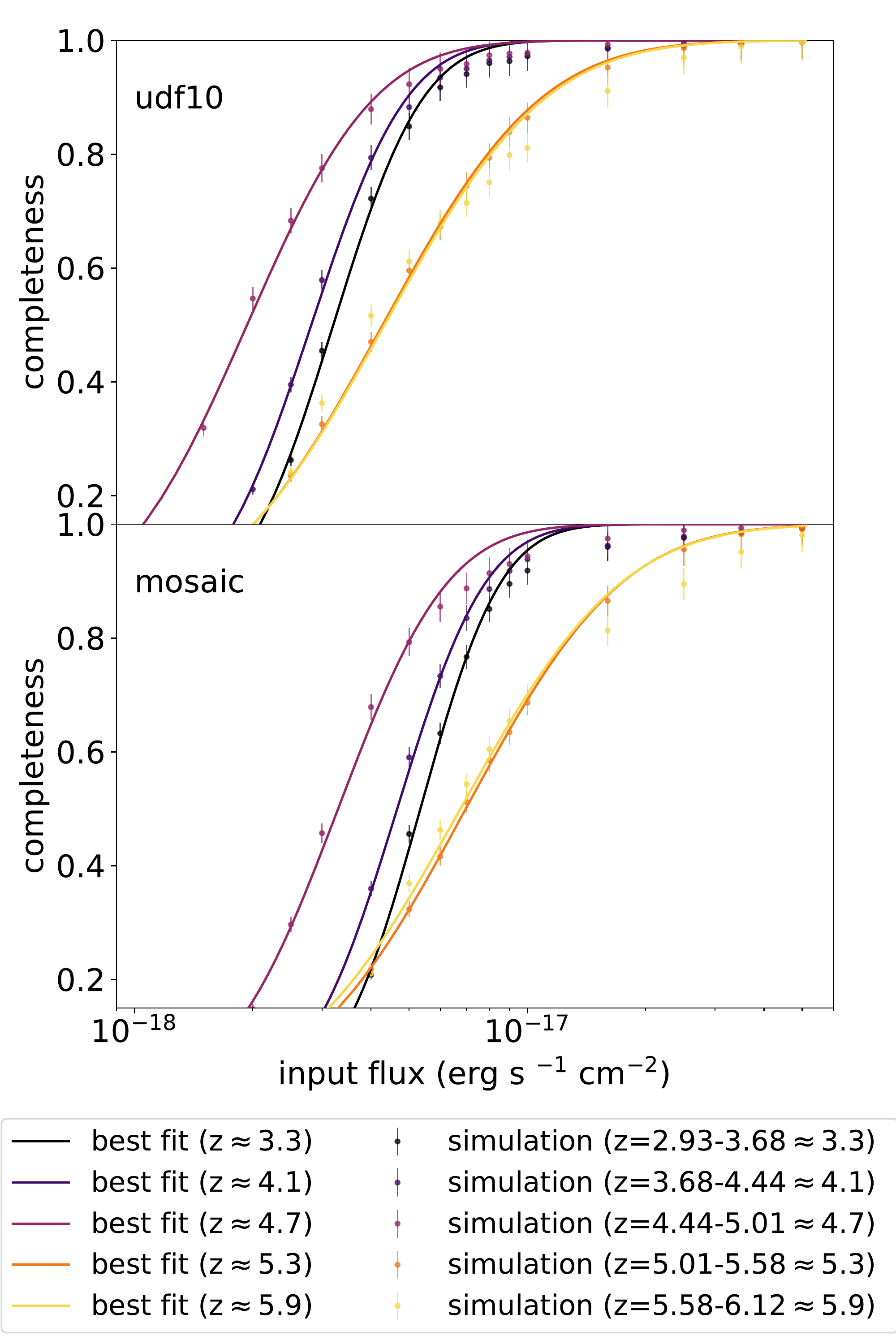}\\
      \caption[Completeness of Ly$\alpha$ emission vs. Ly$\alpha$ flux.]{
      Completeness of Ly$\alpha$ detection as a function of Ly$\alpha$ flux in the \udft\ (upper panel) and \mosaic\ (lower panel) fields. The simulated data points and their best fit completeness functions are indicated by  circles and lines, respectively. Black, purple, violet, orange and yellow colors represent redshifts $z\approx3.3$, $4.1$, $4.7$, $5.3$, and $5.9$, respectively. Error bars are calculated from the Poisson errors of the numbers of the detected fake emission lines.
       \label{fig:lya_comp} }
\end{flushright}
 \end{figure}
 
In order to estimate the detection completeness of Ly$\alpha$ emission for the MUSE HUDF data with \textsf{MARZ}, we insert fake Ly$\alpha$ emission lines into 1D spectra and try to detect them as explained in Sect. \ref{subsec:marz}. 

We take realistic noise into account by creating 1D sky-background spectra from MUSE sub-cubes extracted for different continuum-selected sources as detailed in Sect. \ref{subsec:muse}. We choose a sample of spectra which show a clear Ly$\alpha$ emission line \citep[detected with a very high confidence level of ``Great'' shown in Figure 1 in][]{Inami2017}, and which do not have continuum or other spectroscopic features. We mask the spectral pixels covered by the Ly$\alpha$ emission in these 1D spectra (including $\pm 20$ pixels $\approx \pm 25$ \AA\  around the line center). Note that we do not insert fake Ly$\alpha$ lines in the masked regions. With this procedure, we obtain $131$ ($35$) 1D sky-background spectra in the \mosaic\ (\udft) field.

We add fake emission lines with fluxes taking 18 values regularly spaced in log between $6$ $\times 10^{-19}$ erg s$^{-1}$ cm$^{-2}$ and $50$ $\times 10^{-18}$ erg s$^{-1}$ cm$^{-2}$, and $3102$ redshift values regularly distributed between $z=2.93$ and $z=6.12$. For each flux-redshift pair, we draw 4 lines (yielding a total of 223,344 lines) and add each of them to one of our 131 (35) 1D spectra chosen randomly. Each fake Ly$\alpha$ line, has line-shape parameter values (total FWHM and FWHM ratio of red wing to blue wing) randomly drawn from the measured distribution of Ly$\alpha$ emission line shapes of LAEs used in \citet{Bacon2015} and \citet{Hashimoto2017b}. We use the \textsf{add\_asym\_gaussian} method of \textsf{MPDAF} to generate the fake lines and add them to our test spectra. 

We then repeat the detection procedure of Sect. \ref{subsec:marz}, applying the same cut at a confidence level of ``Good''. In each field (\udft\ or \mosaic), we compute the completeness of Ly$\alpha$ detection as a function of the Ly$\alpha$ flux $f_{\rm Ly\alpha}$ for five redshift bins: $2.91\leq z <3.68$ ($z\approx3.3$), $3.68\leq z <4.44$ ($z\approx4.1$), $4.44\leq z <5.01$ ($z\approx4.7$), $5.01\leq z <5.58$ ($z\approx5.3$), and $5.58\leq z \leq6.12$ ($z\approx5.9$), which are defined from the redshift bins used to derive $M_{1500}$. We fit each simulated completeness curve with a formula based on the error function, \citep[e.g.][]{Rykoff2015arXiv}:
\begin{equation}
C(f_{\rm Ly\alpha})=\frac{1}{2}\left[1-{\rm erf}\left( \frac{-2.5\log_{10}f_{\rm Ly\alpha} -a}{\sqrt{2b}}\right)\right],    
\label{eq:comp_erf}
\end{equation}
where $a$ and $b$ are two free parameters for fitting (see Figure \ref{fig:lya_comp}). We use the function \textsf{curve\_fit} from \textsf{scipy.optimize} to perform the fit. The best fit parameters for the completeness curve in the \mosaic\ and \udft\  are summarized in Table \ref{tbl:comp}. 

At completeness above $\approx0.8$, our best fit relations slightly overestimate the measured completeness. The analytic fit is however at most $\approx5$\% above the $1\sigma$ upper errors, and this does not have a noticeable effect on the calculation of $X_{\rm LAE}$. We nevertheless take this into account in the error propagation of $X_{\rm LAE}$ in Sect. \ref{subsec:uncertainties}.  

Theoretically, completeness functions should just scale with $S/N$ and thus be applicable throughout the wavelength (or redshift) range of the instrument. For the MUSE-WIDE survey, \citet{Herenz2019} indeed find that the shape of their completeness function is  independent of redshift. As expected, we also find very similar behavior of completeness at $z\approx3.3$ and $4.1$, in both fields. At these redshifts, the noise is well behaved and there are only few accidents due to sky-line removal in the spectra. At $z\approx 4.7$, the shape of the completeness curve is still well described by Eq. \ref{eq:comp_erf}, but the curve is shifted to fainter flux with a shallower slope. At $z\approx5.3$ and $5.9$, the shapes of the best-fitting completeness are different from those at lower redshifts: they have a shallower slope, the data points with completeness above $0.8$ are not well fit, and the completeness at a given flux is much lower than that at lower redshifts. The lower normalisation and distorted shape of the completeness may be caused by the many sky emission lines at high redshifts \citep[at $z\gtrsim5$,][]{Drake2017b}. Because \textsf{MARZ} is not a local line detector, as opposed, e.g. to the matched-filtering approach implemented in the tool \textsf{LSDCat} utilized in \citet{Herenz2019} where the filter has a compact support in spectral space, \textsf{MARZ} is affected by relatively long-range noise or distant spectroscopic features in the spectra. Thus, \textsf{MARZ} often does not return a high confidence level (``Great'' and ``Good'') for LAEs at $z\gtrsim5$, and our completeness goes down at $z\gtrsim5$ even for relatively bright Ly$\alpha$ fluxes. We note that the LAE template spectra that we use all have a single-peaked Ly$\alpha$ profile (see Figure \ref{fig:temp_marz}). However, the exact shape of the line profile is shown to have little impact on the detectability with cross-correlation function in general \citep[for instance, see Sect. 4.3 in][]{Herenz2017b}. The shape of the Ly$\alpha$ line of \textsf{MARZ}'s template should not affect significantly the detection rate of LAEs (see also Appendix \ref{ap:temp_marz}). Indeed, for example, our sample contains LAEs with double-peaked lines even though none of our templates have such features. We experimented using all the templates in \textsf{MARZ}, including templates of different galaxy populations such as \OII\ and \OIII\ emitters, and we found only a small impact on our LAE sample, which is well within the error bars. Finally, we checked the dependence of completeness on the FWHM of fake Ly$\alpha$ emission lines at fixed flux, and again found no significant trend.

 \begin{table}
\caption{The best fit parameters of completeness functions}\label{tbl:comp}
\centering
\begin{tabular}{lllll}
\hline         
\noalign{\smallskip}
mean $z$	& $a$ in \udft & $b$ in \udft & $a$ in \mosaic  & $b$ in \mosaic  \\
\noalign{\smallskip}
\hline
$z\approx3.3$ & 43.2 & 0.163 & 43.7 & 0.202 \\
$z\approx4.1$ & 43.3 & 0.197 & 43.9 & 0.228 \\
$z\approx4.7$ & 43.7 & 0.303 & 44.3 & 0.405 \\
$z\approx5.3$ & 42.9 & 0.614 & 43.4 & 0.629 \\
$z\approx5.9$ & 42.9 & 0.653 & 43.4 & 0.647 \\
 \noalign{\smallskip}
\hline
\end{tabular}
\tablefoot{The best fit parameter of a and b of Equation (\ref{eq:comp_erf}) in each redshift bin for the \mosaic\ and \udft\ fields.} 
\end{table}

In the following, $N_{\rm LAE}^{\rm det}(z_{\rm s},\,M_{1500},\, EW)$ denotes the number of galaxies with detected Ly$\alpha$ emission and with spectroscopic redshift $z_{\rm s}$, absolute magnitude $M_{1500}$, and rest-frame equivalent width $EW$ in given ranges. We estimate the true number of LAEs with a corrected value $N_{\rm LAE}^{\rm corr}$ defined as follows. For a given field and redshift bin, we use the fits to the completeness function above to define four Ly$\alpha$ flux bins which correspond to regularly spaced bins in the logarithm of the completeness ($C$), ranging from $C=0.1$ to $C=0.9$. We then count the number of detected LAEs (within a given $z_{\rm s}$, $M_{1500}$, and EW bin) in each flux bin and divide it by the mean completeness (in log) in each of the flux bins. We compute $N_{\rm LAE}^{\rm corr}$ as the sum of these over the four flux bins. When the uncertainties of the LAE fraction is calculated, the completeness correction value in each flux bin is propagated, and the uncertainty of completeness correction itself is taken into account as described in the next section. The number of flux bins is defined through a test described in Appendix \ref{ap:bin_comp}. Four to six bins are a sweet spot where the error bars are small and they appear converged. For a larger number of bins, we often get flux-bins with no object at all, and these bins contribute to a large error bar. For a smaller number of bins, we introduce a larger error on the completeness correction (averaged over the bin). We adopt four bins.

\subsection{$X_{\rm LAE}$ and its error budget} \label{subsec:uncertainties}

Knowing the number of UV-selected galaxies in a volume-limited sample ($N_{\rm 1500}(z_{\rm p},\, M_{1500})$, Sect.\ref{subsec:parent_sample}), and the number of LAE among the {\it inclusive parent sample} ($N_{\rm LAE}^{\rm corr}(z_{\rm s},\,M_{1500},\, EW)$, Sect. \ref{subsec:compsimu}), the fraction of UV-selected galaxies with a Ly$\alpha$ line is simply given as:
\begin{equation}
X_{\rm LAE} = \frac{N_{\rm LAE}^{\rm corr}(z_{\rm s},\,M_{1500},\, EW)}{N_{\rm 1500}(z_{\rm p},\, M_{1500})}. 
\label{eq:xlae}
\end{equation}

The uncertainties on $X_{\rm LAE}$ arise from four components: (1) the uncertainty due to contaminants (type I\hspace{-1pt}I error) and missed objects (type I error) in $N_{\rm 1500}(z_{\rm p},\, M_{1500})$,  (2) the uncertainty due to the completeness correction of $N_{\rm LAE}^{\rm corr}(z_{\rm s},\,M_{1500},\, EW)$, (3) the uncertainty for Bernoulli trials (i.e., fraction of $N_{\rm LAE}^{\rm corr}(z_{\rm s},\,M_{1500},\, EW)$ over $N_{\rm 1500}(z_{\rm p},\, M_{1500})$) measured by a binomial proportion confidence interval and (4) the uncertainty due to cosmic variance. We note that there is no obvious sample selection bias in our UV galaxies as shown in Figure \ref{fig:n_muv}, and our Ly$\alpha$ measurements are  homogeneous over the whole redshift range (see discussion in Sect. \ref{subsec:difference}). We estimate the relative uncertainty of $X_{\rm LAE}$ from error propagation and discuss each of these contributions below.

\begin{description}
\item[(1) {\it Contaminants and missed objects in $N_{\rm 1500}(z_{\rm p},\, M_{1500})$}]\mbox{}\\ 
Sources with $z_{\rm p}$ in a given $z$ range that are truly located outside of the $z$ range are contaminants in the parent continuum-selected sample, while sources with $z_{\rm p}$ outside of the $z$ that are truly located in the $z$ range are missed sources. These mismatches of $z_{\rm p}$ can happen because of confusion between Lyman and $4000$ \AA\ breaks in the SED fitting. In addition, IGM absorption modeling has been suggested to affect $z_{\rm p}$ estimation \citep{Brinchmann2017}. 
As discussed in \citet{Inami2017} and \citet{Brinchmann2017}, the fraction of contaminants is very low for high confidence level objects \citep[with secure redshift, see Figure 20 in][]{Inami2017}. The fraction of missed objects with a relative redshift difference of more than $15$\%, |$z_{\rm s}$-$z_{\rm p}$|/(1+$z_{\rm s}$)$>0.15$, is suggested to be $\approx10$\% \citep[outlier fraction,][]{Brinchmann2017}. Since the missed objects whose $95$\% uncertainty range for $z_{\rm p}$ is outside the $z$ range for MUSE-LAEs are not included in our parent continuum-selected sample, they are also not included in our LAE sample. With an assumption that the fractions of missed objects are the same for the parent and LAE samples, the uncertainties due to missed objects can be neglected as well as those due to contaminants. Note that we do not find any significant relation between the $z_{\rm p}$-$z_{\rm s}$ difference and Ly$\alpha$ EW as well as Ly$\alpha$ flux. 

\item[(2) {\it Completeness correction of $N_{\rm LAE}^{\rm corr}(z_{\rm s},\,M_{1500},\, EW)$}]\mbox{}\\ 
Simulated data of completeness are fitted well with Equation (\ref{eq:comp_erf}) at $z\approx3.3$, $4.1$, and $4.7$. Even at $z\approx5.3$ and $5.9$, the differences in completeness between the simulated data and the best-fitting functions are at most $\approx5$\%, which is much smaller than the uncertainty due to the flux binning. The flux binning for the completeness correction described in Sect. \ref{subsec:compsimu} causes an uncertainty of at most $\approx\pm32$\%. The completeness bins, spaced regularly in log from $0.1$ to $0.9$, correspond to steps of a factor $1.73$, which has a square-root of $\approx 1.32$. 
We use this very conservative estimate of $32$\% for the completeness correction error in our error budget. The completeness correction error is smaller than the error component (3) as described later and does not change the total uncertainty of $X_{\rm LAE}$. We note that the completeness correction value is also taken into account according to error propagation, when the error component (3) is calculated.

\item[ (3) {\it Uncertainties for Bernoulli trials}]\mbox{}\\ 
Measuring the fraction of a sub-sample among a parent sample is a kind of experiment of Bernoulli trials. An uncertainty for a Bernoulli trial is given by a binomial proportion confidence interval (hereafter, BPCI). We use the python module \textsf{binom\_conf\_interval} from \textsf{astropy.stats} and provide an approximate uncertainty for a given confidence interval (=$68$\%, 1$\sigma$, in this work), the number of trials, and the number of successes. Here, the number of trials and the number of successful experiments are $N_{\rm 1500}(z_{\rm p},\, M_{1500})$ and $N_{\rm LAE}^{\rm corr}(z_{\rm s},\,M_{1500},\, EW)$, respectively. However, we cannot obtain $N_{\rm LAE}^{\rm corr}(z_{\rm s},\,M_{1500},\, EW)$ directly from the observations. To include the effect of the completeness correction in each Ly$\alpha$ flux bin (described in Sect. \ref{subsec:compsimu}) in the error propagation, we calculate the uncertainty of the LAE fraction in each Ly$\alpha$ flux bin without applying a completeness correction from \textsf{binom\_conf\_interval} and multiply the uncertainties by the correction value in the flux bin. We choose as an approximation formula the Wilson score interval \citep{Wilson1927}, which is known to return an appropriate output even for a small number of trials and/or success experiments. Our method is confirmed to be accurate by numerical tests described in Appendix \ref{ap:bpci}. For the flux-bins with no LAEs, the average completeness value among the bins and the number of LAEs ($=0$) are used to derive the uncertainties conservatively. When we sum over all the uncertainties in a flux bin to derive the total uncertainties for the component (3), a python module, \textsf{add\_asym} developed in \citet{Laursen2019}, is used to treat asymmetric errors by BPCI. We note that the Poisson errors of $N_{\rm LAE}^{\rm corr}(z_{\rm s},\,M_{1500},\, EW)$ and $N_{\rm 1500}(z_{\rm p},\, M_{1500})$ are commonly used in the literature. However, the error for the LAE fraction should be derived by BPCI like in other fraction studies \citep[e.g. the galaxy merger fraction,][]{Ventou2017} to obtain statistically correct errors. The BPCI method is reviewed for astronomical uses in \citet{Cameron2011}.

\item[(4) {\it Cosmic variance}]\mbox{}\\ 
The survey volume in each redshift range is limited to $\approx 1.5-2.5 \times 10^4$ cMpc$^3$. However, we find that the uncertainty due to cosmic variance is less than the BPCI error and thus not affecting our $X_{\rm LAE}$ significantly (See Sect. \ref{subsubsec:galics_cv} for more details). Since the uncertainty due to cosmic variance  cannot be included in our MUSE measurements, we neglect this error component (4). 

\end{description}
In addition to (1) - (4), uncertainties in photo-$z$ estimations and flux measurements are also potential error components. In Appendix \ref{ap:photoz}, we discuss the impact of uncertainties on $z_{\rm p}$ and conclude that it does not affect the error bar of $X_{\rm LAE}$ significantly. Uncertainties of $M_{1500}$, $\beta$, and the Ly$\alpha$ flux, combine into uncertainties on $EW(\rm Ly\alpha)$. As shown in Figures \ref{fig:uvdep} and \ref{fig:uvdepcompare}, $X_{\rm LAE}$ shows a slight dependence on $M_{1500}$ and $EW(\rm Ly\alpha)$. We thus expect that a small error on these quantities will translate into an even smaller error on $X_{\rm LAE}$.  Note that although some objects have large errors in $M_{1500}$ in Figure \ref{fig:n_muv}, they are not included in our analysis because they are very faint\footnote{Among the subsamples shown in Table \ref{tbl:criteria}, fainter-$M_{1500}$ and higher-$z_{\rm p}$ objects have greater uncertainties in $M_{1500}$. The medians (standard deviations) of the uncertainties for the subsamples with $-18.75\leq M_{1500}\leq-17.75$ mag are $0.05$ mag ($0.02$ mag) at $z\approx3.3$ and $0.15$ mag ($0.12$ mag) at $z\approx5.6$. Those for the subsamples with $-19.0\leq M_{1500}\leq-18.0$ mag is $0.04$ mag ($0.01$ mag) at $z\approx3.3$ and $0.08$ mag ($0.11$ mag) at $z\approx4.1$. These uncertainties are much smaller than the width of $M_{1500}$ bins. With regard to $S/N$ cuts of Ly$\alpha$ fluxes corresponding to $EW(\rm Ly\alpha)$ cuts, those in the \mosaic{} field for $M_{1500}=-17.75$ mag and $EW(\rm Ly\alpha)=65$ \AA\ are estimated to be $\approx17.4$ at $z\approx3.3$ and $\approx4.4$ at $z\approx5.6$, if we assume $\beta=-2$. Similarly, the $S/N$ cuts for $M_{1500}=-18.0$ mag and $EW(\rm Ly\alpha)=50$ \AA\ are $\approx16.9$ at $z\approx3.3$ and $\approx11.1$ at $z\approx4.1$. Here, the noise is the median of those shown in Figure \ref{fig:zevocompare} in each redshift bin.}. Thus, we ignore these two uncertainties, as is commonly done in the rest of the literature.

With these considerations, the uncertainty of the LAE fraction is the quadratic sum of the uncertainty terms (2) and (3). Below, the error bars on $X_{\rm LAE}$ represent the $68$\% confidence intervals around the values calculated by Equation (\ref{eq:xlae}). Note that the dominant error for $X_{\rm LAE}$ derived in this work is component (3), the uncertainties for a Bernoulli trial, which are, for instance, $38$\% and $78$\% of $X_{\rm LAE}$ for $-21.75\leq M_{1500}\leq-18.75$ mag and $EW(\rm Ly\alpha)\geq65$ \AA\ at $z\approx3$ and $5.6$, respectively.

\subsection{Measurement of the slope of $X_{\rm LAE}$ as a function of $z$ and $M_{1500}$} \label{subsec:slope} 
We measure linear slopes of $X_{\rm LAE}$ as a function of $z$ and $M_{1500}$ using a python package for orthogonal distance regression (hereafter, ODR) fitting, \textsf{scipy.odr}, to account for widths of bins in x-axis and uncertainties of $X_{\rm LAE}$ in y-axis. The ODR fitting minimizes the sum of squared perpendicular distances from the data to the fitting model. Since uncertainties of $X_{\rm LAE}$ are not symmetric, a Monte Carlo simulation with $10000$ trials is used. We assume an asymmetric Gaussian profile as a probability distribution function for $X_{\rm LAE}$ with each of the upper and lower uncertainties at a given bin ($z$ or $M_{1500}$) in x-axis. We fit a linear relation ($y=ax+b$) in each trial drawing $X_{\rm LAE}$ randomly with \textsf{scipy.odr}. The best fit values of $a$ and $b$ and their error bars are derived from the median values and the $68$\% confidence intervals around the median values. The results of fitting are shown in Sect. \ref{sec:result}.

\subsection{Extended Ly$\alpha$ emission} \label{subsec:haloDiscussion}
Our aim is to measure how the fraction of UV-selected galaxies showing a strong \lya{} line varies with redshift. We make the choice to discard possible significant contributions to the \lya{} luminosities by extended \lya{} haloes \citep[{\it LAH}, e.g.][Leclercq et al. in prep.]{Wisotzki2016, Drake2017b, Leclercq2017} in our study, though the contribution of LAHs to the total Ly$\alpha$ fluxes is typically more than $\approx50$\% \citep[e.g.][]{Momose2016, Leclercq2017}. There are a number of reasons for this choice. First, it is largely motivated by our lack of understanding of the physical processes lighting up these halos. In particular, it is not clear how they relate to the UV luminosities of their associated galaxies \citep[e.g.][]{Leclercq2017, Kusakabe2019}, to what extent they are associated to star formation \citep[e.g.][their Figure 12]{Yajima2012b}, or whether the nature of this association could vary with redshift. In order to assess the evolution of $X_{\rm LAE}$ with redshift, it thus appears more conservative to limit our measurement of the \lya{} emission from galaxies to the part which most likely has the same origin as the continuum UV light. Any evolution is then more likely to be related to the evolution of the ionisation state of the IGM. With the above procedure, our 1D spectra include as little as possible of the extended \lya{} emission that is found around LAEs \citep{Wisotzki2016,Leclercq2017}. Second, our choice has the advantage of following a similar methodology without including halo fluxes used in other studies \citep[e.g.][]{Stark2011,DeBarros2017, ArrabalHaro2018}, and thus allows for a fair comparison. Third, it is difficult to measure the faintest halos. If we demanded that LAHs were detected around galaxies in our sample, we would limit our sample to the brightest and/or compact halos only \citep[see Sect. 2.2 and Figure 8 of][]{Leclercq2017}. So even though an IFU enables us in principle to separate the central and halo component more clearly than slit and fiber spectrometers, the signal-to-noise ratio required to do so is still prohibitive for statistical studies such as ours. 

Note that because of our choice, the Ly$\alpha$ fluxes and EWs in the present paper are smaller than the total Ly$\alpha$ fluxes and EWs reported by e.g.  \citet[][]{Hashimoto2017b}, \citet[][]{Drake2017b}, and  \citet[][]{Leclercq2017}.

\section{Results}\label{sec:result}

In order to measure the variation of $X_{\rm LAE}$ with redshift or UV absolute magnitude, we design several sub-samples shown in Table \ref{tbl:criteria}. We use $EW(\rm Ly\alpha)$ cuts starting at $25$ \AA, which is a common limit in the literature, and then increase in steps of 20\AA\ to $45$ \AA, $65$ \AA, and $85$ \AA. We also use $50$ \AA\ and $55$ \AA\ cuts, for comparison to \citet{Stark2010} and \citet{Stark2011}. In Sect. \ref{subsec:zevo}, we present our results for the $X_{\rm LAE}$-$z$ relation, going as faint as $M_{1500}=-17.75$ mag for the first time in a homogeneous way over the redshift range $z\approx3$ to $z\approx6$. We discuss how these results compare to existing measurements. In Sect. \ref{subsec:uvdep}, we present the first measurement of the $X_{\rm LAE}$-$M_{1500}$ relation for galaxies as faint as $M_{1500}=-17.00$ mag at $z\approx3$, and compare our findings to other studies. The numerical values of $X_{\rm LAE}$ are summarized in Tables \ref{tbl:x_z} and \ref{tbl:x_muv}. The slopes of the best fit linear relations of $X_{\rm LAE}$ as a function of $z$ and $M_{1500}$ are shown in Figure \ref{fig:xlae_fit} in Appendix \ref{ap:xlae_fit}, and summarised in Tables \ref{tbl:x_z} and \ref{tbl:x_muv}.

\begin{table}[h]
 \caption{LAE fraction as a function of redshift}\label{tbl:x_z}
\begin{tabular}{cccc}
 \hline 
  mean $z$ & $X_{\rm LAE}$ or slope & $1\sigma$ upper error &  $1\sigma$ lower error   \\
 \hline
 \multicolumn{4}{c}{$-21.75\leq M_{1500}\leq-17.75$ mag, $EW(\rm Ly\alpha)\geq65$ \AA} \\
 \hline
   3.3 &     0.04 &                  0.02 &                  0.01 \\
   4.1 &     0.07 &                  0.04 &                  0.02 \\
   4.7 &     0.11 &                  0.07 &                  0.04 \\
   5.6 &     0.20 &                  0.16 &                  0.06 \\
 \multicolumn{4}{l}{modest positive correlation:} \\ 
   slope&    0.07 &                  0.06 &                  0.03 \\ 
\hline
 \multicolumn{4}{c}{$-21.75\leq M_{1500}\leq-18.75$ mag, $EW(\rm Ly\alpha)\geq45$ \AA} \\ 
 \hline
   3.3 &     0.05 &                  0.03 &                  0.02 \\
   4.1 &     0.16 &                  0.10 &                  0.06 \\
   4.7 &     0.21 &                  0.13 &                  0.08 \\
   5.6 &     0.13 &                  0.13 &                  0.05 \\
 \multicolumn{4}{l}{modest positive correlation:} \\ 
   slope&    0.05 &                  0.05 &                  0.03 \\
 \hline
 \multicolumn{4}{c}{$-21.75\leq M_{1500}\leq-18.75$ mag, $EW(\rm Ly\alpha)\geq65$ \AA} \\
 \hline
   3.3 &     0.02 &                  0.02 &                  0.01 \\
   4.1 &     0.09 &                  0.07 &                  0.04 \\
   4.7 &     0.12 &                  0.09 &                  0.05 \\
   5.6 &     0.13 &                  0.13 &                  0.05 \\
 \multicolumn{4}{l}{modest positive correlation:} \\  
   slope&    0.05 &                  0.05 &                  0.03 \\ 
 \hline
 \multicolumn{4}{c}{$-18.75\leq M_{1500}\leq-17.75$ mag, $EW(\rm Ly\alpha)\geq65$ \AA} \\
 \hline
   3.3 &     0.05 &                  0.04 &                  0.02 \\
   4.1 &     0.05 &                  0.05 &                  0.02 \\
   4.7 &     0.10 &                  0.09 &                  0.03 \\
   5.6 &     0.25 &                  0.24 &                  0.09 \\
 \multicolumn{4}{l}{modest positive correlation:} \\    
   slope&    0.09 &                  0.09 &                  0.04 \\ 
 \hline
 \multicolumn{4}{c}{$-20.25\leq M_{1500}\leq-18.75$ mag, $EW(\rm Ly\alpha)\geq25$ \AA} \\
 \hline
   3.3 &     0.13 &                  0.07 &                  0.05 \\
   4.1 &     0.25 &                  0.14 &                  0.09 \\
   4.7 &     0.32 &                  0.22 &                  0.11 \\
   5.6 &     0.13 &                  0.13 &                  0.05 \\
 \multicolumn{4}{l}{no correlation (flat):} \\ 
   slope&    0.01 &                  0.05 &                  0.05 \\
 \hline
 \multicolumn{4}{c}{$-20.25\leq M_{1500}\leq-18.75$ mag, $EW(\rm Ly\alpha)\geq55$ \AA} \\
  \hline
   3.3 &     0.06 &                  0.04 &                  0.03 \\
   4.1 &     0.10 &                  0.09 &                  0.04 \\
   4.7 &     0.18 &                  0.13 &                  0.07 \\
   5.6 &     0.13 &                  0.12 &                  0.05 \\
 \multicolumn{4}{l}{no correlation (almost flat):} \\    
   slope&    0.04 &                  0.05 &                  0.03 \\
 \hline  
\end{tabular}
\tablefoot{The values and 1$\sigma$ uncertainties of the LAE fraction as a function of $z$ and the values of the slope are summarized.}
\end{table}

\begin{table}[h]
 \caption{LAE fraction as a function of UV magnitude}\label{tbl:x_muv}
\begin{tabular}{cccc}
 \hline 
  mean $M_{1500}$ & $X_{\rm LAE}$ or slope & $1\sigma$ upper error &  $1\sigma$ lower error   \\
 \hline
 \multicolumn{4}{c}{$z\approx3.3$, $EW(\rm Ly\alpha)\geq25$ \AA} \\
 \hline
-20.75 &  0.13 &  0.10 &  0.06 \\
-19.50 &  0.10 &  0.07 &  0.04 \\
\multicolumn{4}{l}{no correlation (flat):} \\
 slope &  -0.02&  0.07 &  0.08 \\ 
 \hline
 \multicolumn{4}{c}{$z\approx3.3$, $EW(\rm Ly\alpha)\geq45$ \AA} \\
 \hline
-20.75 &  0.03 &  0.07 &  0.02 \\
-19.50 &  0.03 &  0.03 &  0.02 \\
-18.50 &  0.08 &  0.05 &  0.03 \\
 \multicolumn{4}{l}{no correlation (flat):} \\
 slope &   0.01 &  0.02 &  0.03 \\ 
 \hline
 \multicolumn{4}{c}{$z\approx3.3$, $EW(\rm Ly\alpha)\geq65$ \AA} \\
 \hline
-20.75 &  0.00 &  0.03 &  0.00 \\
-19.50 &  0.02 &  0.04 &  0.01 \\
-18.50 &  0.05 &  0.04 &  0.02 \\
 \multicolumn{4}{l}{modest  positive correlation:} \\
 slope &  0.02 &  0.01 &  0.01\\
 \hline
 \multicolumn{4}{c}{$z\approx3.3$, $EW(\rm Ly\alpha)\geq85$ \AA} \\
 \hline
-20.75 &  0.00 &  0.03 &  0.00 \\
-19.50 &  0.00 &  0.02 &  0.00 \\
-18.50 &  0.01 &  0.02 &  0.00 \\
-17.50 &  0.03 &  0.03 &  0.01 \\
 \multicolumn{4}{l}{no correlation (flat):} \\
 slope &   0.01 & 0.01 &  0.01 \\ 
 \hline
 \multicolumn{4}{c}{$z\approx3.3$, $EW(\rm Ly\alpha)\geq50$ \AA} \\
 \hline
-20.75 &  0.03 &  0.07 &  0.02 \\
-19.50 &  0.03 &  0.03 &  0.02 \\
-18.50 &  0.08 &  0.04 &  0.03 \\
 \multicolumn{4}{l}{no correlation (flat):} \\
 slope &  0.01 &  0.02 &  0.03 \\ 
 \hline
 \multicolumn{4}{c}{$z\approx4.1$, $EW(\rm Ly\alpha)\geq50$ \AA} \\
 \hline
-20.75 &  0.12 &  0.23 &  0.06 \\
-19.50 &  0.04 &  0.09 &  0.02 \\
-18.50 &   0.17 &   0.12 &  0.05 \\
 \multicolumn{4}{l}{no correlation (flat):} \\
 slope &  0.01 &   0.06 &  0.10 \\ 
 \hline
\end{tabular}
\tablefoot{The values and 1$\sigma$ uncertainties of the LAE fraction as a function of $M_{1500}$ and the values of the slope are summarized.}
\end{table}

\subsection{Redshift evolution of $X_{\rm LAE}$}\label{subsec:zevo}

\begin{figure}
 \sidecaption
   \includegraphics[width=9.0cm]{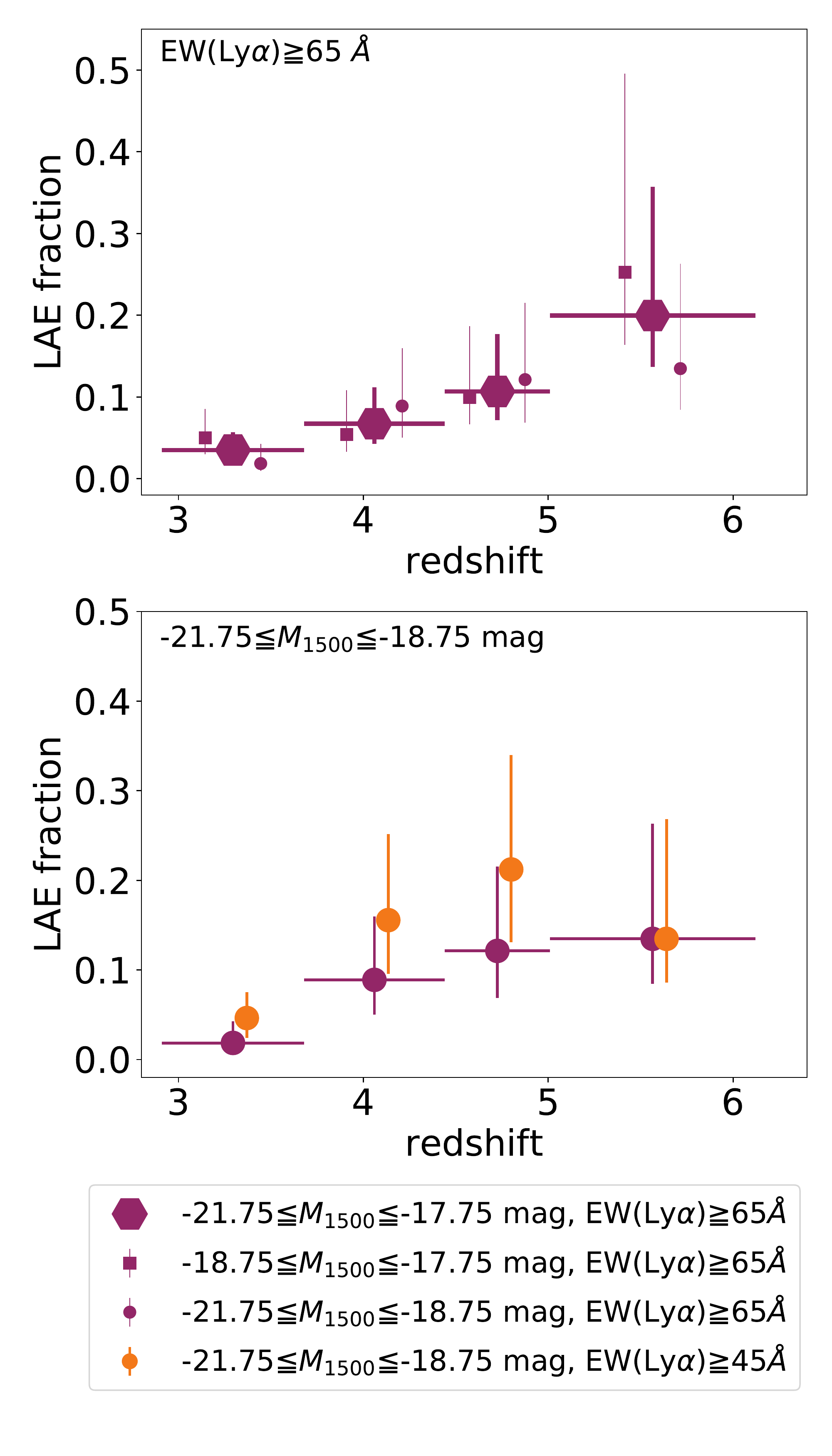}\\
      \caption{ $X_{\rm LAE}$ vs. $z$. In the upper panel, big purple hexagons, small purple circles, and small purple squares indicate LAE fractions for $EW(\rm Ly\alpha)\geq65$ \AA\ derived with MUSE at $-21.75\leq M_{1500}\leq -17.75$ mag, $-21.75\leq M_{1500}\leq -18.75$ mag, and $-18.75\leq M_{1500}\leq -17.75$ mag, respectively. In the lower panel, purple and orange circles represent $X_{\rm LAE}$ for $EW(\rm Ly\alpha)\geq65$ \AA\ and $EW(\rm Ly\alpha)\geq45$ \AA\ at $-21.75\leq M_{1500}\leq -18.75$ mag, respectively. For visualization purposes, we show the width of $z$ only for one symbol in each panel and slightly shift the other points along the abscissa.  
              }
\label{fig:xevo}
\end{figure}

We derive the redshift evolution of $X_{\rm LAE}$ for $EW(\rm Ly\alpha)\geq65$ \AA\ from $M_{1500}=-21.75$ mag to the faint UV magnitude of $-17.75$ mag. We show the results in the upper panel of Figure\, \ref{fig:xevo} with the large filled hexagons. We find a weak rise of $X_{\rm LAE}$ from $z\approx3$ to $z\approx6$, even though poor statistics do not allow us to set a firm constraint at $z\approx5.6$. Breaking our sample into a bright end (with $-21.75\leq M_{1500}\leq-18.75$ mag, purple circles), and a faint end (with $-18.75\leq M_{1500}\leq-17.75$ mag, purple squares), we find similar trends that are consistent within the $1\sigma$ error bars, suggesting that the $X_{\rm LAE}$-$z$ relation does not depend strongly on the rest-frame UV absolute magnitude. The best fit linear relations are $X_{\rm LAE} = 0.07^{+0.06}_{-0.03}z -0.22^{+0.12}_{-0.24}$, $X_{\rm LAE} = 0.05^{+0.05}_{-0.03}z -0.13^{+0.12}_{-0.18}$, and $X_{\rm LAE} = 0.09^{+0.09}_{-0.04}z -0.28^{+0.18}_{-0.36}$ for $-21.75\leq M_{1500}\leq-17.75$ mag, $-21.75\leq M_{1500}\leq-18.75$ mag, and $-18.75\leq M_{1500}\leq-17.75$ mag, respectively. Note that the bright sample here is dominated by the more numerous sub-L$^\ast$ galaxies, which are fainter than the bright samples in the literature.

If we select the brighter part of our sample ($-21.75\leq M_{1500}\leq-18.75$ mag), we can estimate $X_{\rm LAE}$ down to lower equivalent widths. In the lower panel of Figure\,\ref{fig:xevo}, the orange circles show $X_{\rm LAE}$ for galaxies with $EW(\rm Ly\alpha)\geq45$ \AA. This is also found to increase from $z\approx3$ to $z\approx4$--$6$, and is above the relation obtained for $EW(\rm Ly\alpha)\geq65$ \AA{} also shown in the lower panel, as expected. The best fit linear relation is $X_{\rm LAE} = 0.05^{+0.05}_{-0.03}z -0.07^{+0.14}_{-0.20}$. These results are qualitatively consistent with the trend of increasing $X_{\rm LAE}$ with increasing $z$ based on bright samples in the literature (see below). 

The fact that only $\approx0$--$30$\% of galaxies within the UV magnitude range at $z\leq5$ are observed as LAEs with $EW(\rm Ly\alpha)\geq65$ \AA{} requires some explanation of the physical mechanisms. We discuss this further in Sect. \ref{subsec:galics}.

%====
%====

\begin{figure*}
 \sidecaption
   \includegraphics[width=18.0cm]{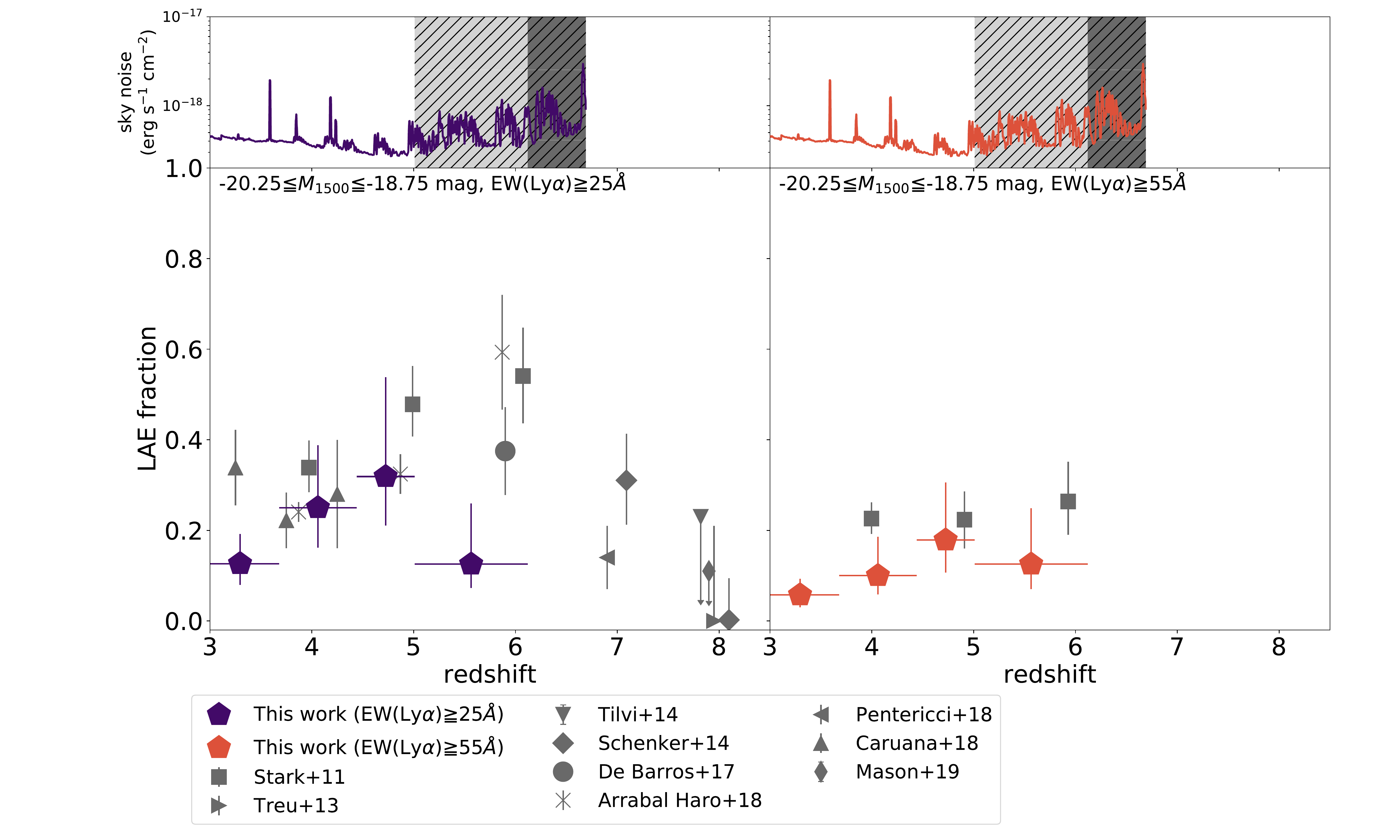}\\
      \caption{$X_{\rm LAE}$ vs. $z$ compared with previous results. Upper left and right panels: sky noise vs. $z$. The purple and orange lines show the $1\sigma$ flux of the sky noise in the  \mosaic\ field (for a wavelength width of $600$ km/s). The sky noise in the \udft\ field is $\approx1.7$ times lower than in the \mosaic\ field. The dark-grey and light-grey hashed areas show redshift ranges that are not included in our sample and that are highly affected by sky lines, respectively. Lower left (right) panel: $X_{\rm LAE}$ vs. $z$ for $-20.25\leq M_{1500}\leq-18.75$ mag and $EW(\rm Ly\alpha)\geq25$ \AA\ ($EW(\rm Ly\alpha)\geq55$ \AA). Purple (orange) pentagons indicate our MUSE results. A grey square, triangle (right), diamond, inverted triangle, circle, triangle (left), cross, triangle,  and thin diamond represent the results by \citet{Stark2011}, \citet{Treu2013}, \citet{Schenker2014}, \citet{Tilvi2014},  \citet{DeBarros2017}, \citet{Pentericci2018}, \citet{ArrabalHaro2018}, \citet{Caruana2018}, and \citet{Mason2019}, respectively. The upper limits in \citet{Tilvi2014} and \citet{Mason2019} show the $86$\% and $68$\% confidence levels of $X_{\rm LAE}$, respectively, while other error bars show 1$\sigma$ uncertainties of $X_{\rm LAE}$. We note that the original $M_{1500}$ ranges in \citet{Treu2013}, \citet{Schenker2014}, \citet{Tilvi2014}, \citet{DeBarros2017}, \citet{ArrabalHaro2018},  and \citet{Mason2019} are $M_{1500}\geq-20.25$ mag. In \citet{Caruana2018}, the original $M_{1500}$ range can be roughly estimated from the apparent $F775W$ cut (see Figure\,\ref{fig:muv_z}), and they include Ly$\alpha$ halos in the flux measurement. For visualization purposes, we slightly shift the data points of \citet{Tilvi2014} and \citet{ArrabalHaro2018} along the abscissae. } 
\label{fig:zevocompare}
\end{figure*}

Next, we compare our MUSE results of $X_{\rm LAE}$ with previous studies. For this purpose, we derive $X_{\rm LAE}$ for $M_{1500}$ in the range $[-20.25;-18.75]$ \citep[which corresponds to the faint UV magnitude range of][see Figure \ref{fig:muv_z}]{Stark2011}, and for $EW(\rm Ly\alpha)\geq25$~\AA\ and $EW(\rm Ly\alpha)\geq55$ \AA. Figure\, \ref{fig:zevocompare} shows our results and those of other studies as a function of redshift. At $z\lesssim5$, we confirm the low values from \citet{ArrabalHaro2018} (grey crosses at $z\approx4$ and $z\approx5$) for $EW(\rm Ly\alpha)\geq25$ \AA. Our median values of $X_{\rm LAE}$ at $z\approx4.1$ and $z\approx4.7$ are somewhat smaller than those at $z\approx4$ and $z\approx5$ from \citet{Stark2011}, although they are compatible within the large error bars. At $z\approx5.6$, our value for $EW(\rm Ly\alpha)\geq25$ \AA\ appears to be significantly lower than those reported in the literature. We note that the discrepancy between our work and \citet{DeBarros2017} is however only $1.14\sigma$, and might thus be caused by the statistical error. However, our result is more than $2\sigma$ away from those of \citet{ArrabalHaro2018} and \citet{Stark2011}, which is less likely to be statistical fluctuation. We discuss in Sect. \ref{subsubsec:lbgbias} potential biases which may explain why these two latter references find large values of $X_{\rm LAE}$. We discuss the effect of cosmic variance in Sect. \ref{subsubsec:galics_cv}, and find that it cannot explain such large differences. Another possibility is that the low value we find reflects a late and/or patchy reionization process, and we discuss that further in Sect. \ref{subsec:implications}.

We also check the best fit linear relation of $X_{\rm LAE}$ as a function of $z$ for $EW(\rm Ly\alpha)\geq25$ \AA, which is $X_{\rm LAE} = 0.01^{+0.05}_{-0.05}z +0.16^{+0.20}_{-0.22}$. The best fit slope is lowered by the point at $z\gtrsim5.6$, and is shallower than $0.11\pm0.04$ in \citet{Stark2011} and $0.18^{+0.06}_{-0.06}$ in \citet[][]{ArrabalHaro2018}. It is consistent with the flat relation reported by \citet{Hoag2019bMNRAS} within the $1\sigma$ error bars ($0.014\pm0.02$) and that by \citet{Caruana2018}. Note that \citet{Caruana2018} discuss the slope of $X_{\rm LAE}$ against $z$ using a sample of the MUSE-Wide GTO survey with an {\it apparent} magnitude cut of $F775<26.5$ mag, which is shown by a grey solid line in Figure\,\ref{fig:muv_z}. They also include the contribution of extended Ly$\alpha$ halos to their Ly$\alpha$ fluxes, which enhances the values, contrary to us. 
With regard to $EW(\rm Ly\alpha)\geq55$ \AA\, the best fit relation is $X_{\rm LAE} = 0.04^{+0.05}_{-0.03}z -0.05^{+0.14}_{-0.19}$, whose slope is consistent with that in \citep[]{Stark2011}, $0.018\pm0.036$.

\subsection{UV magnitude dependence of $X_{\rm LAE}$} \label{subsec:uvdep}

\begin{figure}
 \sidecaption
   \includegraphics[width=9.0cm]{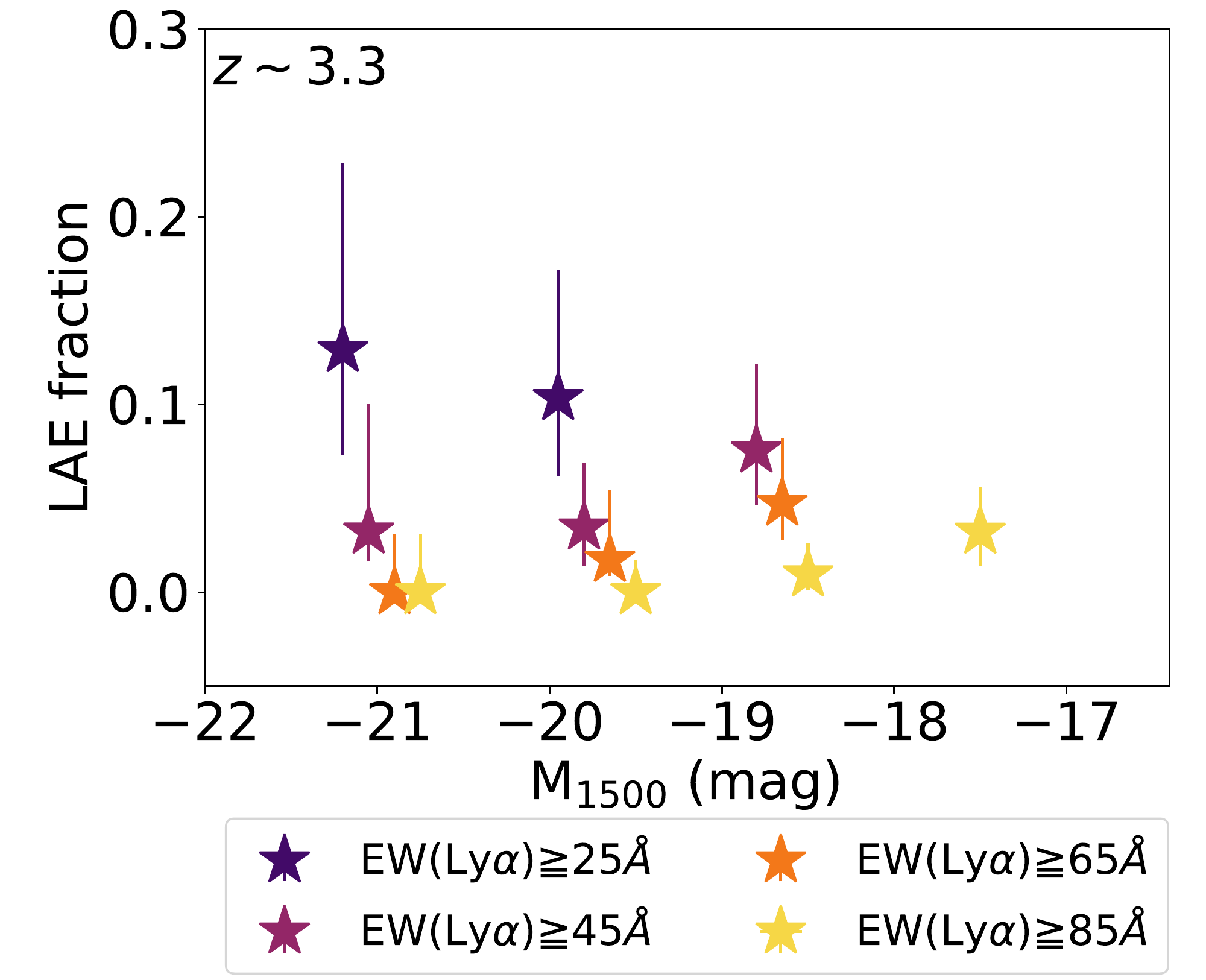}\\
      \caption{ $X_{\rm LAE}$ vs.$M_{1500}$ at $z\approx3.3$ ($=2.91$--$3.68$) for $M_{1500}\in [-21.5; -17.0]$. Purple, violet, orange and yellow stars indicate our MUSE results with $EW(\rm Ly\alpha)\geq25$ \AA, $EW(\rm Ly\alpha)\geq45$ \AA, $EW(\rm Ly\alpha)\geq65$ \AA, and $EW(\rm Ly\alpha)\geq85$  \AA, respectively. For visualization purposes, we show the width of $M_{1500}$ only for the violet stars and slightly shift the other points along the abscissa.}
\label{fig:uvdep}
\end{figure}

Figure\, \ref{fig:uvdep} shows a diagram of $X_{\rm LAE}$ and $M_{1500}$ at $z=2.91$--$3.68$ ($\approx3.3$) for our MUSE sample. This is the first time that the dependence of the LAE fraction on $M_{1500}$ is studied at $M_{1500}\geq-18.5$ mag. The LAE fractions for $EW(\rm Ly\alpha)\geq25$ \AA\ (45 \AA{}, 65 \AA{}, and 85 \AA{}) are shown with the purple (violet, orange, and yellow) stars. The best fit linear relations are 
$X_{\rm LAE} = -0.02^{+0.07}_{-0.08}M_{1500} -0.30^{+1.39}_{-1.62}$,
$X_{\rm LAE} = 0.01^{+0.02}_{-0.03}M_{1500} +0.33^{+0.45}_{-0.54}$,
$X_{\rm LAE} = 0.02^{+0.01}_{-0.01}M_{1500} +0.41^{+0.18}_{-0.30}$, and
$X_{\rm LAE} = 0.01^{+0.01}_{-0.01}M_{1500} +0.12^{+0.13}_{-0.14}$ for EW cuts of $25$ \AA, $45$ \AA, $65$ \AA, and $85$ \AA, respectively. We find no clear dependence of $X_{\rm LAE}$ on $M_{1500}$ in tension with the clear rise of $X_{\rm LAE}$ to faint UV magnitude for an EW cut of $50$ \AA reported in \citep{Stark2010}.

Our results also show that the LAE fraction is sensitive to the equivalent width selection, as expected e.g. from \citet{Hashimoto2017b}. Although this means that the LAE fraction is useful in itself to test cosmological galaxy evolution models \citep[see Sect. \ref{subsec:galics} and][]{Forero-Romero2012, Inoue2018}, it also raises concern for the usage of $X_{\rm LAE}$ as a probe of the IGM neutral fraction at the end of reionization \citep[see also][]{Mason2018}, since homogeneous measurements of Ly$\alpha$ emission over a wide redshift range are required for a fair comparison.

\begin{figure}
 \sidecaption
   \includegraphics[width=9.cm]{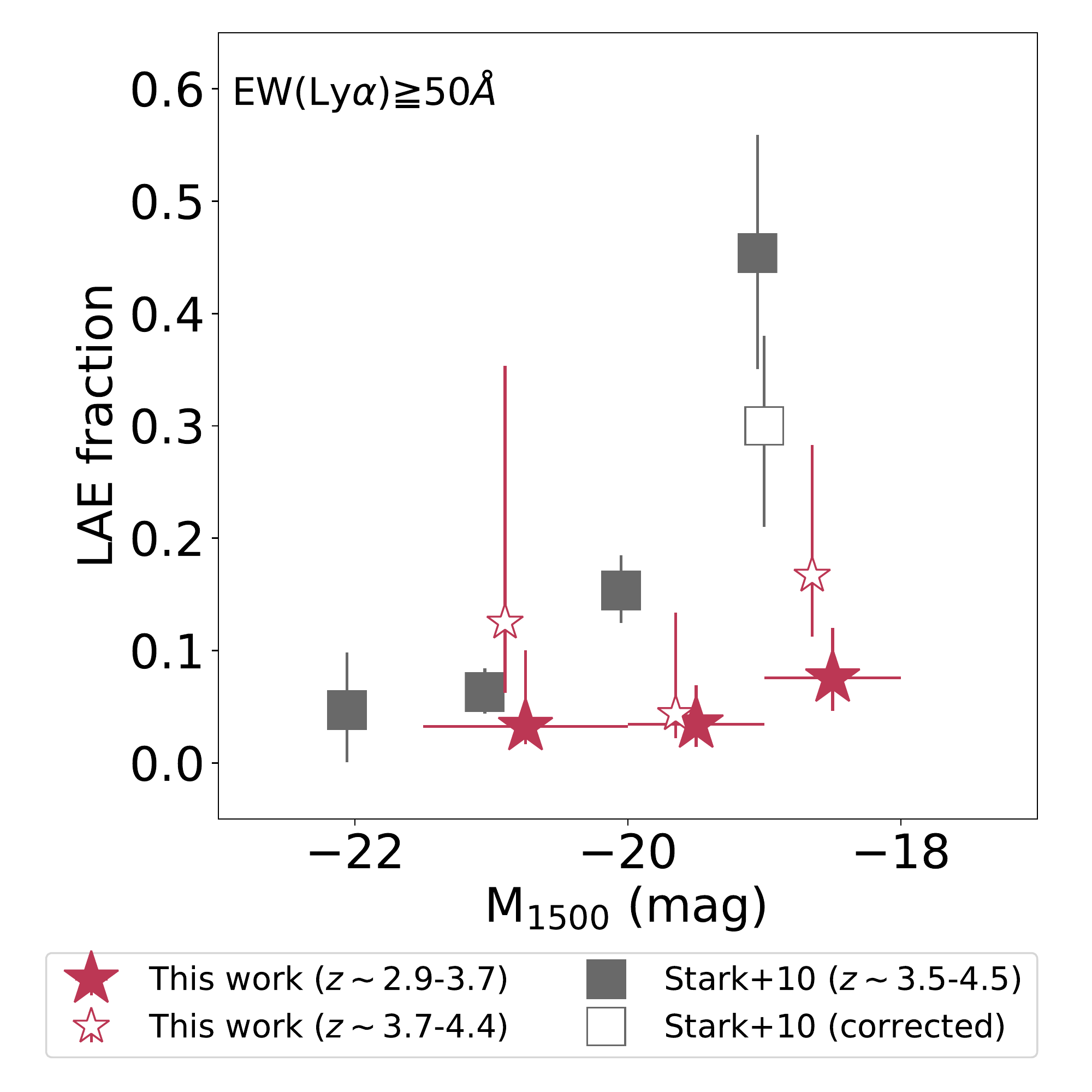}\\
      \caption{$X_{\rm LAE}$ vs. $M_{1500}$ for $EW(\rm Ly\alpha)\geq50$ \AA\ at $z\approx3$-$4$. Our $X_{\rm LAE}$ at $z\approx3.3$ ($\approx2.9$--$3.7$) and $z\approx4.1$ ($\approx3.7$--$4.4$), are indicated by filled and open magenta stars, respectively.  \citet{Stark2010}'s $X_{\rm LAE}$ at $z\approx3.5$--$4.5$ is shown by filled grey squares. The open grey square indicates the corrected $X_{\rm LAE}$ value of \citet{Stark2010} at $M_{1500}=-19$ mag (see Sect. \ref{subsubsec:lbgbias}). For visualization purposes, we slightly shift the magenta open stars and the grey open square along the abscissa and show the width of $M_{1500}$ only for the red filled stars.}
\label{fig:uvdepcompare}
\end{figure}

In Figure\,\ref{fig:uvdepcompare}, our results for the relation between $X_{\rm LAE}$ and $M_{1500}$ for $EW(\rm Ly\alpha)\geq50$ \AA\ at $z\approx3$-$4$ (filled and open red stars) are compared with those in \citet{Stark2010} (filled grey squares). The best fit linear relations for our results at $z\approx2.9$--$3.7$ and at $\approx3.7$--$4.4$ are $X_{\rm LAE} = 0.01^{+0.02}_{-0.03}M_{1500} +0.34^{+0.44}_{-0.55}$ and $X_{\rm LAE} = 0.01^{+0.06}_{-0.10}M_{1500} +0.33^{+1.12}_{-1.80}$, respectively. We find no dependence of $X_{\rm LAE}$ on $M_{1500}$ as opposed to the claim in \citet{Stark2010}, whose best fit relation is $X_{\rm LAE} = 0.13^{+0.03}_{-0.03}M_{1500} +2.87^{+0.74}_{-0.72}$. Our $X_{\rm LAE}$ is lower than that in \citet{Stark2010} at UV magnitude fainter than $M_{1500}\approx-19$ mag, and possibly at $M_{1500}\approx-20$ mag. We discuss the difference in $X_{\rm LAE}$ between this work and \citet{Stark2011} in Sect. \ref{subsec:difference}.

\section{Discussion}\label{sec:discussion}

In this section, we assess the cause of the differences between our results and previous results, compare our results with predictions from a cosmological galaxy formation model, and discuss the evolution of the LAE fraction and implication for reionization.

\subsection{Possible causes of the differences between our MUSE results and previous results}\label{subsec:difference}

\begin{figure}
 \sidecaption
   \includegraphics[width=9.0cm]{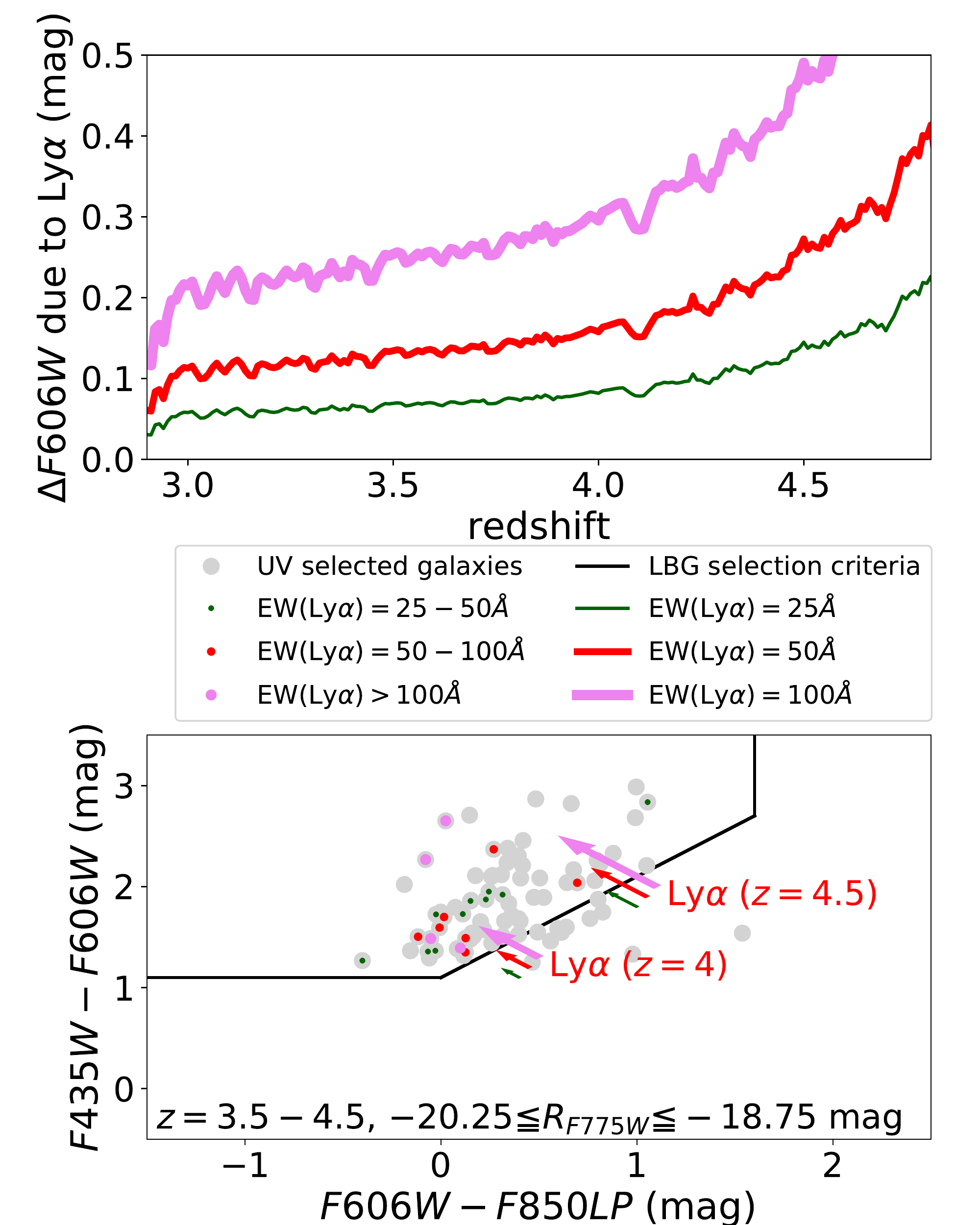}\\
      \caption{Tests of a possible LBG selection bias. 
Upper panel: The shift of $F606W$ magnitude due to contamination of Ly$\alpha$ emission as a function of redshift. Green, red, and violet lines show the shifts for $EW(\rm Ly\alpha)=25$ \AA, $EW(\rm Ly\alpha)=50$ \AA, and $EW(\rm Ly\alpha)=100$ \AA, respectively.
Lower panel: Color-color diagram for $B$ ($F435W$)-dropouts: $F435W$-$F606W$ as a function of $F606W$-$F850LP$. The grey, green, red, and violet points indicate UV selected galaxies with $-20.25 \leq M_{1500}\leq-18.75$ mag at $z_{\rm p}=3.5$--$4.5$, those with $EW(\rm Ly\alpha)=20$--$50$ \AA, $EW(\rm Ly\alpha)=50$--$100$ \AA, and $EW(\rm Ly\alpha)\geq100$  \AA, respectively. The black line represents the color-color criteria for $B$-dropout. Green, red, and violet arrows show the shifts of colors due to contamination of Ly$\alpha$ emission to $F606W$ at $z\approx4$ and $z\approx4.5$ for $EW(\rm Ly\alpha)=25$ \AA, $EW(\rm Ly\alpha)=50$  \AA, and $EW(\rm Ly\alpha)=100$ \AA, respectively. }
\label{fig:lbgtest}
\end{figure}

In Figure\,\ref{fig:zevocompare}, we find that our measurements of $X_{\rm LAE}$ are systematically lower than those of \citet{Stark2011}, although consistent within the error bars. This tension supports the results of \citet{ArrabalHaro2018}, who also find low median values of $X_{\rm LAE}$ at $z\approx4$--$5$. It is worth discussing the potential origins of this tension, since the median values of $X_{\rm LAE}$ have been used to assess cosmic reionization in theoretical studies \citep[e.g.,][]{Dijkstra2014}. The difference between our study and that of \citet{Stark2010} is more striking in Figure\,\ref{fig:uvdepcompare} which shows $X_{\rm LAE}$ as a function of $M_{1500}$. Here, our results are inconsistent with theirs at a faint UV magnitude, even when taking into account the large error bars. Below we discuss two possible origins of this discrepancy in the plot of $X_{\rm LAE}$ as a function of $M_{1500}$: the LBG selection bias, and systematics due to different observing methods. 

\subsubsection{LBG selection bias}\label{subsubsec:lbgbias}
There is a possibility that the LBG sample of \citet{Stark2010} is biased towards bright Ly$\alpha$ emission i.e., higher $X_{\rm LAE}$, as pointed out in previous studies \citep[LBG selection bias, e.g., ][see also \citealt{Cooke2014} for another potential bias of LBGs due to LyC leakers]{Stanway2008, DeBarros2017,Inami2017}. In other words, LBG selections could be biased towards having higher $X_{\rm LAE}$ if they preferentially miss low-EW sources. The LBG selection consists of a set of color-color criteria and signal-to-noise cuts \citep[e.g.,][]{Stark2009}. \citet{DeBarros2017} obtain a relatively low median $X_{\rm LAE}$ at $z\approx6$ and discuss the causes. They use common selection criteria for the i-dropout \citep{Bouwens2015b}, but add an additional criterion of the H-band ($F160W$, rest-frame UV at $\approx6$) magnitude cut. When Ly$\alpha$ emission is located in the red band, strong Ly$\alpha$ emission in a UV spectrum can significantly enhance the Lyman break. The additional criterion in \citet{DeBarros2017} can suppress the LBG selection bias, which increases $X_{\rm LAE}$ for faint UV sources as will be discussed below.

Here we estimate the effect on $X_{\rm LAE}$ of two aspects of the LBG selection bias for $B$ ($F435W$)-dropouts: the impact of Ly$\alpha$ contamination on the signal-to-noise ratio cut in the $V$ ($F606W$) band, and on the color-color criteria in a diagram of $F435W$-$F606W$ vs. $F606W$-$F850LP$ (black solid line in Figure\,\ref{fig:lbgtest}). We estimate $F606W$ magnitudes assuming a power-law spectrum with a UV slope of $-2$ ($\lambda\geq912$ \AA\ in rest frame) and IGM transmission following \citet{Madau1995}. As shown in the upper panel in Figure\,\ref{fig:lbgtest}, strong Ly$\alpha$ emission can increase the flux in $F606W$ noticeably, especially at $z\gtrsim4$. The magnitude shift becomes larger at higher redshifts because of the increasingly smaller rest-frame wavelength range of the UV continuum that is covered by $F606W$ as the redshift goes up. Even at $z\approx4$, however, a moderate Ly$\alpha$ emission line with $EW(\rm Ly\alpha)=50$ \AA\ can cause a $\approx0.16$ mag difference in $F606W$. This affects both the signal-to-noise ratio and the colors. 

We first illustrate the effect on the signal-to-noise ratio cut of $F606W$ by considering an object with $M_{1500}=-19$ mag. At $z=4.5$ ($z=4.0$), a source without Ly$\alpha$ emission has $F606W=28.3$ mag ($F606W=27.5$ mag), while a source with $EW(\rm Ly\alpha)=50$ \AA\ has a $0.27$ ($0.16$) brighter magnitude. Since these $F606W$ magnitudes are close to the 5$\sigma$ limiting magnitude of $28.0$ mag in \citet{Stark2009}, which corresponds to a completeness of $50$\% in the case of a $S/N\geq5$ cut, the completeness for their $B$-dropout changes drastically around $28.0$ mag.

Second, the effect on the two colors for the $B$-dropout selection is shown in the lower panel of Figure\,\ref{fig:lbgtest}. The green, red and violet arrows show color-color shifts in the diagram of $F435W$-$F606W$ and $F606W$-$F850LP$ due to the magnitude shift of $F606W$ in the case of $EW(\rm Ly\alpha)=25$ \AA, $50$ \AA, and $100$ \AA, respectively. The shift of $F606W$ enhances the possibility of meeting the dropout criteria. Indeed, at $z=3.5$--$4.5$, all of our LAEs with an absolute magnitude of $F775W=-20.25$--$-18.75$ mag are located in the dropout selection region (upper left region of bottom panel in Figure\, \ref{fig:lbgtest}). However, $\approx10$\% of continuum-selected galaxies do not meet the dropout criteria. Therefore, strong Ly$\alpha$ emission can enhance the probability to meet LBG-selection criteria both in terms of the signal-to-noise cut and of the color-color criteria. 

We estimate the completeness of the $B$-dropout galaxies in \citet{Stark2009} using a plot of surface number density as a function of UV magnitude for $B$-dropouts in \citet{Bouwens2007}.
\citet{Stark2009} use the same color-color criteria as \citet{Bouwens2007} but with $\approx0.6$ mag shallower data sets than those in \citet{Bouwens2007}.  Figure 1 in \citet{Bouwens2007} shows the surface number density of the $B$-dropouts as a function of apparent $F775W$ magnitude (i.e., apparent rest-frame UV magnitude). At absolute UV magnitudes of $\approx-18.8$ and $-18.3$ at $z\approx4$, the completeness values are $\approx25$\% and $\approx5$\%, respectively. The completeness values for $\approx0.6$ mag shallower data in \citet{Stark2009} are estimated to be $\approx25$\% and $\approx5$\% at $M_{1500}\approx-19.4$ mag and $M_{1500}\approx-18.9$ mag, respectively, if the behavior of completeness as a function of $S/N$ is similar. As shown in the lower panels of Figure \ref{fig:n_muv}, the $B$-dropout galaxies in \citet{Stark2009} are not complete at $M_{1500}\approx-19.0$ mag.  Moreover, \citet{Stark2009} adopt stricter signal-to-noise ratio cuts than \citet{Bouwens2007}'s criteria for $B$-dropouts, and the completeness values in \citet{Stark2009} may be lower than estimated here,  especially for faint sources. 

Following the discussions above, at $z\approx3.5$--$4.5$, the observed number (the denominator of $X_{\rm LAE}$) for their B-dropouts with $M_{1500}\approx-19$ mag is estimated to be less than $25$\% of the true value, while the observed number (the numerator of $X_{\rm LAE}$) for LAEs with $EW(\rm Ly\alpha)=50$ \AA\ is estimated to be larger than $50$\% of the true value under the assumption that all the LAEs meet the color-color criteria. This means that $X_{\rm LAE}$ for their B-dropout sample may be more than $\approx1.5$ times larger than the $X_{\rm LAE}$ for a complete sample, $0.3^{+0.08}_{-0.09}$. If the overestimate would be corrected for $X_{\rm LAE}$ at $M_{1500}\approx-19$ mag, their $X_{\rm LAE}$ would be consistent with ours within the $1\sigma$ error bars as shown by the open grey square in Figure\,\ref{fig:uvdepcompare}.

Therefore, the $B$-dropout selection bias may be the dominant cause of the difference in $X_{\rm LAE}$ between LBGs and photo-$z$ selected galaxies at $z\approx4$ at faint UV magnitudes. This may also have an effect on the difference in $X_{\rm LAE}$ at $z\approx4$ shown in Figure\,\ref{fig:zevocompare}. Indeed, the difference for the $55$ \AA\ cut is more pronounced than that for the $25$ \AA\ cut. 
Although we do not discuss quantitatively biases of dropout selections at other redshifts, strong Ly$\alpha$ emission will cause similar effects, as discussed in the references. We note that the LBG sample in \citet{ArrabalHaro2018} consists of $\approx70$\%  photometric-redshift selected objects and of $\approx30$\% dropout selected objects based on dropout selection criteria for their medium-band filters. Since we measure $X_{\rm LAE}$ for a photometric-redshift selected sample, this may result in the similarity of $X_{\rm LAE}$ to ours at $z\lesssim5$, though their sample is not complete in UV at $M_{1500}\gtrsim-20$ mag. 

Note that \citet{Oyarzun2017} mention yet another potential bias for LBG samples which are incomplete in UV, due to a correlation between the $EW(\rm Ly\alpha)$ and $M_{1500}$. This bias leads to an underestimate of $X_{\rm LAE}$ because large equivalent width objects are preferentially missed when faint-UV galaxies drop out of the sample. Our results are not affected by this bias, but it could affect the results of other work shown in Figure \ref{fig:zevocompare}, and could have compensated the LBG selection bias we discussed above. In Figure \ref{fig:uvdepcompare}, the bias from \citet{Oyarzun2017} has no impact because we are looking at $X_{\rm LAE}$ as a function of UV magnitude.

\subsubsection{Different observational methods}\label{subsubsec:method}

The Ly$\alpha$ emission in our sample is measured with an IFU (without including the Ly$\alpha$ halo) and is thus less affected by uncertainties due to slit-loss and aperture corrections. \citet{Hoag2019bMNRAS} measure the spatial offset between the Ly$\alpha$ emission and the UV continuum. They find a typical standard deviation for the offset which decreases towards higher redshifts ($2.17^{+0.19}_{-0.14}$ kpc ($\approx0\farcs3$) at $z\approx3.25$ to $1.19^{+1.29}_{-0.33}$ kpc ($\approx0\farcs2$) at $z\approx5.25$). They argue that the evolution of the spatial offset contributes to the increasing trend of $X_{\rm LAE}$ with $z$ measured with slit spectroscopy with $1\farcs$ slits such as in \citet{Stark2011}. According to \citet{Hoag2019bMNRAS} Figure 7, the simulated cumulative distribution function (CDF) of slit-loss is similar from $z\approx3.5$ to $5.5$ but is shifted at $z\approx3$--$3.5$ to a larger slit-loss for a $1\farcs$ slit. At $z\approx3.5-4.5$, the CDF reaches $\approx90$\% at a slitloss of $\approx10$\%. Moreover, their measured offsets are much larger than that for a lensed LAE at $z\approx1.8$ \citep[$0.65$ kpc,][]{Erb2019arXiv} and typical values for LAEs at $z\approx3$--$6$ \citep[$\lesssim0\farcs1$,][]{Leclercq2017}. Hence, this is probably not the dominant cause of the high $X_{\rm LAE}$ values of \citet{Stark2010,Stark2011} at $z\approx3.5$--$4.5$. Note that the aperture diameters (convolved mask diameters) for our Ly$\alpha$ measurement with IFU data are typically larger than $1''$ (see Sect. \ref{subsec:muse}). Our measurements are less affected by the spatial offset between the Ly$\alpha$ emission and the UV continuum. Meanwhile, \citet{Hoag2019bMNRAS} estimate slitlossess based on the spatial component of slit spectra.

\subsubsection{Summary}\label{subsubsec:summary}

We find indications that the dominant cause of the difference in $X_{\rm LAE}$ measured in \citet{Stark2010,Stark2011} and presented here is the LBG selection bias. Strong Ly$\alpha$ emission can enhance the probability to meet the LBG-selection criteria both in terms of the signal-to-noise ratio and color. The LBG selection bias has a strong effect on $X_{\rm LAE}$ especially for faint UV magnitudes, where LBG samples are not complete. Possible discrepancies arising from different observational methods probably affect $X_{\rm LAE}$ to a lesser extent. Thanks to the MUSE observations and to the HST photo-$z$ sample, our $X_{\rm LAE}$ are derived from the most homogeneous and complete sample to date.

\subsection{Comparison with the GALICS model}\label{subsec:galics}
Using a homogeneous and complete UV sample and MUSE spectroscopic data, we have measured the LAE fraction for the first time at very faint magnitudes ($M_{1500}\leq-17.0$ mag). While we confirm a weak increase of $X_{\rm LAE}$ as a function of redshift at $3 \lesssim z \lesssim$ $5$, we find that LAEs with $EW(\rm Ly\alpha)\geq45$ \AA\  make up a relatively low fraction of the underlying rest-frame UV-detected galaxy population, $\approx0$--$20$\%. This implies the existence of a duty cycle either for the star formation activity or for the escape and/or production of Ly$\alpha$ photons. Another possibility is that only a small fraction of all galaxies can evolve into LAEs or can be observed as LAEs in a limited range of inclinations \citep[e.g.][]{Verhamme2012}. Our results suggest no dependence of $X_{\rm LAE}$ on $M_{1500}$ at $z\approx3$. Keeping in mind that we want to assess the merits of the redshift evolution of $X_{\rm LAE}$ to probe the IGM neutral fraction at $z\gtrsim 6$, it is essential to understand these trends after reionization. To do so, we compare our results with predictions from the semi-analytic model of \citet{Garel2015}. 

\subsubsection{Description of the model}\label{subsubsec:galics_descrip}

\citet{Garel2015} present an updated version of the GALICS hybrid model \citep[Galaxies In Cosmological Simulations,][]{Hatton2003} which is designed to study the formation and evolution of galaxies in the high redshift Universe. GALICS relies on an N-body cosmological simulation to follow the hierarchical growth of dark matter structures and on semi-analytic prescriptions to describe the physics of the baryonic component. The box size of the simulation is $100$ $h^{-1}$ cMpc on a side, and the dark-matter particle mass is $\approx8.5\times10^7\, M_\odot$ (with $1024^3$ particles) \citep{Garel2012}. In \citet{Garel2015}, stars are formed according to a Kennicutt-Schmidt law when the galaxy’s gas surface density $\Sigma_{\rm gas}$ is larger than a threshold value, $\Sigma_{\rm gas}^{\rm thresh}$ \citep[e.g.,][]{Schmidt1959,Kennicutt1998}, and the intrinsic \lya{} emission from galaxies is computed assuming case B recombination as $L_{Ly\alpha}^{\rm intr} \propto 0.67 Q(H)$. Here, $Q(H)$ is the production rate of hydrogen-ionising photons estimated from the stellar spectral energy distributions.
In order to predict the observed \lya{} properties of galaxies, \citet{Garel2015} combine GALICS with the library of radiative transfer (RT) simulations of \citet{Schaerer2011} which predict the escape fraction of \lya{} photons through galactic winds \citep[$f_{\rm esc}$; see ][for more details]{Verhamme2006,Verhamme2008,Garel2012}. $f_{\rm esc}$ depends on the wind parameters (the wind expansion velocity, velocity dispersion, dust opacity, neutral hydrogen column density) which are computed by GALICS. The \lya{} luminosity emerging from each individual galaxy is then given by $L_{\rm Ly\alpha} = L_{\rm Ly\alpha}^{\rm intr} \times f_{\rm esc}$. 

The GALICS model was tuned to reproduce the UV and Ly$\alpha$ luminosity functions at $z\approx3$--$6$ in \citet{Garel2012}. This model was then shown to also reproduce accurately the observed stellar mass functions and star-formation-rate to stellar mass relations at $z\approx3$--$6$ \citep{Garel2016}. While their fiducial model can match these observational constraints at $3 \lesssim$ $z \lesssim$ $6$, it fails to reproduce the wide distribution of \lya{} EWs, in particular the high EW values (at $EW(\rm Ly\alpha)\gtrsim50$ \AA). \citet{Garel2015} discuss the possibility that this mismatch is linked to the lack of burstiness of star formation in GALICS given that the \lya{} EW is primarily set by the combination of (i) the production rate of \lya{} photons, dominated by short-lived stars, and (ii) the stellar UV emission which traces star formation over longer timescales \citep[see also][]{Charlot1993,Madau1998}. In the fiducial model, most galaxies keep forming stars at a rather constant rate because the surface gas density threshold is almost always met. In an alternative model (labeled bursty SF), they increase this threshold by a factor $10$, such that gas needs to accrete onto galaxies for longer periods before reaching the required surface density. This naturally gives rise to a star formation duty cycle and \citet{Garel2015} show that their bursty SF model predicts EW distributions in much better agreement with observations than the fiducial model does. 

Following the procedure of \citet{Garel2016}, we create mock surveys for both the fiducial and bursty SF models using the Mock Map Facility (MOMAF) tool \citep{Blaizot2005}. In practice, for each model, we generate 100 lightcones that mimic the geometry and redshift range of the MUSE HUDF survey, i.e. a square field of $\approx10$ arcmin$^2$ and $2.8 \lesssim$ $z \lesssim$ $6.7$, and we compute $X_{\rm LAE}$ from the mocks in the same bins of redshift and UV magnitude as for the observational measurements.  

\subsubsection{Measured LAE fraction vs. GALICS predictions}\label{subsubsec:galics_results}

\begin{figure}
 \sidecaption
   \includegraphics[width=9cm]{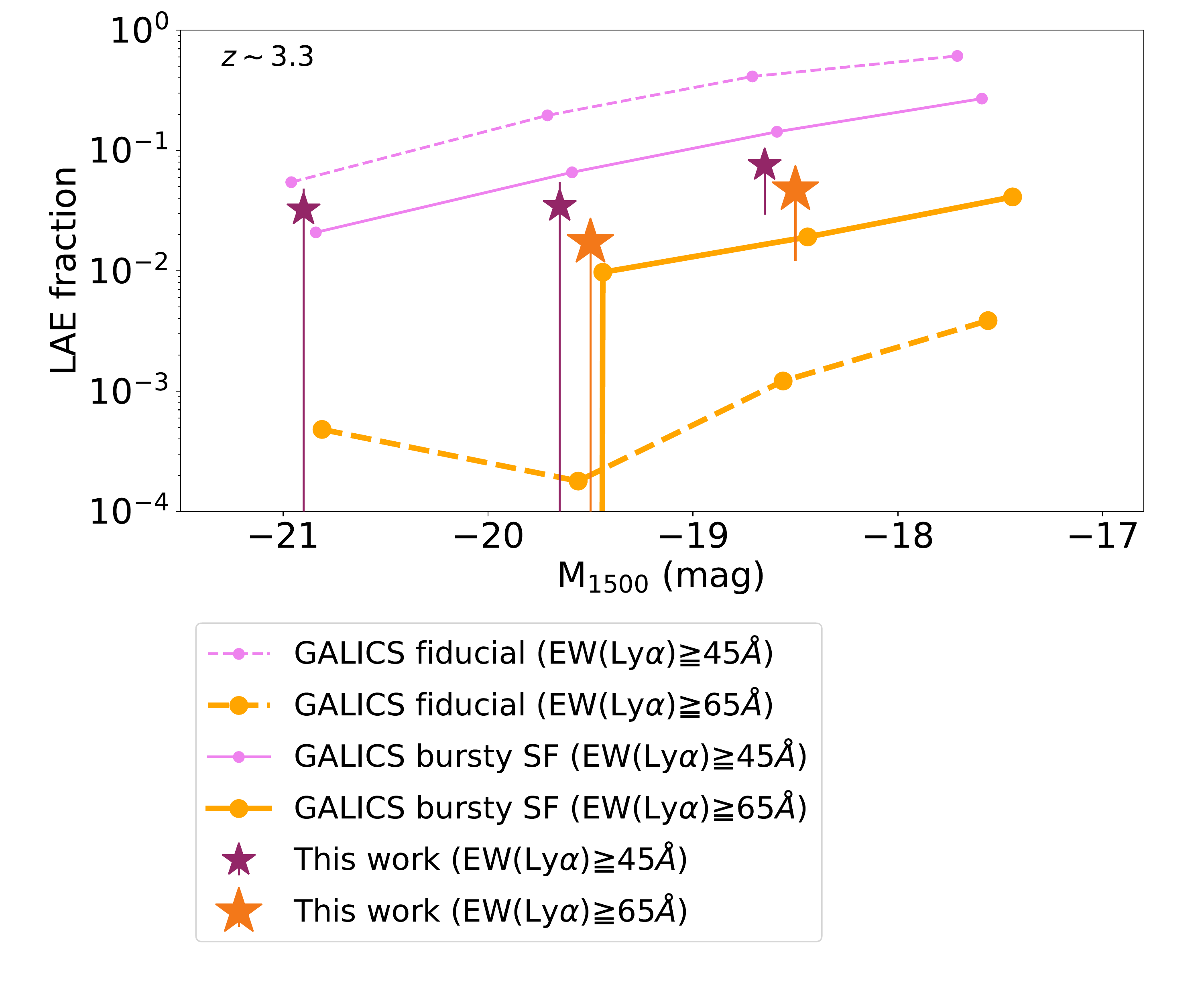}\\
      \caption{ 
      $X_{\rm LAE}$ vs. $M_{1500}$ at $z\approx3.3$ from our MUSE results compared to predictions from the GALICS mocks. The MUSE results at $z\approx3.3$ for $EW(\rm Ly\alpha)\geq45$ \AA\ and $EW(\rm Ly\alpha)\geq65$ \AA\ are indicated by violet and orange stars, respectively. The magenta and orange {\it dashed} lines with dots show the average $X_{\rm LAE}$ computed from 100 mocks of the fiducial GALICS model \citep[][]{Garel2015} at the same redshift for $EW(\rm Ly\alpha)\geq45$ \AA\ and $EW(\rm Ly\alpha)\geq65$ \AA, respectively. Those from the bursty SF model are shown by {\it solid} lines with dots. For visualization purposes, we slightly shift the points along the abscissa. Note that $X_{\rm LAE}$ for $EW(\rm Ly\alpha)\geq65$ \AA\ from MUSE and from the bursty SF model is $0$ at $M_{1500}\approx-21$ mag.} 
\label{fig:uvdep_galics}
\end{figure}

In Figure\,\ref{fig:uvdep_galics}, we show our MUSE measurement of $X_{\rm LAE}$ as a function of $M_{1500}$ at $z\approx3.3$ for $EW(\rm Ly\alpha)\geq45$ \AA\ and $EW(\rm Ly\alpha)\geq65$ \AA{}. These EW cuts correspond to our most secure measurements. We also show in this Figure the predictions from the GALICS model. 
Both fiducial and bursty SF GALICS models (dashed and solid lines, respectively) show an increase of $X_{\rm LAE}$ towards faint UV magnitudes with a slope which is in good agreement with the data (star symbols). As discussed in \citet{Verhamme2008} and \citet{Garel2015}, this trend may be the result of two factors. First, UV bright sources have intrinsically smaller $EW(\rm Ly\alpha)$ due to less significant or less recent bursts of star formation. Second, these galaxies are often more massive with higher \HI\ and dust contents which can dramatically reduce the escape fraction of Ly$\alpha$ photons and therefore the observed EW. Hence, few bright UV galaxies display a strong \lya{} emission line. 

For a more detailed comparison, we see that the fiducial GALICS model (dashed lines) does not reproduce the observed $X_{\rm LAE}$. This model overestimates (underestimates) the LAE fraction at almost all UV magnitudes for $EW(\rm Ly\alpha)\geq45$ \AA{} ($EW(\rm Ly\alpha)\geq65$ \AA). This is a consequence of the too narrow EW distribution predicted by this model (see Sect. \ref{subsubsec:galics_descrip}), which \citet{Garel2015} attribute to overly smooth star formation histories. In the bursty SF model however (solid lines), galaxies have more diverse recent star formation histories which result in wider EW distributions, and consequently this model is able to reproduce our measured $X_{\rm LAE}$ much better. 

\begin{figure}
 \sidecaption
   \includegraphics[width=9cm]{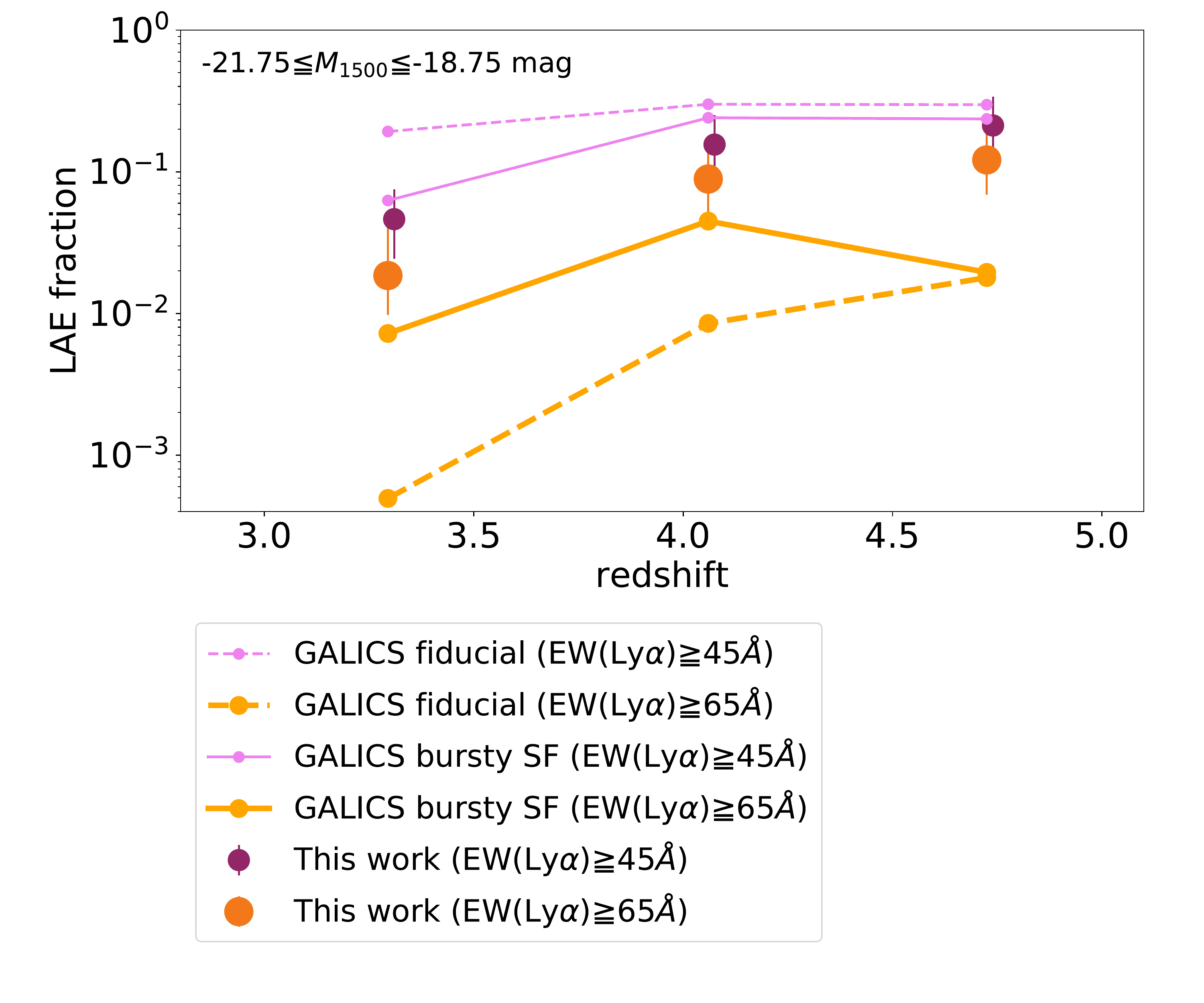}\\
      \caption{
      $X_{\rm LAE}$ vs. $z$ for $M_{1500}\in[-21.75;-18.75]$, at $z<5$, from our MUSE results compared to predictions from the GALICS model. The MUSE results for $EW(\rm Ly\alpha)\geq45$ \AA\ and $EW(\rm Ly\alpha)\geq65$ \AA\ are indicated by violet and orange circles, respectively. The magenta and orange dashed (solid) lines with dots show the average $X_{\rm LAE}$ computed from 100 mocks of the fiducial (bursty SF) GALICS model \citep[][]{Garel2015} for $EW(\rm Ly\alpha)\geq45$ \AA\ and $EW(\rm Ly\alpha)\geq65$ \AA, respectively. For visualization purposes, we slightly shift the points along the $x$-axis.}
\label{fig:xevo_galics}
\end{figure}

In Figure\,\ref{fig:xevo_galics}, we compare the GALICS predictions of $X_{\rm LAE}$ as a function of $z$ with $X_{\rm LAE}$ from the MUSE data at $z\lesssim5$, i.e. where our observational measurements are most robust. For the same reasons as above, we find that the bursty SF model is much more successful at reproducing the observations than the fiducial model. This is particularly true for the lowest EW cut (i.e. $45$ \AA) where the agreement is quite good (solid orange curve). For $EW(\rm Ly\alpha)\geq65$ \AA{} however, we note that the bursty model slightly underpredicts the observed LAE fraction (solid magenta curve), especially at z $\gtrsim 4.5$. Additional ingredients could possibly be missing from this model that would help produce more galaxies with large EWs (in particular in the higher redshift bin) such as radiative transfer in asymmetric geometries, or Ly$\alpha$ production from other channels like collisions (gravitational cooling) or fluorescence \citep[see e.g. ][for a more detailed discussion on these aspects]{Verhamme2012,Garel2015,Dijkstra2017}. Also, the assumed IMF or the metallicity evolution of model galaxies may not be realistic and lead to low EWs \citep[e.g.][]{Hashimoto2017a,Hashimoto2017b}.

Overall, these comparisons suggest that the measurements of $X_{\rm LAE}$ by MUSE in the post-reionization epoch can be reasonably well interpreted with current models of high-$z$ galaxies such as GALICS. We find that the observed trends between $X_{\rm LAE}$ and redshift/UV magnitude are mainly shaped by the burstiness of star formation in GALICS. It is also caused by the variation of $f_{\rm esc}$ with respect to the physical properties of the galaxies as discussed in \citet{Garel2015}. In GALICS, these two aspects modulate the observed \lya{} EWs of galaxies and therefore the LAE fraction at $z \lesssim 5$.

%##############################################

%##############################################
\subsubsection{Cosmic variance}\label{subsubsec:galics_cv}

\begin{figure}
 \sidecaption
   \includegraphics[width=8.8cm]{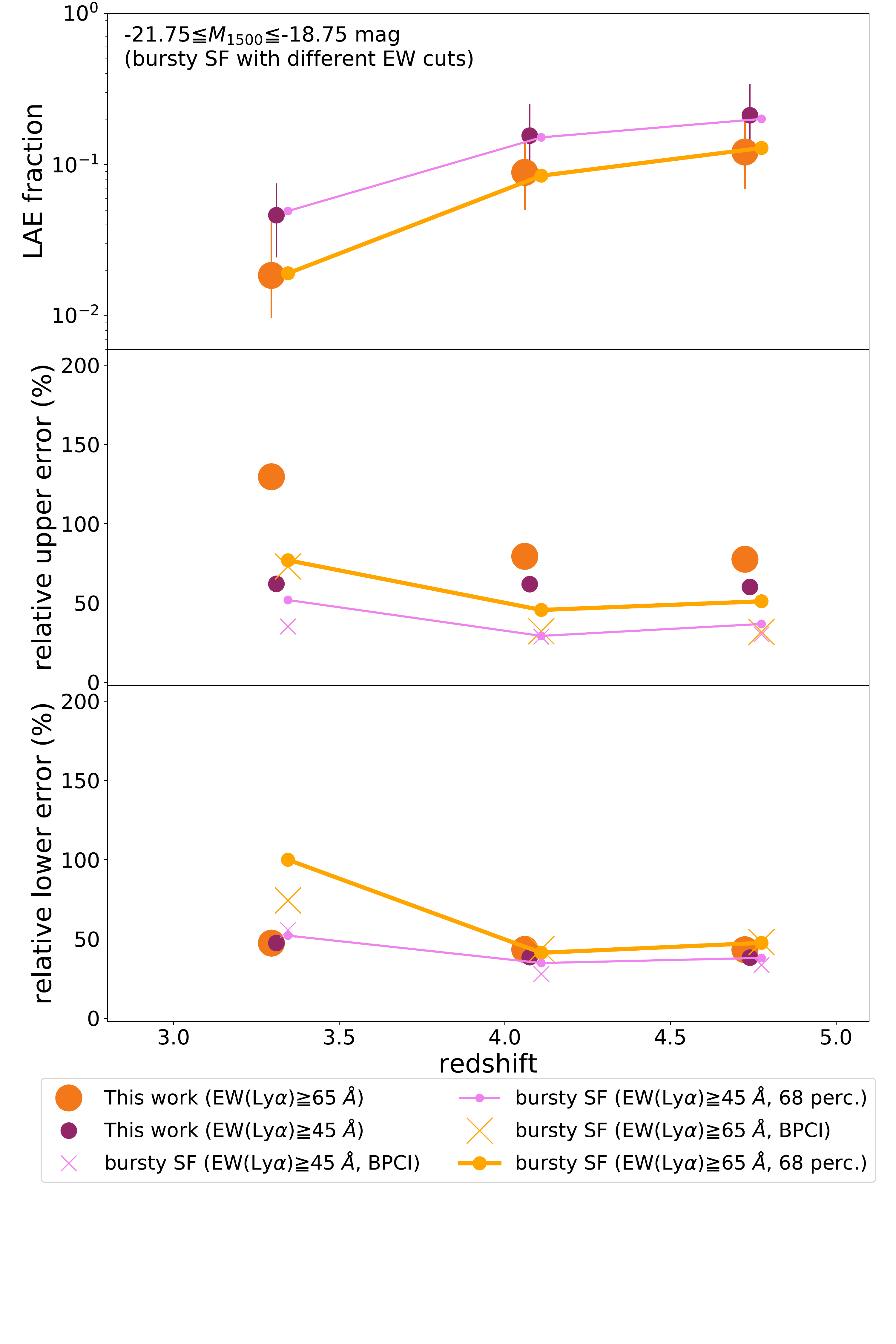}\\
      \caption{
      Test of cosmic variance and uncertainties of $X_{\rm LAE}$ for our MUSE observations using GALICS mocks of the bursty SF model. {\it top panel}: $X_{\rm LAE}$ vs. $z$ for $M_{1500}\in[-21.75;-18.75]$, at $z<5$. In order to better compare GALICS with our observations and to provide a more accurate estimate of cosmic variance, we use slightly different EW cuts for the model. We replace the 45\AA{} cut with $46$\AA, $48$\AA, and $46$\AA{} cuts at $z\approx3.3$, $4.1$, and $4.7$, and we replace the 65\AA{} cut with 53\AA, 52\AA, and 49\AA{} at the same redshifts. With these cuts, the values of $X_{\rm LAE}$ from MUSE (violet and orange circles at for $EW(\rm Ly\alpha)\geq45$ \AA\ and $EW(\rm Ly\alpha)\geq65$ \AA{}) and GALICS (solid lines) match. {\it middle panel}: Relative upper $1\sigma$ uncertainties of $X_{\rm LAE}$ vs. $z$ for $M_{1500}\in[-21.75;-18.75]$, at $z<5$. The relative $68$\% percentiles of $X_{\rm LAE}$ (field-to-field variance) measured among $100$ GALICS mocks are indicated by violet and orange circles with soloid lines for $EW(\rm Ly\alpha)\geq45$ \AA\ and $EW(\rm Ly\alpha)\geq65$ \AA, respectively. The $68$\% percentile includes both of the cosmic variance and statistical error. The statistical errors estimated from BPCI are shown by violet and orange crosses. The MUSE uncertainties (estimated from BPCI including completeness correction effects) for $EW(\rm Ly\alpha)\geq45$ \AA\ and $EW(\rm Ly\alpha)\geq65$ \AA\ are indicated by violet and orange circles, respectively. {\it bottom panel}: Relative lower $1\sigma$ uncertainties of $X_{\rm LAE}$ vs. $z$ for $M_{1500}\in[-21.75;-18.75]$, at $z<5$. The symbols are the same as those in the middle panel. For visualization purposes, we slightly shift the points along the $x$-axis.
      }
\label{fig:xevo_galicscv}
\end{figure}

The area of the MUSE HUDF survey is limited to $9.92$ arcmin$^2$ which translates into comoving volumes of $\approx 1.5-2.5 \times 10^4$ cMpc$^3$ for the redshift ranges we are considering here. As explained in Sect. \ref{subsec:uncertainties}, we have accounted for several sources of uncertainty to compute the error on $X_{\rm LAE}$ but so far we ignored cosmic variance (see Sect. \ref{subsec:uncertainties} for discussion). To assess the significance of this effect, we can estimate the cosmic variance from the GALICS mock lightcones, which are cut out from the $\approx3\times10^6$ cMpc$^3$ simulation box \citep{Garel2016}. We compute the $1\sigma$ standard deviation as the field-to-field variation, which includes the effects of both cosmic variance and the binomial proportion confidence interval. Note that our estimate of cosmic variance with GALICS only accounts for the clustering of galaxies and not for the possible contribution of large-scale variations in the IGM transmissivity due to a patchy reionization. 

In the middle and bottom panels of Figure\,\ref{fig:xevo_galicscv}, we compare the relative uncertainty of our MUSE $X_{\rm LAE}$ estimates (circles) with the relative uncertainties due to the field-to-field variation of $X_{\rm LAE}$ (solid lines with dots), which is estimated from the $100$ mocks based on the bursty SF model. For a fair comparison, we match $X_{\rm LAE}$ of GALICS to that of MUSE for $EW(\rm Ly\alpha)\geq45$ \AA\ ($EW(\rm Ly\alpha)\geq65$ \AA), by adopting slightly different $EW(\rm Ly\alpha)$ cuts in the model catalogs, of $46$ \AA, $48$ \AA, and $46$\AA\ ($53$ \AA, $52$ \AA, and $49$ \AA) at $z\approx3.3$, $4.1$, and $4.7$,  respectively (see the top panel of Figure\,\ref{fig:xevo_galicscv}). Note that the total uncertainty for the MUSE $X_{\rm LAE}$ is calculated by summing the statistical error (binomial proportional confidence interval) multiplied by completeness corrections in each flux bin for completeness correction (see Sect. \ref{subsec:uncertainties} for more details). 
For both EW cuts, the relative upper errors of our MUSE $X_{\rm LAE}$ are much larger than those of the field-to-field variance for the bursty GALICS model. The relative lower errors of our MUSE $X_{\rm LAE}$ are in the same level of those of the field-to-field variance at $z\approx4$--$5$. The contribution of cosmic variance to the field-to-field variance in the GALICS mock is negligible, since statistical errors (crosses) are dominant. It suggests that the cosmic variance is a subdominant source of uncertainties in our measurement of $X_{\rm LAE}$. Therefore, we conclude that our MUSE results are not strongly affected by cosmic variance. We note that the field-to-field variance may be slightly underestimated in the mock catalogs because the fluctuations on scales larger than the simulated box ($150$ cMpc) are not sampled. The size of the simulated volume is however significantly larger than our MUSE survey volume ($\approx 1.5-2.5 \times 10^4$ cMpc$^3$) and so this underestimate should be weak.

\subsection{The redshift evolution of the LAE fraction and implications for reionization}\label{subsec:implications}

The main purpose of this paper is to measure the evolution of $X_{\rm LAE}$ in the post-reionization epoch, at $z\lesssim6$ (as shown in Figures \ref{fig:xevo}, \ref{fig:zevocompare}, and \ref{fig:xevo_galics}). Our results confirm the rise of $X_{\rm LAE}$ with redshift found in the literature between $z\approx3$ and $6$ for $-21.75\leq M_{1500}\leq-17.75$ mag, $X_{\rm LAE} = 0.07^{+0.06}_{-0.03}z -0.22^{+0.12}_{-0.24}$, in Figure \ref{fig:xevo}. Meanwhile, the trend stopps at $z\sim5.6$ for $-20.25\leq M_{1500}\leq-18.75$ mag in Figure \ref{fig:zevocompare}.  As discussed in Sect. \ref{subsubsec:galics_cv}, this evolution at $z\approx3$-$5$ is not caused by the cosmic variance of the limited survey field of the MUSE HUDF Survey. Instead, it is probably caused by higher intrinsic $EW(\rm Ly\alpha)$ and/or higher Ly$\alpha$ escape fractions at higher redshift in a given $M_{1500}$ range, due to less massive and less dusty galaxies at higher redshift \citep[e.g.,][]{Speagle2014a,Bouwens2016a, Santini2017}. It is also very important to understand the co-evolution of the Ly$\alpha$ and UV luminosity functions \citep[e.g.,][]{Ouchi2008,Dunlop2013, Konno2016,Konno2018,Ono2018}. 

\citet{DeBarros2017} obtain a relatively low $X_{\rm LAE}$ ($\approx40$\%) at $z\approx6$, which implies a less dramatic turn-over at $z>6$ than previously found \citep[e.g.,][]{Stark2011,Pentericci2014, Schenker2014,Tilvi2014}. If  we interpret our point at $z\approx 5.5$ as a statistical fluctuation $1\sigma$ below a true value $\approx 0.35$ as found by \citet{DeBarros2017}, we confirm this shallower increase of the LAE fraction towards $z\approx5$ and $z\approx6$. This could indicate the stop of the evolution of the Ly$\alpha$ escape fraction, possibly related to the plateau evolution of the star formation main sequence at $z\approx5$--$6$ suggested by \citet{Speagle2014a} and \citet{Salmon2015}, and implying a constant stellar mass at a given $M_{1500}$ \citep[see however,][]{Santini2017}.

Another possibility is that our low value of $X_{\rm LAE}$ at $z\approx5.5$ is genuine and indeed indicates a late transition in the ionisation state of the IGM. However, our data at $z\leq5$ combined the work of \citet{DeBarros2017} and \citet{Pentericci2018}, which is not affected either by the LBG selection bias discussed above, suggest an earlier reionization, at $z\approx6-7$. Our measurement at $z\approx5.5$ may thus indicate a patchy reionization process. \citet{Bosman2018} measure the mean and scatter of the IGM Ly$\alpha$ opacity with the largest sample of quasars so far. They confirm the existence of tails towards high values in the Ly$\alpha$ opacity distributions, which may persist down to $z\approx5.2$. They find a linear increase in the mean Ly$\alpha$ opacity from $\approx1.8$ at $z\approx5$ to $\approx3.8$ at $z\approx6$. These results also imply a late and/or patchy reionization scenario, in which reionization ends at $z\approx5.2$--$5.3$ \citep[e.g.,][see also \citealt{Kashino2019arXiv}]{Kulkarni2019,Keating2019}. The Gunn-Peterson absorption trough in quasar spectra is only sensitive to a low Ly$\alpha$ opacity (very low \HI\ gas fraction) and the LAE fraction is therefore a complementary tool. In the near future, the James Web Space Telescope (JWST)/Near Infrared Spectrograph (NIRspec)  will enable us to observe Ly$\alpha$ emission at $z\approx5$ to $z\gtrsim10$ homogeneously and help make significant progress. We can also use the WST/Near Infrared Imager and Slitless Spectrograph (NIRISS) for Ly$\alpha$ spectroscopy. Most importantly, one can measure H$\alpha$ emission at $z\approx0$ to $z\approx7$ with JWST/NIRspec, and subsequently the line ratio of Ly$\alpha$ to H$\alpha$ to disentangle between intrinsic evolution and escape fraction.

Another important point raised in this paper is that $X_{\rm LAE}$ estimates are sensitive to $EW(\rm Ly\alpha)$ selections (see Figures \ref{fig:xevo} and \ref{fig:zevocompare}). Although a general consensus seems to emerge from previous work, a quantitative interpretation of the evolution of $X_{\rm LAE}$ with redshift requires more accurately constructed samples. In addition, the contribution of extended Ly$\alpha$ emission to the total Ly$\alpha$ budget is typically large and has a large scatter \citep[typically more than $\approx$50\%,][]{Momose2016, Leclercq2017}. Methods for measuring Ly$\alpha$ flux have a large effect on $EW(\rm Ly\alpha)$ and then $X_{\rm LAE}$. It means that accurate and homogeneous measurements of Ly$\alpha$ emission are required to use $X_{\rm LAE}$ as a tracer of the \HI\ gas fraction of the IGM. In Figures \ref{fig:uvdep} and \ref{fig:uvdepcompare}, the $X_{\rm LAE}$-$z$ relation does not depend strongly on the rest-frame UV absolute magnitude. This suggests that at $z<6$, combining UV-bright and faint samples can give us better statistics for $X_{\rm LAE}$ measurements. In addition, as discussed in Sect. \ref{subsubsec:lbgbias}, a firm definition of parent samples avoiding a selection bias is also required to assess the evolution of $X_{\rm LAE}$. The uncertainties of $X_{\rm LAE}$ have to be calculated with BPCI (Bernoulli trials, see Sect. \ref{subsec:uncertainties}). Moreover, \citet{Mason2018} warn of the interpretation of the evolution of $X_{\rm LAE}$ with the same UV magnitude, since galaxies with the same UV magnitude have very different stellar and halo masses at different redshifts \citep[e.g.,][]{Speagle2014a, Behroozi2013}. Because of such effects, sophisticated models of galaxy formation are needed to robustly interpret variations of $X_{\rm LAE}$ with cosmic time. 
We propose a method using a UV complete sample including faint galaxies, based on a photo-$z$ selection and an absolute magnitude cut, together with Ly$\alpha$ measurements by an IFU with a high sensitivity and a wide-wavelength coverage in a large field-of-view like VLT/MUSE and VLT/BlueMUSE \citep{Richard2019BlueMUSE} at $z\approx2$--$6.6$.

%===========

%%%
%%%

\section{Summary and Conclusions}\label{sec:conclusions}
We have investigated the LAE fraction at $z\approx3$--$6$ using the second data release of the  MUSE {\it Hubble} Ultra Deep Field Survey and the HST catalog of the UVUDF. Thanks to the unprecedented depth of the MUSE and HST data for Ly$\alpha$ and UV, respectively, we have studied the LAE fraction for galaxies as faint as $M_{1500}=-17.0$ mag at $z\approx3$ for the first time with a UV-complete sample. We have also derived the LAE fraction as a function of redshift homogeneously from $z=3$ to $6$, down to $M_{1500}=-17.75$ mag. Our results are summarized as follows: 

\begin{enumerate}
\item We derived the redshift evolution of $X_{\rm LAE}$ for a number of EW and UV magnitude selections, including the first estimate down to $-17.75$ mag. These results are summarised in Table \ref{tbl:x_z}. For all selections, we find low values of $X_{\rm LAE}$ $\approx0.04$-$0.3$, and a weak rise of $X_{\rm LAE}$ with $z$, qualitatively consistent with the trend reported for brighter samples in the literature.

\item We compared our MUSE results with those in the literature for $M_{1500}\in [-20.25;-18.75]$. At $z\lesssim5$, our values of $X_{\rm LAE}$ are consistent with those in \citet{ArrabalHaro2018} and \citet{Stark2011} within $1\sigma$ error bars for $EW(\rm Ly\alpha)\geq25$ \AA{} (see left panel of Figure\,\ref{fig:zevocompare}). Our $X_{\rm LAE}$ at $z\approx5.6$ is lower than those in the literature, which may be caused by statistical errors, or a late and/or patchy reionization process.

\item We measured the dependence of $X_{\rm LAE}$ on $M_{1500}$ at $z=2.9$--$3.7$ for $EW(\rm Ly\alpha)\geq25$ \AA, 45 \AA, 65 \AA, and 85 \AA{} (see Figure\,\ref{fig:uvdep}). This is the first time this has been measured down to $M_{1500}=-17.0$ mag (for the largest EWs of our sample), and for a volume-limited sample. We found no clear dependence of $X_{\rm LAE}$ on $M_{1500}$, in contrast to previous reports. 

\item We compared the dependence of $X_{\rm LAE}$ on $M_{1500}$ for $EW(\rm Ly\alpha)\geq50$ \AA\ at $z\approx3$-$4$ derived from MUSE with results from the literature (Figure\,\ref{fig:uvdepcompare}). Again we found no dependence of $X_{\rm LAE}$ on $M_{1500}$. Our slopes of $0.01^{+0.02}_{-0.03}$ and $0.01^{+0.06}_{-0.10}$ at $z\approx2.9$--$3.7$ and at $\approx3.7$--$4.4$, respectively, are shallower than that in \citet{Stark2010}, $0.13^{+0.03}_{-0.03}$ at $\approx3.5$--$4.5$. We also found lower values of our $X_{\rm LAE}$ at a faint UV magnitude of $M_{1500}\gtrsim-19$ mag. 

\item The dominant causes of the difference of $X_{\rm LAE}$ in our work and previous studies appear to be LBG selection biases in those studies. We showed how these can lead to an over-estimate by a factor $\approx1.5$ of $X_{\rm LAE}$ at $z\approx4$ for galaxies with $M_{1500}=-19$ mag and $EW(\rm Ly\alpha)\geq50$ \AA.

\item We compared our MUSE results with predictions from a cosmological semi-analytic galaxy evolution model \citep[GALICS,][]{Garel2015}. When GALICS uses a bursty star formation model, it can reproduce our measurement of $X_{\rm LAE}$ as a function of $M_{1500}$ at $z\approx3$. The fiducial GALICS model however cannot. The bursty model can also reproduce $X_{\rm LAE}$ as a function of $z$ at $z\lesssim4$. We assessed cosmic variance for our MUSE results using the bursty SF model and found that it does not have a significant effect on our results.

\item Overall, we found that $X_{\rm LAE}$ is lower than $\approx30$\%. This implies a low duty cycle of LAEs, suggesting bursty star formation or strong time variations in the production of Ly$\alpha$ photons and/or in their escape fraction.

\end{enumerate}

Despite the difficulties of the method, the dominant source of uncertainties in our work is the Poisson noise due to the small number of objects in our samples. This is encouraging and suggests that future deep surveys with e.g., MUSE and JWST will enable us to produce accurate measurements of $X_{\rm LAE}$ with secure samples and to extend our understanding of the evolution of $X_{\rm LAE}$ at all redshifts, after and during the epoch of reionization.

\begin{acknowledgements}
We thank the anonymous referee for constructive comments and suggestions. We would like to express our gratitude to Stephane De Barros and Pablo Arrabal Haro for kindly providing their data plotted in Figures \ref{fig:muv_z}, \ref{fig:n_muv},  and \ref{fig:zevocompare}. We are grateful to Pascal Oesch, Kazuhiro Shimasaku, Masami Ouchi, Rieko Momose, Daniel Schaerer, Hidenobu Yajima, Taku Okamura, Makoto Ando, and Hinako Goto for giving insightful comments and suggestions. This work is based on observations taken by VLT, which is operated by European Southern Observatory. This research made use of Astropy\footnote{\url{http://www.astropy.org}}, which is a community-developed core Python package for Astronomy \citep{TheAstropyCollaboration2013,TheAstropyCollaboration2018}, \textsf{MARZ}, \textsf{MPDAF}, and \textsf{matplotlib} \citep{Hunter2007}. H.K. acknowledges support from Japan Society for the Promotion of Science (JSPS) through the JSPS Research Fellowship for Young Scientists and Overseas Challenge Program for Young Researchers. This work was supported by the project FOGHAR (Agence Nationale de la Recherche, ANR-13-BS05-0010-02). JB acknowledges support from the ORAGE project from the Agence Nationale de la Recherche under grant ANR-14-CE33-0016-03. JR acknowledges support from the ERC starting grant 336736-CALENDS. T. H. acknowledges supports by the Grant-inAid for Scientic Research 19J01620. 

\end{acknowledgements}

\begin{appendix}
\section{Uncertainties of $z_{\rm p}$}\label{ap:photoz}
\subsection{Impact of lacking IRAC data on $z_{\rm p}$ estimation in \citet{Rafelski2015}}\label{ap:irac}
It is well known that LBG samples and photo-$z$ samples at $z\approx3$--$7$ can be contaminated by lower-$z$ galaxies with a $3646$ \AA\ Balmer or $4000$ \AA\ break at $z\approx0$--$1$, especially at faint magnitude. Spitzer/IRAC data can provide rest-frame optical data which are useful to break the degeneracy of $z_{\rm p}$ \citep[e.g.,][]{Bradac2019}. In this work, we use the catalog from R15, where $z_{\rm p}$ are derived using HST data alone. The advantage of this choice is discussed in \citet{Brinchmann2017}, where they show that IRAC data can in fact worsen photo-$z$ performance for faint galaxies in the MUSE HUDF sample (see their Appendix A). It might be a reflection of the difficulty of providing reliable IRAC photometry for sources as faint as most of our sample.

As discussed in Sect. \ref{subsec:uncertainties}, the fraction of low-$z$ contaminants among R15 galaxies is found to be low within MUSE samples \citep[Figure 20 in][]{Inami2017}. To avoid contaminants due to poor photo-$z$ estimations, we apply an $S/N>2$ cut for our sample and then apply a stricter cut on $M_{1500}$ (see Sect. \ref{subsec:parent_sample} and Figure \ref{fig:muv_z} for more details).

\subsection{Effects of $z_{\rm p}$ errors on redshift binning}\label{ap:check_photoz}

To check the effect of the error on $z_{\rm p}$ on redshift binning, we compare the median of upper and lower $95\%$ errors of $z_{\rm p}$ to the half-width of redshift bins for $-20.25\leq M_{1500}\leq-18.75$ mag and $-18.75\leq M_{1500}\leq-17.75$ mag shown in Figures \ref{fig:xevo} and \ref{fig:zevocompare}. The upper (lower) $95\%$ errors are calculated from a difference between $95$\% upper (lower) limit of photo-$z$ and photo-$z$ where likelihood is maximized in BPZ. The results are summarized in Table \ref{tbl:zperror}. For $-20.25\leq M_{1500}\leq-18.75$ mag, the median values of the $95$\% $z_{\rm p}$ errors are smaller than the width of redshift bins at $z\approx3.3$ to $5.6$. For $-18.75\leq M_{1500}\leq-17.75$ mag, the errors are larger than those for the brighter $M_{1500}$, but the median of $95$\% errors are still smaller than or comparable to the width of redshift bins at the all over redshift range. Therefore, the $95$\% errors of $z_{\rm p}$ are typically smaller than the width of redshift bins in Figures \ref{fig:xevo} and \ref{fig:zevocompare}. Note that we include widths of the redshift bins in the linear relation fitting (see Sect. \ref{subsec:slope}).

Next, we check the fraction of possible low-$z$ contaminants, $f_{\rm large\ error}$. Although the probability distribution functions (PDFs) of $z_{\rm p}$ in R15 catalog are not published, galaxies with a bimodal PDF of $z_{\rm p}$ show a large lower $95$\% error, which reaches at $z\approx0$--$1$ for a sample at $z\approx3$--$7$. We calculate $f_{\rm large\ error}$ from the wavelengths of breaks, $z_{\rm p}$, and its $95$\% errors and summarize the results in Table \ref{tbl:zperror}. 
For $-20.25\leq M_{1500}\leq-18.75$ mag, $f_{\rm large\ error}$ is $0.02$ to $0.09$, implying that our $X_{\rm LAE}$ in Figure \ref{fig:zevocompare} is not affected significantly by contaminants. Meanwhile, for $-18.75\leq M_{1500}\leq-17.75$ mag, $f_{\rm large\ error}$ is $0.05$ at $z\approx3.3$ so that our $X_{\rm LAE}$ is not suppressed significantly by contaminants in Figure \ref{fig:uvdepcompare}. At $z\approx4.1$ to $5.6$, $f_{\rm large\ error}$ is relatively high, $0.13$ to $0.35$. However, these are conservative estimations of the upper limit of low-$z$ contaminated fraction, since not all of the galaxies with a large photo-$z$ error locate at $z\approx0$--$1$. In fact, our sample shows the maximum likelihood at $z\approx3$--$6$. Our $X_{\rm LAE}$ at $z\gtrsim4.1$ and $5.6$ in Figure \ref{fig:xevo} should not be affected significantly by low-$z$ contaminants significantly.

 \begin{table*}
\caption{ Comparison of the median of $95\%$ upper and lower errors of $z_{\rm p}$ to the half-width of redshift bins and the fraction of galaxies with a large lower error.}\label{tbl:zperror}
\centering
\begin{tabular}{ccccc}
\hline         
\noalign{\smallskip}
mean $z_{\rm p}$	& half-width of $z$ bins & median of $95$\% upper error & median of $95$\% lower error & $f_{\rm large\ error}$  \\
\noalign{\smallskip}
\hline
 \multicolumn{5}{c}{$-20.25\leq M_{1500}\leq-18.75$ mag} \\
\hline
$z\approx3.3$ & 0.39 & 0.19 & 0.20 & 0.02\\
$z\approx4.1$ & 0.38 & 0.24 & 0.25 & 0.05\\
$z\approx4.7$ & 0.29 & 0.28 & 0.27 & 0.0\\
$z\approx5.6$ & 0.56 & 0.30 & 0.32 & 0.09\\
\hline
 \multicolumn{5}{c}{$-18.75\leq M_{1500}\leq-17.75$ mag} \\
\hline
$z\approx3.3$ & 0.39 & 0.21 & 0.23 & 0.05\\
$z\approx4.1$ & 0.38 & 0.30 & 0.34 & 0.2\\
$z\approx4.7$ & 0.29 & 0.33 & 0.34 & 0.13\\
$z\approx5.6$ & 0.56 & 0.35 & 0.44 & 0.35\\
\hline
\end{tabular}
\tablefoot{ The mean redshift, the half-width of the redshift bin, the median of upper $95\%$ errors of $z_{\rm p}$, the median of lower $95\%$ errors, and the fraction of galaxies with a large lower error suggesting a bimodal $z_{\rm p}$ PDF are shown.} 
\end{table*}

\section{The template spectra used in \textsf{MARZ}}\label{ap:temp_marz} 
\begin{figure*}[h]
 \sidecaption
   \includegraphics[width=18cm]{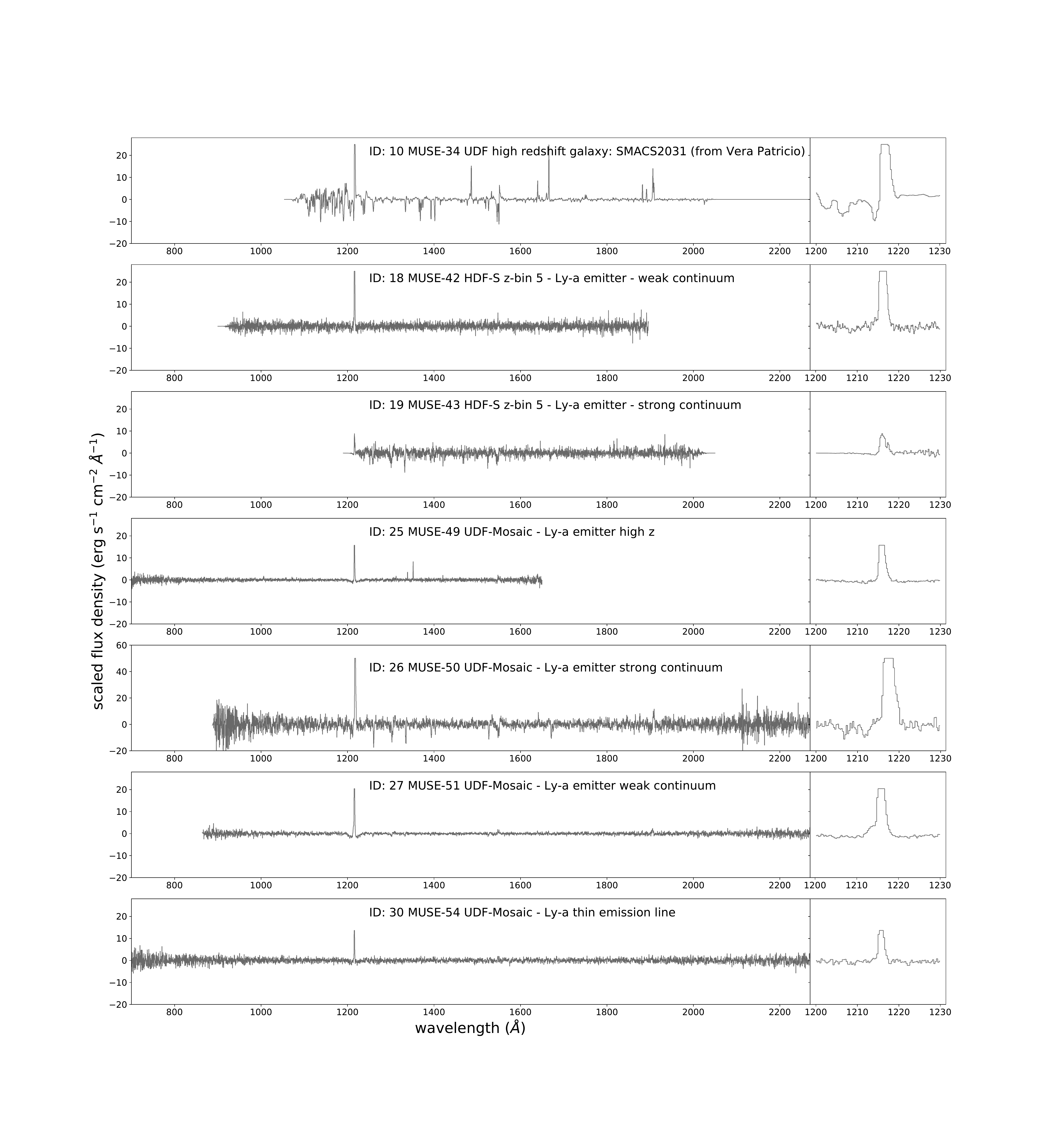}\\
   \vspace{-20pt}
      \caption{The LAE template spectra of \textsf{MARZ} used in this work: those of ID$=10$, $18$, and $19$ are used in \citet{Inami2017}, while those of ID$=25$, $26$, $27$, and $30$ are newly created from MUSE data (Bacon et al. in prep.).  Left panels: scaled spectra of the templates in the rest frame. Right panels: zooms of the Ly$\alpha$ emission line in each left panel.
      }
\label{fig:temp_marz}
\end{figure*}

Figure \ref{fig:temp_marz} shows the template spectra of LAEs in \textsf{MARZ} used in this work. The continuum are subtracted in the templates and the bright Ly$\alpha$ lines are clipped to lie between $-30$ and $+30$ times the mean absolute deviation in the similar manner to those used in the original \textsf{MARZ} \citep{Hinton2016} and \textsf{AUTOZ} \citep[][]{Baldry2014} as shown in the right panels in Figure \ref{fig:temp_marz}. Cross-correlation functions indicate locations of lines in the fitting range and are not affected significantly by the line shape of templates in general \citep[see Sect. 4.3 in][]{Herenz2017b}. In fact, the completeness of \textsf{MARZ} does not depend on the FWHM of fake lines in completeness simulations as we mention in Sect. \ref{subsec:compsimu}. Although our templates do not cover various types of Ly$\alpha$ lines, it does not have an effect on detection of Ly$\alpha$ emission.

\section{An example of contamination of Ly$\alpha$ detection}\label{ap:contami}
\begin{figure}[h]
 \sidecaption
   \includegraphics[width=9cm]{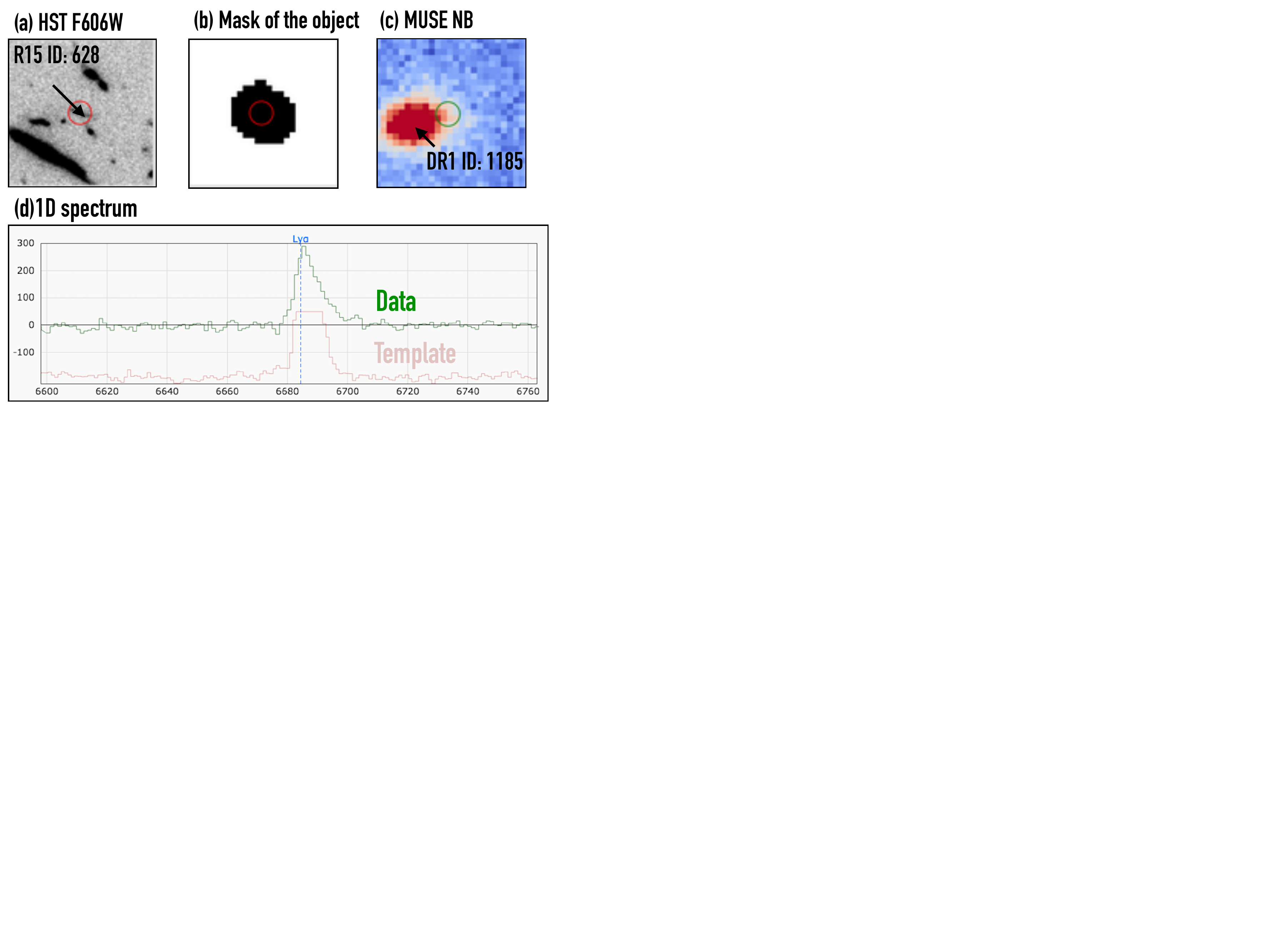}\\
      \caption{
      An example of contamination of Ly$\alpha$ emission from a neighboring object. Panels (a) - (d) show sub-panels in \textsf{MARZ}'s screen \citep{Inami2017} for a UV-selected source with \citet{Rafelski2015} ID $=628$: (a) HST F606W cutout, (b) mask of the object for the extraction of the 1D spectrum, (c) MUSE NB cutout, and (d) 1D spectrum. The red and green circle in the images show the position of the UV-selected galaxy. The green and red lines in panel (d) indicate observational data and the best fit template. 
      }
\label{fig:contami}
\end{figure}

Figure \ref{fig:contami} shows an example of contamination of Ly$\alpha$ emission from a neighboring object. Panels (a) - (d) show sub-panels in \textsf{MARZ} for a UV-selected source \citep[see][for more details of \textsf{MARZ}'s screen]{Inami2017}. The HST cutout around a UV-galaxy with \citet{Rafelski2015} ID $=628$ is shown in panel (a). The 1D spectrum shown in panel (d) clearly shows a strong Ly$\alpha$ emission line with the highest confidence level by \textsf{MARZ}. The spectrum is extracted from the object mask (panel (b)), and clearly contaminated by diffuse Ly$\alpha$ emission from a neighboring object (MUSE ID $=1185$ in the DR1 catalog) as shown in the narrow band in panel (c). We can remove these contaminated objects from our sample of Ly$\alpha$ emitter candidates in visual inspection.

\section{The number of flux bins used to correct incompleteness of Ly$\alpha$ detection}\label{ap:bin_comp}

\begin{figure}
 \sidecaption
   \includegraphics[width=12cm]{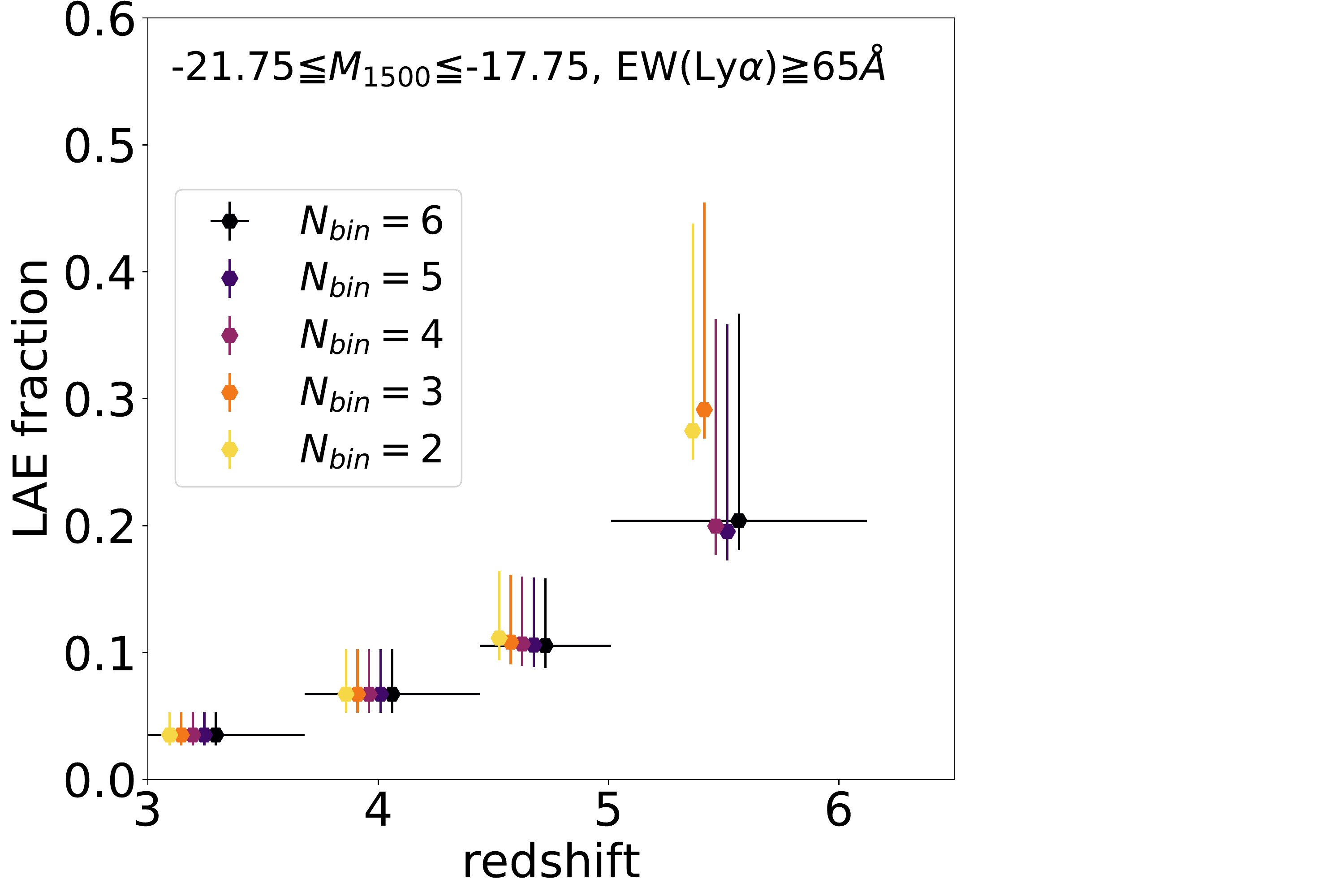}\\
      \caption{A test for the effect of binning of Ly$\alpha$ flux to correct incompleteness of the number of LAEs. The median values and error bars of $X_{\rm LAE}$ for the plot of the evolution of $X_{\rm LAE}$ (Figure \ref{fig:xevo}) are shown. The black, purple, violet, orange, and yellow hexagons indicate $N_{\rm bin}=6$, $5$, $4$, $3$, and $2$, respectively. For visualization purposes, we slightly shift the points along the $x$-axis and show the width of $z$ only for $N_{\rm bin}=6$ .
      }
\label{fig:binning}
\end{figure}

We also examine the effect of binning to correct incompleteness of the number of LAEs. Because the number of objects we have is small, somewhat arbitrary binning may affect $X_{\rm LAE}$. For a large number of bins, we often get flux-bins with no object at all, and these bins increase the error bars. For a small number of bins, we introduce a large error on the completeness correction as described above. To test the effect of binning, we vary the number of bins ($N_{\rm bin}$) we use, from $2$ to $6$, and see how the median values and error bars change. In Figure \ref{fig:binning}, we show the median values and error bars of $X_{\rm LAE}$ for $N_{\rm bin}=2$--$6$ for the plot of the evolution of $X_{\rm LAE}$ (Figure\ref{fig:xevo}). The uncertainties of the completeness correction are $20$\%, $25$\%, $32$\%, $44$\%, and $73$\% for $N_{\rm bin}=6$, $5$, $4$, $3$, and $2$ (see Sect. \ref{subsec:uncertainties}).  We find that $4$ to $6$ bins are a sweet spot where the error bars are small and appear converged and adopt $N_{\rm bin}=4$ in Sect. \ref{subsec:uncertainties}.

\section{Error propagation of completeness correction values in a binomial proportion confidence interval}\label{ap:bpci}

We test the applicability of a binomial proportion confidence interval (BPCI) for the case of a completeness correction to calculate error bars of the LAE fraction. First, we examine how to propagate completeness correction values in the error calculation with BPCI numerically. To do mock observations, we generate $N_{\rm LAE}^{\rm true}(z_{\rm s},\,M_{1500},\, EW)$ randomly with $100000$ trials using the python module \textsf{numpy.random.binomial} for each $N_{\rm 1500}(z_{\rm p},\, M_{1500})$ and the true LAE fraction ($X_{\rm LAE}^{\rm true}$). Then we generate $N_{\rm LAE}^{\rm det}(z_{\rm s},\,M_{1500},\, EW)$ randomly for each $N_{\rm LAE}^{\rm true}(z_{\rm s},\,M_{1500},\, EW)$ again using \textsf{numpy.random.binomial} for a given completeness correction value as a probability of detection. We can obtain the probability distribution of $X_{\rm LAE}$ with a given completeness correction value, $N_{\rm 1500}(z_{\rm p},\, M_{1500})$, and $X_{\rm LAE}^{\rm true}$. We compare the $1\sigma$ upper and lower uncertainties from the probability distribution function of the mock $X_{\rm LAE}$ with those derived from two methods of BPCI (\textsf{binom\_conf\_interval}) with input parameters of the number of LAEs and $N_{\rm 1500}(z_{\rm p},\, M_{1500})$ if the incompleteness of the number of LAEs are corrected ($N_{\rm LAE}^{\rm true}(z_{\rm s},\,M_{1500},\, EW)$) or not ($N_{\rm LAE}^{\rm det}(z_{\rm s},\,M_{1500},\, EW)$). We confirm that it is better to input $N_{\rm LAE}^{\rm det}(z_{\rm s},\,M_{1500},\, EW)$ and to multiply the obtained uncertainties by the correction value (see also Sect. \ref{subsec:uncertainties}). 

\begin{figure}
 \sidecaption
   \includegraphics[width=9cm]{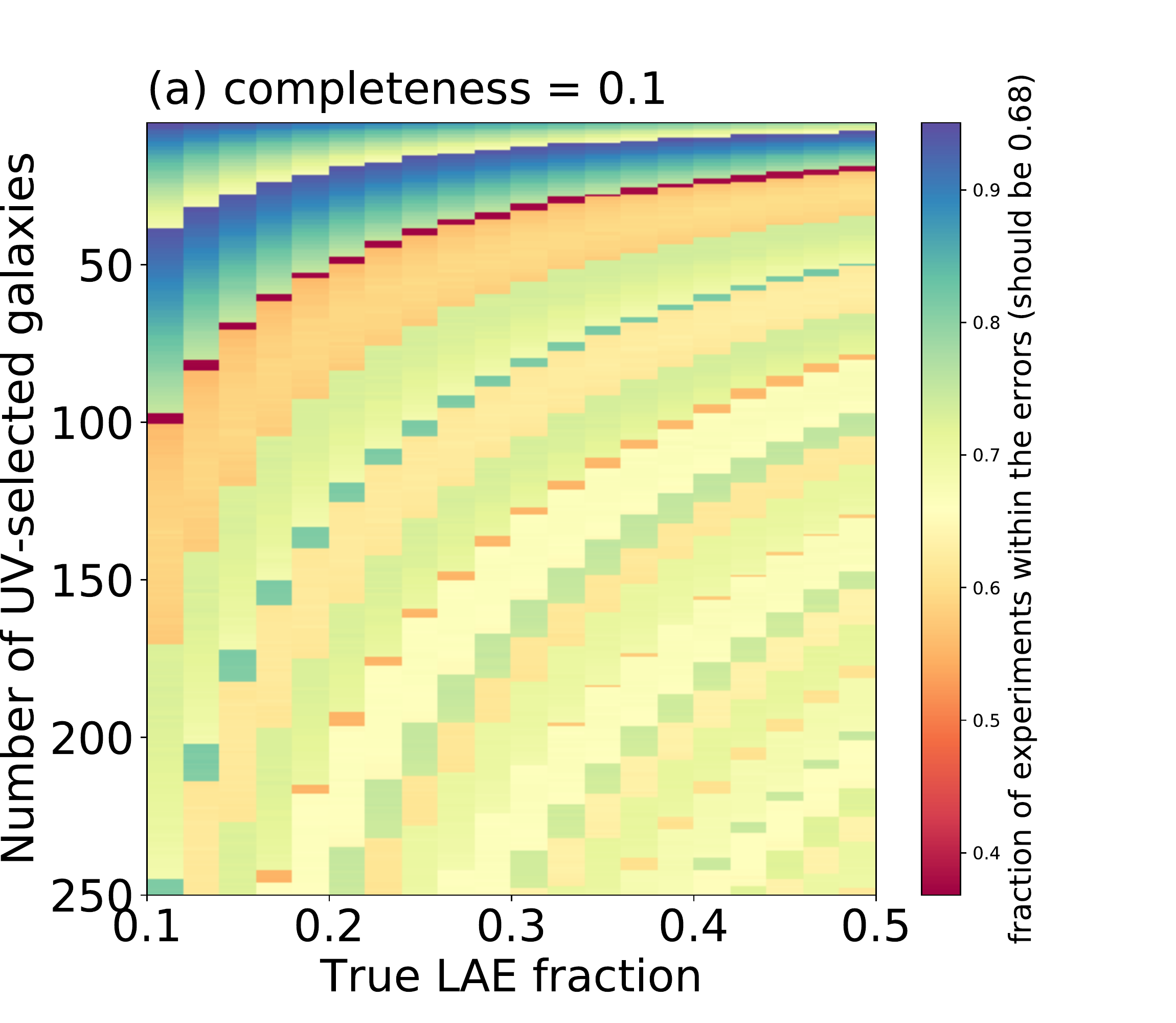}\\
   \includegraphics[width=9cm]{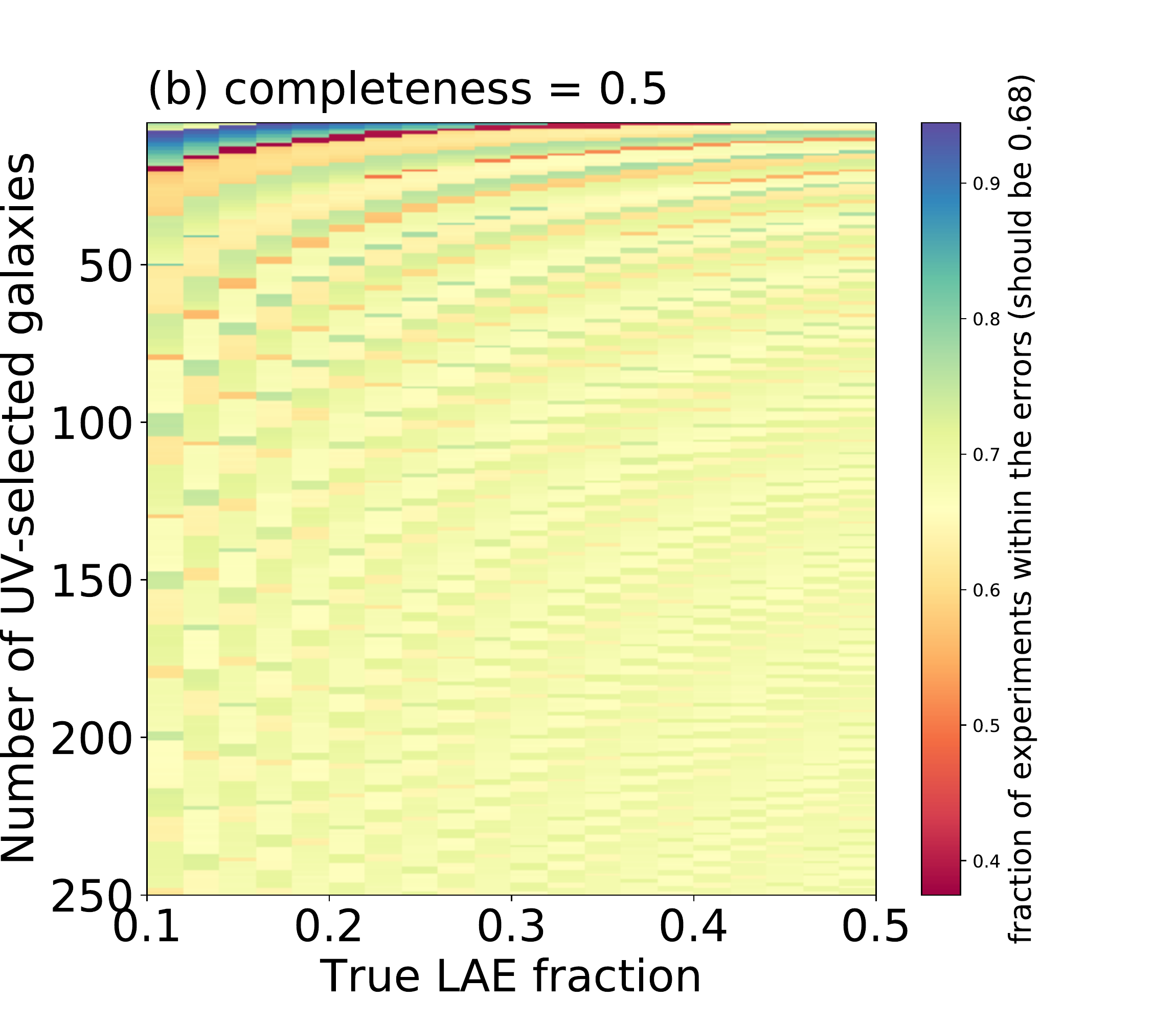}\\   
      \caption{
      Test of the accuracy of our uncertainty estimation of $X_{\rm LAE}$. We generate mock $X_{\rm LAE}$ distribution numerically for each $N_{\rm 1500}(z_{\rm p},\, M_{1500})$ and $X_{\rm LAE}^{\rm true}$ with a given completeness of $0.1$ in panel (a) and $0.5$ in panel (b). We then derive the fraction of experiments within our upper and lower limits calculated with BPCI. The colors encode the fraction of the experiment for each $N_{\rm 1500}(z_{\rm p},\, M_{1500})$ and $X_{\rm LAE}^{\rm true}$ in the x and y axes, respectively. 
      }
\label{fig:bpci_test}
\end{figure}

Second, we check the accuracy of the method above in the plane of $N_{\rm 1500}(z_{\rm p},\, M_{1500})$ and $X_{\rm LAE}^{\rm true}$ for a given completeness correction value. We calculate the $1\sigma$ upper and lower limits of $X_{\rm LAE}$ with this method. Again we generate $100000$ mock values of $N_{\rm LAE}^{\rm det}(z_{\rm s},\,M_{1500},\, EW)$ and then $X_{\rm LAE}$ numerically. The fractions of the mock $X_{\rm LAE}$ within the range of the $1\sigma$ upper and lower limits among all the experiments are calculated. If the method is accurate, this fraction would be $\approx68$\%. We check the fraction of experiments for $N_{\rm 1500}(z_{\rm p},\, M_{1500})$ from $0$ to $250$ with a step of $1$ and for $X_{\rm LAE}^{\rm true}$ from $0.1$ to $0.5$ with a step of $0.02$. In Figure \ref{fig:bpci_test}, the example with low completeness correction values, $0.1$ and $0.5$, is shown. In the panel (a) for completeness correction $=0.1$, most of the plane is colored with yellow green corresponding to a fraction close to $0.68$, while it is colored with blue or red corresponding to overestimation and underestimation of the errors, respectively, for cases with poor statistics (i.e., low $N_{\rm 1500}(z_{\rm p},\, M_{1500})$ and $X_{\rm LAE}^{\rm true}$). In panel (b) for completeness correction $=0.5$, the method is found to produce errors accurately except for the very poor statistics cases (at the upper left region in the panel). We can estimate the error more accurately with this method for a higher completeness case. Even with a low completeness and smallest numbers, the uncertainties are overestimated. Therefore, we adopt this method conservatively.

\section{The best fit linear relations of $X_{\rm LAE}$ as a function of $z$ and $M_{1500}$}\label{ap:xlae_fit}
\begin{figure}
 \sidecaption
   \includegraphics[width=8cm]{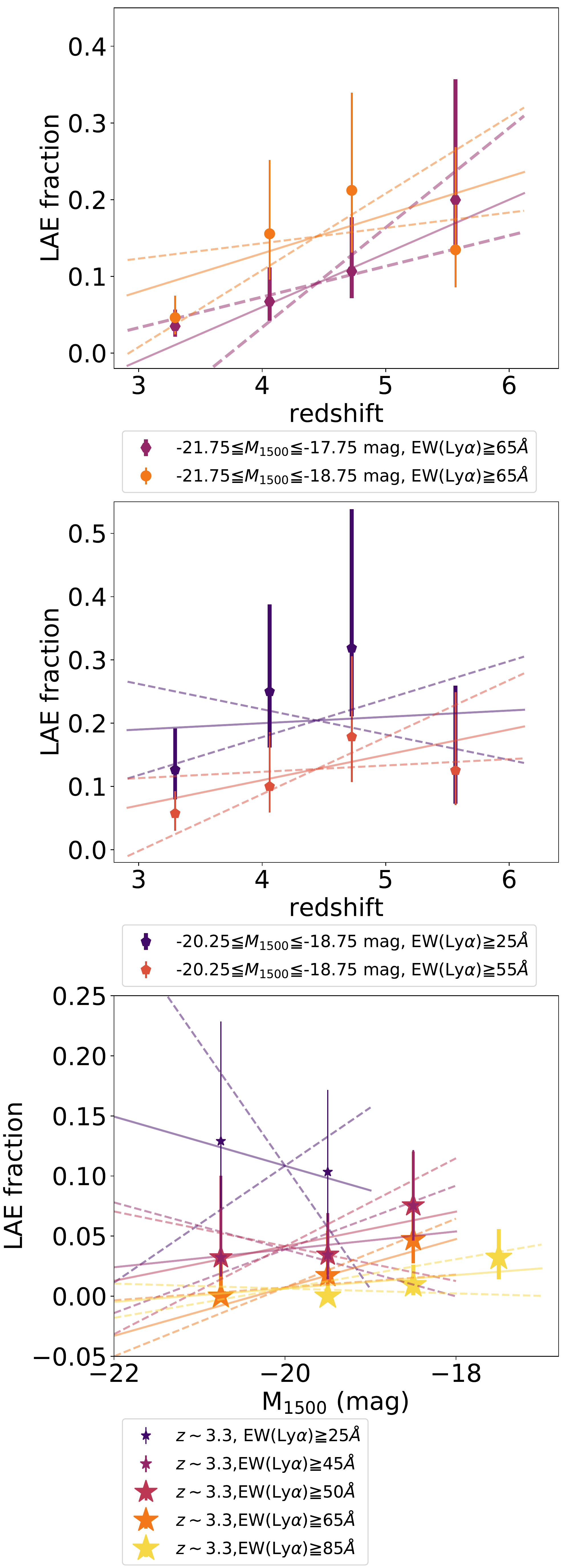}\\
      \caption{ The slopes of the best fit linear relations of of $X_{\rm LAE}$ as a function of $z$ (top and middle panels) and $M_{1500}$ (bottom panel). Symbols are the same as those in Figures \ref{fig:xevo}, \ref{fig:zevocompare}, \ref{fig:uvdep}, and \ref{fig:uvdepcompare}. The best fit linear relations and the $\pm1\sigma$ slopes are shown by the solid and dashed lines, respectively, with lighter colors of those for the symbols. 
      }
\label{fig:xlae_fit}
\end{figure}

 We show the best fit linear relations of $X_{\rm LAE}$ as a function of $z$ and $M_{1500}$ in Figure \ref{fig:xlae_fit}. The equations of relations are shown in Sects. \ref{subsec:zevo} and \ref{subsec:uvdep}. 

\end{appendix}

\bibliographystyle{aa}

\end{document}